\documentclass[11pt]{article}

\usepackage{multirow,booktabs,threeparttable,color}
\usepackage{setspace,graphicx,epstopdf,amsmath,amsfonts,amssymb,amsthm,subcaption}
\usepackage{verbatim}
\usepackage{marginnote,datetime,enumitem,rotating,fancyvrb}
\usepackage{hyperref}
\usdate
\usepackage{adjustbox}
\usepackage{siunitx}
\usepackage{float}
\usepackage[longnamesfirst]{natbib}

\usepackage{amssymb}
\usepackage{amsmath}
\usepackage{bm}
\usepackage{bbm}
\usepackage{float}
\usepackage{fancyhdr}
\usepackage{booktabs}
\usepackage{graphicx,rotating}
\usepackage{color}
\usepackage{setspace}
\usepackage{enumitem}
\usepackage{pdfpages}

\usepackage{caption}
\usepackage{subcaption}
\usepackage{appendix}
\usepackage[margin=1in]{geometry}%

\usepackage[verbose]{placeins}

\usepackage[normalem]{ulem}
\useunder{\uline}{\ul}{}

\usepackage{indentfirst} 
\usepackage{endnotes}    
\usepackage{jf}          
\usepackage[labelfont=bf,labelsep=period]{caption}   
\usepackage{ragged2e}
\usepackage{booktabs, ltablex, makecell, tabularx}

\newcolumntype{R}{>{\RaggedRight}X}

\newcommand{\E}{\mathbbm E_{t}}
\newcommand{\R}{{R^e_{t+1}}}
\newcommand{\Ri}{{R^e_{t+1,i}}}
\newcommand{\Rj}{{R^e_{t+1,j}}}
\newcommand{\var}{ \text{Var}_{t}}
\newcommand{\cov}{ \text{Cov}_{t}}
\newcommand{\M}{M_{t+1}}
\newcommand{\Err}{\mathbbm E_{t}[\R \R^{\top}]}
\newcommand{\bi}{\beta_{t,i}}
\newcommand{\ei}{\epsilon_{t+1,i}}
\newcommand{\w}{\omega_{t}}
\newcommand{\wi}{\omega_{t,i}}
\newcommand{\sig}{{\Sigma^{\epsilon}_t}}
\newcommand{\be}{\beta_{t}}

\newcommand{\Nb}{{\tilde N^{\text{basis}}}}
\newcommand{\Et}{\mathbbm E_{t}}

\newcommand{\Rt}{{R^e_{t}}}

\usepackage[margin=10pt,labelfont=bf]{caption}
\newcommand\tcaptab[1]{\captionsetup{position=top, font=normalsize, labelfont=bf, textfont=normalfont, justification=centering, margin=0mm, aboveskip=1mm, belowskip=0mm, labelsep=colon, singlelinecheck=false}\caption{#1}}
\newcommand\bnotetab[1]{\captionsetup{position=bottom, font=footnotesize,  textfont=normalfont, margin=1mm, skip=2mm, justification=justified, singlelinecheck=false}\caption*{#1}}
\newcommand\tcapfig[1]{\captionsetup{position=top, font=normalsize, labelfont=bf, textfont=normalfont, justification=centering, margin=0mm, aboveskip=2mm, belowskip=0mm, labelsep=colon, singlelinecheck=false}\caption{#1}}
\newcommand\bnotefig[1]{\captionsetup{position=bottom, font=footnotesize,  textfont=normalfont, margin=1mm, skip=2mm, justification=justified, singlelinecheck=false}\caption*{#1}}
\newcommand\subcap[1]{\captionsetup{position=bottom, font=small, labelfont=bf, textfont=normalfont, justification=RaggedRight, margin=0mm, aboveskip=-5mm, belowskip=0mm, labelsep=space, singlelinecheck=false}\caption{#1}}
\newcommand\subcaptab[1]{\captionsetup{position=bottom, font=small, labelfont=bf, textfont=normalfont, justification=centering, margin=0mm, aboveskip=4mm, belowskip=0mm, labelsep=space, singlelinecheck=false}\caption{#1}}

\setlist{noitemsep}

\begin{document}

\title{Deep Learning in Asset Pricing\thanks{\scriptsize We thank Doron Avramov, Ravi Bansal, Daniele Bianchi (discussant), Svetlana Bryzgalova, Agostino Capponi, Xiaohong Chen, Anna Cieslak, John Cochrane, Lin William Cong, Victor DeMiguel, Jens Dick-Nielsen (discussant), Kay Giesecke, Stefano Giglio, Goutham Gopalakrishna (discussant), Robert Hodrick, Bryan Kelly (discussant), Serhiy Kozak, Martin Lettau, Anton Lines, Marcelo Medeiros (discussant), Scott Murray (discussant), Stefan Nagel, Andreas Neuhierl (discussant), Kyoungwon Seo (discussant), Gustavo Schwenkler, Neil Shephard and Guofu Zhou and seminar and conference participants at Yale SOM, Stanford, UC Berkeley, Washington University in St. Louis, Temple University, Imperial College London, University of Zurich, UCLA, Bremen University, Santa Clara University, King's College London, the Utah Winter Finance Conference, the GSU-RSF FinTech Conference, the New Technology in Finance Conference, the LBS Finance Summer Symposium, the Fourth International Workshop in Financial Econometrics, Triangle Macro-Finance Workshop, GEA Annual Meeting, the Western Mathematical Finance Conference, INFORMS, SIAM Financial Mathematics, CMStatistics, Shanghai Edinburgh Fintech Conference, Annual NLP and Machine Learning in Investment Management Conference,  Annual Conference on Asia-Pacific Financial Markets, European Winter Meeting of Econometric Society, AI in Asset Management Day, Winter Research Conference on Machine Learning and Business, French Association of Asset and Liability Manager Conference, Midwest Finance Association Annual Meeting, Annual Meeting of the Swiss Society for Financial Market Research, Society for Financial Econometrics Annual Conference and China Meeting of the Econometric Society for helpful comments. We thank the China Merchants Bank for generous research support. We gratefully acknowledge the best paper awards at the Utah Winter Finance Conference and Asia-Pacific Financial Markets Conference, the 2nd place at the CQA Academic Paper Competition and the honorable mention at the AQR Capital Insight Award.}}
\date{\today}
\author{Luyang Chen\thanks{\scriptsize Stanford University, Institute for Computational and Mathematical Engineering, Email: lych@alumni.stanford.edu.}
\and
Markus Pelger\thanks{\scriptsize Stanford University, Department of Management Science \& Engineering, Email: mpelger@stanford.edu.}
\and
Jason Zhu\thanks{\scriptsize Stanford University, Department of Management Science \& Engineering, Email: jzhu121@stanford.edu.}}

\onehalfspacing

\begin{titlepage}
\maketitle
\thispagestyle{empty}


\begin{abstract}

We use deep neural networks to estimate an asset pricing model for individual stock returns that takes advantage of the vast amount of conditioning information, while keeping a fully flexible form and accounting for time-variation. The key innovations are to use the fundamental no-arbitrage condition as criterion function, to construct the most informative test assets with an adversarial approach and to extract the states of the economy from many macroeconomic time series. Our asset pricing model outperforms out-of-sample all benchmark approaches in terms of Sharpe ratio, explained variation and pricing errors and identifies the key factors that drive asset prices.

\vspace{1cm}


\end{abstract}
\end{titlepage}


The fundamental question in asset pricing is to explain differences in average returns of assets. No-arbitrage pricing theory provides a clear answer - expected returns differ because assets have different exposure to the stochastic discount factor (SDF) or pricing kernel. The empirical quest in asset pricing for the last 40 years has been to estimate a stochastic discount factor that can explain the expected returns of all assets. There are four major challenges that the literature so far has struggled to overcome in a unified framework: First, the SDF could by construction depend on all available information, which means that the SDF is a function of a potentially very large set of variables. Second, the functional form of the SDF is unknown and likely complex. Third, the SDF can have a complex dynamic structure and the risk exposure for individual assets can vary over time depending on economic conditions and changes in asset-specific attributes. Fourth, the risk premium of individual stocks has a low signal-to-noise ratio, which complicates the estimation of an SDF that explains the expected returns of all stocks.

In this paper we estimate a general non-linear asset pricing model with deep neural networks for all U.S. equity data based on a substantial set of macroeconomic and firm-specific information. Our crucial innovation is the use of the no-arbitrage condition as part of the neural network algorithm. We estimate the stochastic discount factor that explains all stock returns from the conditional moment constraints implied by no-arbitrage.
It is a natural idea to use machine learning techniques like deep neural networks to deal with the high dimensionality and complex functional dependencies of the problem. However, machine-learning tools are designed to work well for prediction tasks in a high signal-to-noise environment. As asset returns in efficient markets seem to be dominated by unforecastable news, it is hard to predict their risk premia with off-the-shelf methods. We show how to build better machine learning estimators by incorporating economic structure.
Including the no-arbitrage constraint in the learning algorithm significantly improves the risk premium signal and makes it possible to explain individual stock returns. Empirically, our general model outperforms out-of-sample the leading benchmark approaches and provides a clear insight into the structure of the pricing kernel and the sources of systematic risk.

Our model framework answers three conceptional key questions in asset pricing. (1) What is the functional form of the SDF based on the information set? Popular models, for example the Fama-French five-factor model, impose that the SDF depends linearly on a small number of characteristics. However, the linear model seems to be misspecified and the factor zoo suggests that there are many more characteristics with pricing information. Our model allows for a general non-parametric form with a large number of characteristics. (2) What are the right test assets? The conventional approach is to calibrate and evaluate asset pricing models on a small number of pre-specified test assets, for example the 25 size and book-to-market double-sorted portfolios of \cite{10.2307/2329112}. However, an asset pricing model that can explain well those 25 portfolios does not need to capture the pricing information of other characteristic sorted portfolios or individual stock returns. Our approach constructs in a data-driven way the most informative test assets that are the hardest to explain and identify the parameters of the SDF. (3) What are the states of the economy? Exposure and compensation for risk should depend on the economic conditions. A simple way to capture those would be for example NBER recession indicators. However, this is a very coarse set of information given the hundreds of macroeconomic time series with complex dynamics. Our model extracts a small number of state processes that are based on the complete dynamics of a large number of macroeconomic time series and are the most relevant for asset pricing.

Our estimation approach combines no-arbitrage pricing and three neural network structures in a novel way. Each network is responsible for solving one of the three key questions outlined above. First, we can explain the general functional form of the SDF as a function of the information set using a feedforward neural network. Second, we capture the time-variation of the SDF as a function of macroeconomic conditions with a recurrent Long-Short-Term-Memory (LSTM) network that identifies a small set of macroeconomic state processes. Third, a generative adversarial network constructs the test assets by identifying the portfolios and states with the most unexplained pricing information. These three networks are linked by the no-arbitrage condition that helps to separate the risk premium signal from the noise and serves as a regularization to identify the relevant pricing information.

Our paper makes several methodological contributions. 
First, we introduce a non-parametric adversarial estimation approach to finance and show that it can be interpreted as a data-driven way to construct informative test assets. Estimating the SDF from the fundamental no-arbitrage moment equation is conceptionally a generalized method of moment (GMM) problem. The conditional asset pricing moments imply an infinite number of moment conditions. Our generative adversarial approach provides a method to find and select the most relevant moment conditions from an infinite set of candidate moments. 
Second, we introduce a novel way to use neural networks to extract economic conditions from complex time series. We are the first to propose LSTM networks to summarizes the dynamics of a large number of macroeconomic time series in a small number of economic states. More specifically, our LSTM approach aggregates a large dimensional panel cross-sectionally in a small number of time-series and extracts from those a non-linear time-series model. The key element is that it can capture short and long-term dependencies which are necessary for detecting business cycles.
Third, we propose a problem formulation that can extract the risk premium in spite of its low signal-to-noise ratio. The no-arbitrage condition identifies the components of the pricing kernel that carry a high risk premia but have only a weak variation signal. Intuitively, most machine learning methods in finance\footnote{These models include \cite{gu2018}, \cite{messmer2017} or \cite{kelly2018}.} fit a model that can explain as much variation as possible, which is essentially a second moment object. The no-arbitrage condition is based on explaining the risk premia, which is based on a first moment. We can decompose stock returns into a predictable risk premium part and an unpredictable martingale component. Most of the variation is driven by the unpredictable component that does not carry a risk premium. When considering average returns the unpredictable component is diversified away over time and the predictable risk premium signal is strengthened. However, the risk premia of individual stocks is time-varying and an unconditional mean of stock returns might not capture the predictable component. Therefore, we consider unconditional means of stock returns instrumented with all possible combinations of firm-specific characteristics and macroeconomic information. This serves the purpose of pushing up the risk premium signal while taking into account the time-variation in the risk premium.

Our adversarial estimation approach is economically motivated by the seminal work of \cite{hansen1997}. They show that estimating an SDF that minimizes the largest possible pricing error is closest to an admissible true SDF in a least square distance. Our generative adversarial network builds on this idea and creates characteristic managed portfolios with the largest pricing errors for a candidate SDF which are then used to estimate a better SDF.  Our approach also builds on the insight of \cite{bansal1993no} who propose the use of neural networks as non-parametric estimators for the SDF from a given set of moment equations. Hence, the adversarial network element of our paper combines ideas of \cite{hansen1997} and \cite{bansal1993no}. 

Our empirical analysis is based on a data set of all available U.S. stocks from CRSP with monthly returns from 1967 to 2016 combined with 46 time-varying firm-specific characteristics and 178 macroeconomic time series. It includes the most relevant pricing anomalies and forecasting variables for the equity risk premium. 
Our approach outperforms out-of-sample all other benchmark approaches, which include linear models and deep neural networks that forecast risk premia instead of solving a GMM type problem. We compare the models out-of-sample with respect to the Sharpe ratio implied by the pricing kernel, the explained variation and explained average returns of individual stocks. Our model has an annual out-of-sample Sharpe ratio of 2.6 compared to 1.7 for the linear special case of our model, 1.5 for the deep learning forecasting approach and 0.8 for the Fama-French five-factor model. At the same time we can explain 8\% of the variation of individual stock returns and explain 23\% of the expected returns of individual stocks, which is substantially higher than the other benchmark models. On standard test assets based on single- and double-sorted anomaly portfolios our asset pricing model reveals an unprecedented pricing performance. In fact, on all 46 anomaly sorted decile portfolios we achieve a cross-sectional $R^2$ higher than 90\%. 


Our empirical main findings are five-fold. 
First, economic constraints improve flexible machine learning models. We confirm \cite{gu2018}'s insight that deep neural network can explain more structure in stock returns because of their ability to fit flexible functional forms with many covariates. However, when used for asset pricing, off-the-shelf simple prediction approaches can perform worse than even linear no-arbitrage models. It is the crucial innovation to incorporate the economic constraint in the learning algorithm that allows us to detect the underlying SDF structure. Although our estimation is only based on the fundamental no-arbitrage moments, our model can explain more variation out-of-sample than a comparable model with the objective to maximize explained variation. This illustrates that the no-arbitrage condition disciplines the model and yields better results among several dimensions.

Second, we confirm that non-linear and interaction effects matter as pointed out among others by \cite{gu2018} and \cite{bryzgalova2019}. Our finding is more subtle and also explains why linear models, which are the workhorse models in asset pricing, perform so well. We find that when considering firm-specific characteristics in isolation, the SDF depends approximately linearly on most characteristics. Thus, specific linear risk factors work well on certain single-sorted portfolios. The strength of the flexible functional form of deep neural networks reveals itself when considering the interaction between several characteristics. Although in isolation firm characteristics have a close to linear effect on the SDF, the multi-dimensional functional form is complex. Linear models and also non-linear models that assume an additive structure in the characteristics (for example, additive splines or kernels) rule out interaction effects and cannot capture this structure.

Third, test assets matter. Even a flexible asset pricing model can only capture the asset pricing information that is included in the test assets that it is calibrated on. An asset pricing model estimated on the optimal test assets constructed by the adversarial network has a 20\% higher Sharpe ratio than one calibrated on individual stock returns without characteristic managed portfolios. By explaining the most informative test assets we achieve a superior pricing performance on conventional sorted portfolios,  e.g. size and book-to-market single- or double-sorted portfolios. In fact, our model has an excellent pricing performance on all 46 anomaly sorted decile portfolios. 

Fourth, macroeconomic states matter. Macroeconomic time series data have a low dimensional ``factor'' structure, which can be captured by four hidden state processes. The SDF structure depends on these economic states that are closely related to business cycles and times of economic crises. In order to find these states we need to take into account the full time series dynamics of all macroeconomic variables. The conventional approach to deal with non-stationary macroeconomic time series is to use differenced data that capture changes in the time series. However, using only the last change as an input loses all dynamic information and renders the macroeconomic time series essentially useless. Even worse, prediction based on only the last change in a large panel of macroeconomic variables leads to worse performance than leaving them out overall, because they have lost most of their informational content and thus making it harder to separate the signal from the noise. 

Fifth, our conceptional framework is complementary to multi-factor models. Multi-factor models are based on the assumption that the SDF is spanned by the those factors. We provide a unified framework to construct the SDF in a conditional multi-factor model, which in general does not coincide with the unconditional mean-variance efficient combination of the factors. 
We combine the general conditional multi-factor model of \cite{kelly2018} with our model. We show that using the additional economic structure of spanning the SDF with IPCA factors and combining it with our SDF framework can lead to an even better asset pricing model.


Our findings are robust to the time periods under consideration, small capitalization stocks, the choice of the tuning parameters, and limits to arbitrage. The SDF structure is surprisingly stable over time. We estimate the functional form of the SDF with the data from 1967 to 1986, which has an excellent out-of-sample performance for the test data from 1992 to 2016. The risk exposure to the SDF for individual stocks varies over time because the firm-specific characteristics and macroeconomic variables are time-varying, but the functional form of the SDF and the risk exposure with respect to these covariates does not change. When, allowing for a time-varying functional form by estimating the SDF on a rolling window, we find that it is highly correlated with the benchmark SDF and only leads to minor improvements. The estimation is robust to the choice of the tuning parameters. All of the best performing models selected on the validation data capture essentially the same asset pricing model. Our asset pricing model also performs well after excluding small and illiquid stocks from the test assets or the SDF.

 \subsection*{Related Literature}

Our paper contributes to an emerging literature that uses machine learning methods for asset pricing. In their pioneering work \cite{gu2018} conduct a comparison of machine learning methods for predicting the panel of individual US stock returns and demonstrate the benefits of flexible methods. Their estimates of the expected risk premia of stocks map into a cross-sectional asset pricing model. We use their best prediction model based on deep neural networks as a benchmark model in our analysis. We show that including the no-arbitrage constraint leads to better results for asset pricing and explained variation than a simple prediction approach. Furthermore, we clarify that it is essential to identify the dynamic pattern in macroeconomic time series before feeding them into a machine learning model and we are the first to do this in an asset pricing context. \cite{messmer2017} and \cite{feng2018deep2} follow a similar approach as \cite{gu2018} to predict stock returns with neural networks. \cite{bianchi2019} provide a comparison of machine learning method for predicting bond returns in the spirit of \cite{gu2018}.\footnote{Other related work includes \cite{sirignano2016deep} who estimate mortgage prepayments, delinquencies, and foreclosures with deep neural networks, \cite{moritz2016} who apply tree-based models to portfolio sorting and \cite{heaton2017deep} who automate portfolio selection with a deep neural network. \cite{horel2019} propose a significance test in neural networks and apply it to house price valuation.} \cite{freyberger2017dissecting} use Lasso selection methods to estimate the risk premia of stock returns as a non-linear additive function of characteristics. \cite{feng2018deep} impose a no-arbitrage constraint by using a set of pre-specified linear asset pricing factors and estimate the risk loadings with a deep neural network. 
\cite{rossi2018} uses Boosted Regression Trees to form conditional mean-variance efficient portfolios based on the market portfolio and the risk-free asset. Our approach also yields the conditional mean-variance efficient portfolio, but based on all stocks. \cite{kelly2019} extend the linear conditional factor model of \cite{kelly2018} to a non-linear factor model using an autoencoder neural network.\footnote{The intuition behind their and our approach can be best understood when considering the linear special cases. Our approach can be viewed as a conditional, non-linear generalization of \cite{kozak2017} with the additional elements of finding the macroeconomic states and identifying the most robust conditioning instruments. Fundamentally, our object of interest is the pricing kernel. \cite{kelly2018} obtain a multi-factor factor model that maximizes the explained variation. The linear special case applies PCA to a set of characteristic based factors to obtain a linear lower dimensional factor model, while their more general autoencoder obtains the loadings to characteristic based factors that can depend non-linearly on the characteristics. We show in Section \ref{sec:IPCA} how our SDF framework and their conditional multi-factor framework can be combined to obtain an even better asset pricing model.} We confirm their crucial insight that imposing economic structure on a machine learning algorithm can substantially improve the estimation. 
\cite{bryzgalova2019} use decision trees to build a cross-section of asset returns, that is, a small set of basis assets that capture the complex information contained in a given set of stock characteristics. Their asset pricing trees generalize the concept of conventional sorting and are pruned by a novel dimension reduction approach based on no-arbitrage arguments.
 \cite{avramov2020} raise the concern that the performance of machine learning portfolios could deteriorate in the presence of trading frictions.\footnote{We have shared our data and estimated models with \cite{avramov2020}. In their comparison study \cite{avramov2020} also include a portfolio derived from our GAN model. However, they do not consider our SDF portfolio based on $\omega$ but use the SDF loadings $\beta$ to construct a long-short portfolio based on prediction quantiles. Similarly, they use extreme quantiles of a forecasting approach with neural networks to construct long-short portfolios, which is again different from our SDF framework. Thus, they study different portfolios.} 
We discuss how their important insight can be taken into account when constructing machine learning investment portfolios. A promising direction is presented in \cite{bryzgalova2019} and \cite{cong2020} who estimate optimal machine learning portfolios subject to trading friction constraints.


The workhorse models in equity asset pricing are based on linear factor models exemplified by \cite{fama1993,fama20151}. Recently, new methods have been developed to study the cross-section of returns in the linear framework but accounting for the large amount of conditioning information. \cite{lettaupelger2018} extend principal component analysis (PCA) to account for no-arbitrage. They show that a no-arbitrage penalty term makes it possible to overcome the low signal-to-noise ratio problem in financial data and find the information that is relevant for the pricing kernel. Our paper is based on a similar intuition and we show that this result extends to a non-linear framework. \cite{kozak2017} estimate the SDF based on characteristic sorted factors with a modified  elastic net regression.\footnote{We show that the special case of a linear formulation of our model is essentially a version of their model and we include it as the linear benchmark case in our analysis.} \cite{kelly2018} apply PCA to stock returns projected on characteristics to obtain a conditional multi-factor model where the loadings are linear in the characteristics. \cite{pelger2019} combines high-frequency data with PCA to capture non-parametrically the time-variation in factor risk. \cite{xiong2018} show that macroeconomic states are relevant to capture time-variation in PCA-based factors.

%

Our approach uses a similar insight as \cite{bansal1993no} and \cite{chen2009}, who propose using a given set of conditional GMM equations to estimate the SDF with neural networks, but restrict themselves to a small number of conditioning variables. In order to deal with the infinite number of moment conditions we extend the classical GMM setup of \cite{hansen1982} and \cite{chamberlain1987} by an adversarial network to select the optimal moment conditions. A similar idea has been proposed by \cite{greg2018} for non-parametric instrumental variable regressions. Our problem is also similar in spirit to the Wasserstein GAN in \cite{arjovsky2017} that provides a robust fit to moments. The Generative Adversarial Network (GAN) approach was first proposed by Goodfellow et al. (\citeyear{goodfellow2014generative}) for image recognition. In order to find the hidden states in macroeconomic time series we propose the use of Recurrent Neural Networks with Long-Short-Term-Memory (LSTM). LSTMs are designed to find patterns in time series data and have been first proposed by \cite{hochreiter1997}. They are among the most successful commercial AIs and are heavily used for sequences of data such as speech (e.g. Google with speech recognition for Android, Apple with Siri and the ``QuickType'' function on the iPhone or Amazon with Alexa).

The rest of the paper is organized as follows. Section \ref{sec:model} introduces the model framework and Section \ref{sec:estimation} elaborates on the estimation approach. 
The empirical results are collected in Section \ref{sec:empirical}. Section \ref{sec:conclusion} concludes. The simulation results and implementation details are delegated to the Appendix while the Internet Appendix collects additional empirical robustness results.


\section{Model}\label{sec:model}

\subsection{No-Arbitrage Asset Pricing}\label{sec:noarbitrage}

Our goal is to explain the differences in the cross-section of returns $R$ for individual stocks. Let $R_{t+1,i}$ denote the return of asset $i$ at time $t+1$. 
The fundamental no-arbitrage assumption is equivalent to the existence of a strictly positive stochastic discount factor (SDF) $M_{t+1}$ such that for any return in excess of the risk-free rate $R_{t+1,i}^e=R_{t+1,i} - R_{t+1}^f$, it holds
\begin{align*}
\mathbbm E_t \left[ M_{t+1} R^e_{t+1,i}\right] = 0 \qquad \Leftrightarrow \qquad
\E[\Ri] = \underbrace{\left(-\frac{\cov(\Ri,M_{t+1})}{\var(M_{t+1})}\right)}_{\beta_{t,i}} \cdot \underbrace{\frac{\var(\M)}{\E[\M]}}_{\lambda_t},
\end{align*}
where $\beta_{t,i}$ is the exposure to systematic risk and $\lambda_t$ is the price of risk. $E_t[.]$ denotes the expectation conditional on the information at time $t$. The SDF is an affine transformation of the tangency portfolio.\footnote{ See Back (2010) for more details. As we work with excess returns we have an additional degree of freedom. Following Cochrane (2003) we use the above normalized relationship between the SDF and the mean-variance efficient portfolio. We consider the SDF based on the projection on the asset space.} Without loss of generality we consider the SDF formulation
\begin{align*}
\M=1-  \sum_{i=1}^N \wi \Ri = 1-\w^{\top} \R .
\end{align*}
The fundamental pricing equation $\E[\R \M] = 0$ implies the SDF weights 
\begin{align}\label{eqn:weights}
\w= \Err^{-1} \E[\R]    ,
\end{align} 
which are the portfolio weights of the conditional mean-variance efficient portfolio.\footnote{Any portfolio on the globally efficient frontier achieves the maximum Sharpe ratio. These portfolio weights represent one possible efficient portfolio.} We define the tangency portfolio as $F_{t+1}=\w ^{\top} \R$ and will refer to this traded factor as the SDF. The asset pricing equation can now be formulated as
\begin{align*}
\E[\Ri]  = \frac{\cov(\Ri,F_{t+1})}{\var(F_{t+1})}  \cdot \E[F_{t+1}] = \bi \E[F_{t+1}].
\end{align*}
Hence, no-arbitrage implies a one-factor model
\begin{align*}
\Ri= \bi F_{t+1} + \ei
\end{align*}
with $\E[\ei]=0$ and $\cov(F_{t+1},\ei)=0$. Conversely, the factor model formulation implies the stochastic discount factor formulation above. Furthemore, if the idiosyncratic risk $\ei$ is diversifiable and the factor $F$ is systematic,\footnote{Denote the conditional residual covariance matrix by $\sig=\var(\epsilon_t)$. Then, sufficient conditions are $\| \sig \|_2 < \infty$ and  $\frac{\beta^{\top} \beta}{N} >0 $ for $N \rightarrow \infty$, i.e. $\sig$ has bounded eigenvalues and $\be$ has sufficiently many non-zero elements. Importantly, the dependency in the residuals is irrelevant for our SDF estimation. For our estimator we obtain the SDF weights $\omega_{t,i}$ and the stock loadings $\beta_{t,i}$. Neither step requires a weak dependency in the residuals. For the asset pricing analysis, we separate the return space into the part spanned by loadings and its orthogonal complement which does not impose assumptions on the covariance matrix of the residuals.} then knowledge of the risk loadings is sufficient to construct the SDF:
\begin{align*}
 \left( \be^{\top} \be \right)^{-1} \be^{\top} \R = F_{t+1} + \left( \be^{\top} \be \right)^{-1} \be^{\top} \epsilon_{t+1} = F_{t+1} + o_p(1).
\end{align*}

The fundamental problem is to find the SDF portfolio weights $\w$ and risk loadings $\beta_t$. Both are time-varying and general functions of the information set at time $t$. The knowledge of $\w$ and $\be$ solves three problems: (1) We can explain the cross-section of individual stock returns. (2) We can construct the conditional mean-variance efficient tangency portfolio. (3) We can decompose stock returns into their predictable systematic component and their non-systematic unpredictable component.

While equation \ref{eqn:weights} gives an explicit solution for the SDF weights in terms of the conditional second and first moment of stock returns, it becomes infeasible to estimate without imposing strong assumptions. Without restrictive parametric assumptions it is practically not possible to estimate reliably the inverse of a large dimensional conditional covariance matrix for thousands of stocks. Even in the unconditional setup the estimation of the inverse of a large dimensional covariance matrix is already challenging. In the next section we introduce an adversarial problem formulation which allows us to side-step solving explicitly an infeasible conditional mean-variance optimization.

\subsection{Generative Adversarial Methods of Moments}

Finding the SDF weights is equivalent to solving a method of moment problem.\footnote{No-arbitrage requires the conditional moment equations implied by the Law of One Price and a strictly positive SDF. Our estimation is based on the conditional moments without directly enforcing the positivity of the SDF. Note that in our empirical analysis our estimated SDF is always positive on the in-sample data and hence does not require the additional positivity constraint.} 
The conditional no-arbitrage moment condition implies infinitely many unconditional moment conditions
\begin{align}
\mathbbm E[\M \Ri g(I_t,I_{t,i})] = 0 \label{eqn:apt}
\end{align}
for any function $g(.) : \mathbbm R^p \times \mathbbm R^q  \rightarrow \mathbbm R^D$, where $I_t \times I_{t,i} \in \mathbbm R^p \times \mathbbm R^q$ denotes all the variables in the information set at time $t$ and $D$ is the number of moment conditions. We denote by $I_t$ all $p$ macroeconomic conditioning variables that are not asset specific, e.g. inflation rates or the market return, while $I_{t,i}$ are $q$ firm-specific characteristics, e.g. the size or book-to-market ratio of firm $i$ at time $t$. The unconditional moment conditions can be interpreted as the pricing errors for a choice of portfolios and times determined by $g(.)$. The challenge lies in finding the relevant moment conditions to identify the SDF.

A well-known formulation includes 25 moments that corresponds to pricing the 25 size and value double-sorted portfolios of \cite{10.2307/2329112}. For this special case each $g$ corresponds to an indicator function if the size and book-to-market values of a company are in a specific quantile. Another special case is to consider only unconditional moments, i.e. setting $g$ to a constant. This corresponds to minimizing the unconditional pricing error of each stock.

The SDF portfolio weights $\omega_{t,i}=\omega(I_t,I_{t,i})$ and risk loadings $\beta_{t,i}=\beta(I_t,I_{t,i})$ are general functions of the information set, that is,  $\omega: \mathbbm R^p \times \mathbbm R^q  \rightarrow \mathbbm R$ and $\beta:  \mathbbm R^p \times \mathbbm R^q  \rightarrow \mathbbm R$.
For example, the SDF weights and loadings in the Fama-French 3 factor model are a special case, where both functions are approximated by a two-dimensional kernel function that depends on the size and book-to-market ratio of firms. The Fama-French 3 factor model only uses firm-specific information but no macroeconomic information, e.g. the loadings cannot vary based on the state of the business cycle.

We use an adversarial approach to select the moment conditions that lead to the largest mis-pricing:

\begin{align}\label{eqn:unconditional}
\min_{\omega} \max_{g} \frac{1}{N} \sum_{j=1}^N\left \| \mathbbm E \left[  \left(1- \sum_{i=1}^N \omega(I_t,I_{t,i}) \Ri \right)  \Rj g(I_t,I_{t,j}) \right] \right \|^2,
\end{align}
where the function $\omega$ and $g$ are normalized functions chosen from a specified functional class. This is a minimax optimization problem. These types of problems can be modeled as a zero-sum game, where one player, the asset pricing modeler, wants to choose an asset pricing model, while the adversary wants to choose conditions under which the asset pricing model performs badly. This can be interpreted as first finding portfolios or times that are the most mispriced and then correcting the asset pricing model to also price these assets. The process is repeated until all pricing information is taking into account, that is the adversary cannot find portfolios with large pricing errors. Note that this is a data-driven generalization for the research protocol conducted in asset pricing in the last decades. Assume that the asset pricing modeler uses the Fama-French 5 factor model, that is $M$ is spanned by those five factors. The adversary might propose momentum sorted test assets, that is $g$ is a vector of indicator functions for different quantiles of past returns. As these test assets have significant pricing errors with respect to the Fama-French 5 factors, the asset pricing modeler needs to revise her candidate SDF, for example, by adding a momentum factor to $M$. Next, the adversary searches for other mispriced anomalies or states of the economy, which the asset pricing modeler will exploit in her SDF model.

Our adversarial estimation with a minimax objective function is economically motivated and based on the insights of \cite{hansen1997}. They show that if the SDF implied by an asset pricing model is only a proxy that does not price all possible assets in the economy, then minimizing the largest possible pricing error corresponds to estimating the SDF that is the closest to an admissible true SDF in a least square distance.\footnote{In more detail, Hansen and Jagannathan (1997) study the minimax estimation problem for the SDF in general (infinite dimensional) Hilbert spaces. Minimizing the pricing error is equivalent to minimizing the distance between the empirical pricing functional and an admissable true unknown pricing functional. The Riesz Representation Theorem provides a mapping between the pricing functional and the SDF and hence minimizing the worst pricing error implies convergence to an admissable SDF in a specific norm. Hence, our SDF is the solution to problem 1 and 2 in \cite{hansen1997}. The minimax framework to estimate or evaluate an SDF has also been used in \cite{bakshi1997}, \cite{chen1995} and \cite{bansal1993international} among others.} In our case the SDF is implicitly constrained by the fact that it can only depend on stock specific characteristics $I_{i,t}$ but not the identity of the stocks themselves and by a regularization in the estimation as specified in Section \ref{sec:implementation}. Hence, even in-sample the SDF will have non-zero pricing errors for some stocks and their characteristic managed portfolios, which puts us into the setup of \cite{hansen1997}.

Choosing the conditioning function $g$ correspond to finding optimal instruments in a GMM estimation. 
The conventional GMM approach assumes a finite number of moments that identify a finite dimensional set of parameters. The moments are selected to achieve the most efficient estimator within this class. \cite{NagelSingleton2011} use this argument to build optimal managed portfolios for a particular asset pricing model. Their approach assumes that the set of candidate test assets identify all the parameters of the SDF and they can therefore focus on which test asset provide the most efficient estimator.  Our problem is different in two ways that rule out using the same approach. First, we have an infinite number of candidate moments without the knowledge of which moments identify the parameters. Second, our parameter set is also of infinite dimension, and we consequently do not have an asymptotic normal distribution with a feasible estimator of the covariance matrix. In contrast, our approach selects the moments based on robustness.\footnote{See \cite{blanchet2016} for a discussion on robust estimation with an adversarial approach.} By controlling the worst possible pricing error we aim to choose the test assets that can identify all parameters of the SDF and provide a robust fit. \cite{hansen1997} discuss the estimation of the SDF based on the minimax objective function and compare it with the conventional efficient GMM estimation for parametric models with a low dimensional parameter set. They conclude that the minimax estimation has desirable properties when models are misspecified and the resulting SDFs have substantially less variation than with the conventional GMM approach.

Our conditioning function $g$ generates a very large number of test assets to identify a complex SDF structure. The cross-sectional average is taken over the moment deviations, that is, pricing errors for $\Rj g(I_t,I_{t,j})$, instead of considering the moment deviation for the $D$ time-series $\frac{1}{N} \sum_{i=1}^N \Rj g(I_t,I_{t,j})$ which would correspond to the traditionally characteristic managed test assets. In the second case we would only have $D$ test assets while our approach yields $D \cdot N$ test assets. Note that there is no gain from using only $D$ portfolios as the average squared moment deviations of all $D \cdot N$ instrumented stocks provide an upper bound for the pricing errors of the $D$ portfolios. However, it turns out that the use of the larger number of test assets substantially accelerates the convergence and the minimax problem can empirically already converge after three steps.\footnote{The number of test assets limits the number of parameters and hence the complexity of the SDF that is identified in each step of the iterative optimization. Our approach yields $D \cdot N$ test assets, which allows us to fit a complex structure for the SDF with few iteration steps. Obviously, instrumenting each stock with $g$ only leads to more test assets if the resulting portfolios are not redundant. Empirically, we observe that there is a large variation in the vectors of characteristics cross-sectionally and over time. For example, if $g$ is an indicator function for small cap stocks, the returns of $\Rj g(I_t,I_{t,j})$ provide different information for different stocks $j$ as the other characteristics are in most cases not identical and stocks have a small market capitalization at different times $t$.}

Our objective function in Equation \ref{eqn:unconditional} uses only an approximate arbitrage condition in the sense of \cite{ross1976} and \cite{chamberlain1983}. Our moment conditions are averaged over the sample of all instrumented stocks, that is the objective is $1/N \sum_{i=1}^{N} \sum_{d=1}^D  \alpha_{i,d}^2$, where the moment deviation $\alpha_{i,d}= E[M_{t+1} R_{i} g_d(I_t,I_{t,i}))]$ can be interpreted as the pricing error of stock $i$ instrumented by the element $g_d$ of the vector valued function $g(.)$. Note that the instruments $g_d$ are normalized to be in $[-1,1]$. In our benchmark model we consider $N=10,000$ stocks and $D=8$ instruments and therefore average in total over 80,000 instrumented assets. Hence, our SDF will depend only on information that affects a very large proportion of the stocks, that is, systematic mispricing. This also implies that the adversarial approach will only select instruments that lead to mispricing for most stocks. 

Once we have obtained the SDF factor weights, the loadings are proportional to the conditional moments $\E[F_{t+1} R_{t+1,i}^e]$. A key element of our approach is to avoid estimating directly conditional means of stock returns. Our empirical results show that we can better estimate the conditional co-movement of stock returns with the SDF factors, which is a second moment, than the conditional first moment.  
Note, that in the no-arbitrage one-factor model, the loadings are proportional to $\cov(\Ri,F_{t+1})$ and $\E[F_{t+1} R_{t+1,i}^e]$, where the last one has the advantage that we avoid estimating the first conditional moment.

\subsection{Alternative Models}

We consider two special cases as alternatives: One model can take a flexible functional form but does not use the no-arbitrage constraint in the estimation. The other model is based on the no-arbitrage framework, but considers a linear functional form.

Instead of minimizing the violation of the no-arbitrage condition, one can directly estimate the conditional mean. Note that the conditional expected returns $\mu_{t,i}$ are proportional to the loadings in the one-factor formulation:
\begin{align*}
\mu_{t,i} :=\E[\Ri] = \bi \E[F_{t+1}].
\end{align*}
Hence, up to a time-varying proportionality constant the SDF weights and loadings are equal to $\mu_{t,i}$. This reduces the cross-sectional asset pricing problem to a simple forecasting problem. Hence, we can use the forecasting approach pursued in \cite{gu2018} for asset pricing.

The second benchmark model assumes a linear structure in the factor portfolio weights $\wi=\theta^{\top} I_{t,i}$ and linear conditioning in the test assets:
\begin{align*}
\frac{1}{N}\sum_{j=1}^N \mathbbm E \left[ \left(1- \frac{1}{N}\sum_{i=1}^N \theta^{\top} I_{t,i} \Ri \right) \Rj I_{t,j} \right] =0 \qquad \Leftrightarrow \qquad \mathbbm E \left[ \left(1-  \theta^{\top} \tilde F_{t+1} \right) \tilde F_{t+1}^{\top} \right] =0,
\end{align*}
where $\tilde F_{t+1}= \frac{1}{N}\sum_{i=1}^N I_{t,i} \Ri$ are $q$ characteristic managed factors. Such characteristic managed factors based on linearly projecting onto quantiles of characteristics are exactly the input to PCA in \cite{kelly2018} or the elastic net mean-variance optimization in \cite{kozak2017}.\footnote{\cite{kozak2017} consider also cross-products of the characteristics. They show that the PCA rotation of the factors improves the pricing performance. \cite{lettaupelger2018} extend this important insight to RP-PCA rotated factors. We consider PCA based factors in \ref{sec:IPCA}. Our main analysis focuses on conventional long-short factors as these are the most commonly used models in the literature.} The solution to minimizing the sum of squared errors in these moment conditions is a simple mean-variance optimization for the $q$ characteristic managed factors that is, $\theta = \left( \mathbbm E \left[ \tilde F_{t+1}\tilde F_{t+1}^{\top}\right] \right)^{-1}\mathbbm E \left[\tilde F_{t+1} \right]$ are the weights of the tangency portfolio based on these factors.\footnote{As before we define as tangency portfolio one of the portfolios on the global mean-variance efficient frontier.} We choose this specific linear version of the model as it maps directly into the linear approaches that have already been successfully used in the literature. This linear framework essentially captures the class of linear factor models. Appendix \ref{sec:appcondSDF} provides a detailed overview of the various models for conditional SDFs and their relationship to our framework. 

\section{Estimation}\label{sec:estimation}

\subsection{Loss Function and Model Architecture}


The empirical loss function of our model minimizes the weighted sample moments which can be interpreted as weighted sample mean pricing errors:
\begin{equation}
L(\omega | \hat g, I_t, I_{t,i})=\frac{1}{N}\sum_{i=1}^N\frac{T_i}{T}\  \left \|\frac{1}{T_i}\sum_{t\in T_i}M_{t+1}R_{t+1,i}^e\hat g(I_t,I_{t,i}) \right \|^2.  \label{eqn:loss}
\end{equation}
for a given conditioning function $\hat g(.)$ and information set. We deal with an unbalanced panel in which the number of time series observations $T_i$ varies for each asset. As the convergence rates of the moments under suitable conditions is $1/\sqrt{T_i}$, we weight each cross-sectional moment condition by $\sqrt{T_i}/\sqrt{T}$, which assigns a higher weight to moments that are estimated more precisely and down-weights the moments of assets that are observed only for a short time period.

For a given conditioning function $\hat g(.)$ and choice of information set the SDF portfolio weights are estimated by a feedforward network that minimizes the pricing error loss
\begin{align*}
\hat \omega = \min_{\omega} L(\omega | \hat g, I_t, I_{t,i}).
\end{align*}
We refer to this network as the SDF network. 

\begin{figure}[htbp]
\centering
\tcaptab{GAN Model Architecture}\label{fig:networks}
\includegraphics[width=0.8\textwidth]{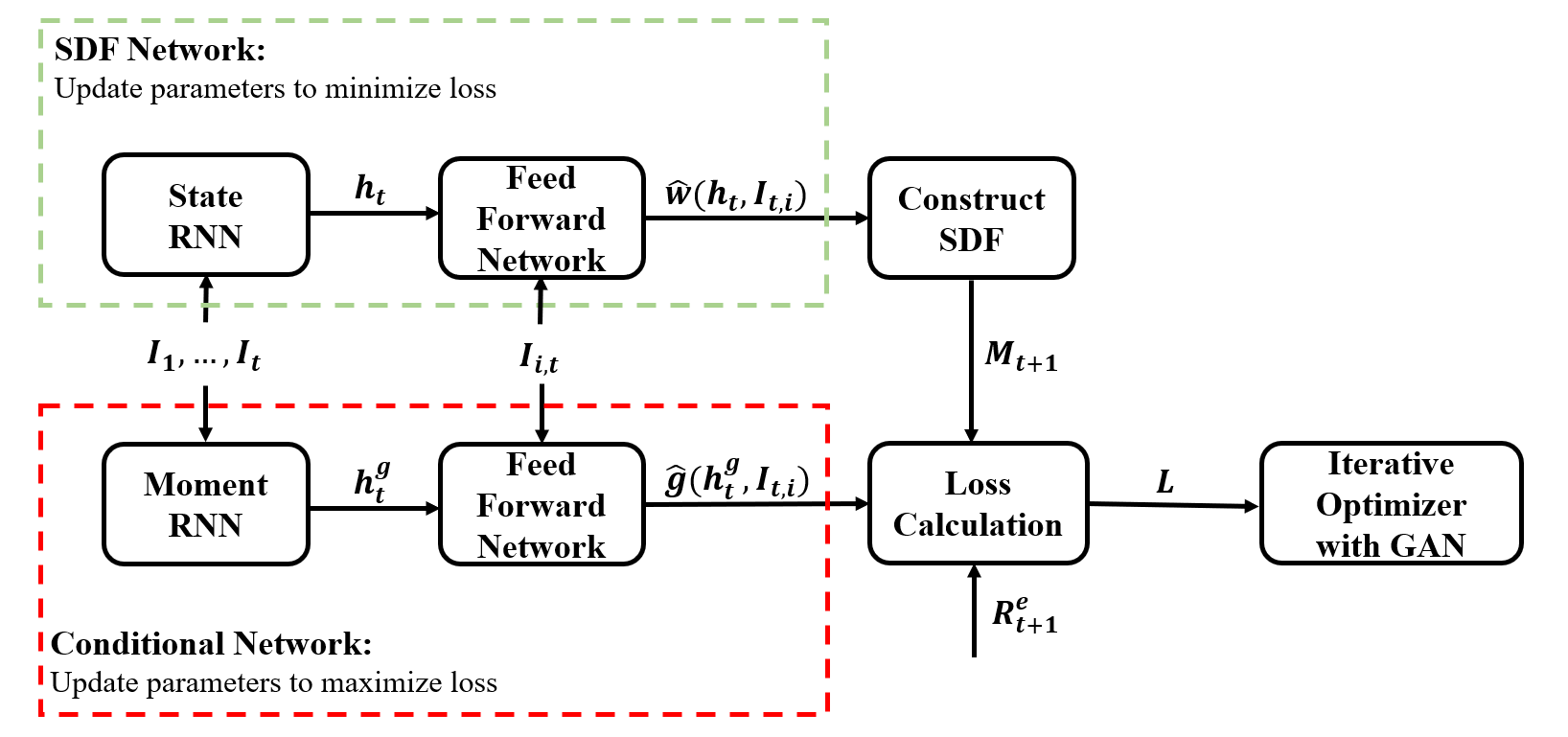}
\bnotetab{This figures shows the model architecture of GAN (Generative Adversarial Network) with RNN (Recurrent Neural Network) with LSTM cells. The SDF network has two parts: (1) A LSTM estimates a small number of macroeconomic states. (2) These states together with the firm-characteristics are used in a FFN to construct a candidate SDF for a given set of test assets. The conditioning network also has two networks: (1) It creates its own set of macroeconomic states, (2) which it combines with the firm-characteristics in a FFN to find mispriced test assets for a given SDF $M$. These two networks compete until convergence, that is neither the SDF nor the test assets can be improved.}
\end{figure}

We construct the conditioning function $\hat g$ via a conditional network with a similar neural network architecture.  The conditional network serves as an adversary and competes with the SDF network to identify the assets and portfolio strategies that are the hardest to explain. The macroeconomic information dynamics are summarized by macroeconomic state variables $h_t$ which are obtained by a Recurrent Neural Network (RNN) with Long-Short-Term-Memory units. The model architecture is summarized in Figure \ref{fig:networks} and each of the different components are described in detail in the next subsections.

In contrast, forecasting returns similar to \cite{gu2018} uses only a feedforward network and is labeled as FFN. It estimates conditional means $\mu_{t,i}=\mu(I_t,I_{t,i})$ by minimizing the average sum of squared prediction errors:
\begin{align*}
\hat \mu = \min_{\mu} \frac{1}{T} \sum_{t=1}^T \frac{1}{N_t} \sum_{i=1}^{N_t} \left( R_{t+1,i}^e - \mu(I_t,I_{t,i}) \right)^2 .
\end{align*}
We only include the best performing feedforward network from \cite{gu2018}'s comparison study. Within their framework this model outperforms tree learning approaches and other linear and non-linear prediction models.
In order to make the results more comparable with \cite{gu2018} we follow the same procedure as outlined in their paper. Thus, the simple forecasting approach does not include an adversarial network or LSTM to condense the macroeconomic dynamics.



\subsection{Feedforward Network (FFN)}

A feedforward network (FFN)\footnote{FFN are among the simplest neural networks and treated in detail in standard machine learning textbooks, e.g. \cite{goodfellow2016}.} is a flexible non-parametric estimator for a general functional relationship $y = f(x)$ between the covariates $x$ and a variable $y$. In contrast to conventional non-parametric estimators like kernel regressions or splines, FFNs do not only estimate non-linear relationships but are also designed to capture interaction effects between a large dimensional set of covariates. We will consider four different FFNs: For the covariates $x=[I_t,I_{t,i}]$ we estimate (1) the optimal weights in our GAN network ($y=\omega$), (2) the optimal instruments for the moment conditions in our GAN network ($y=g$), (3) the conditional mean return ($y=\mathbbm E_t[R^e_{t+1,i}]$) and (4) the second moment ($y=\mathbbm E_t[R^e_{t+1,i}F_{t+1}$]) to obtain the SDF loadings $\beta_{t,i}$. 

\begin{figure}[htbp]
\centering
\tcapfig{Illustration of Feedforward Network with Single Hidden Layer}
\includegraphics[width=0.75\textwidth]{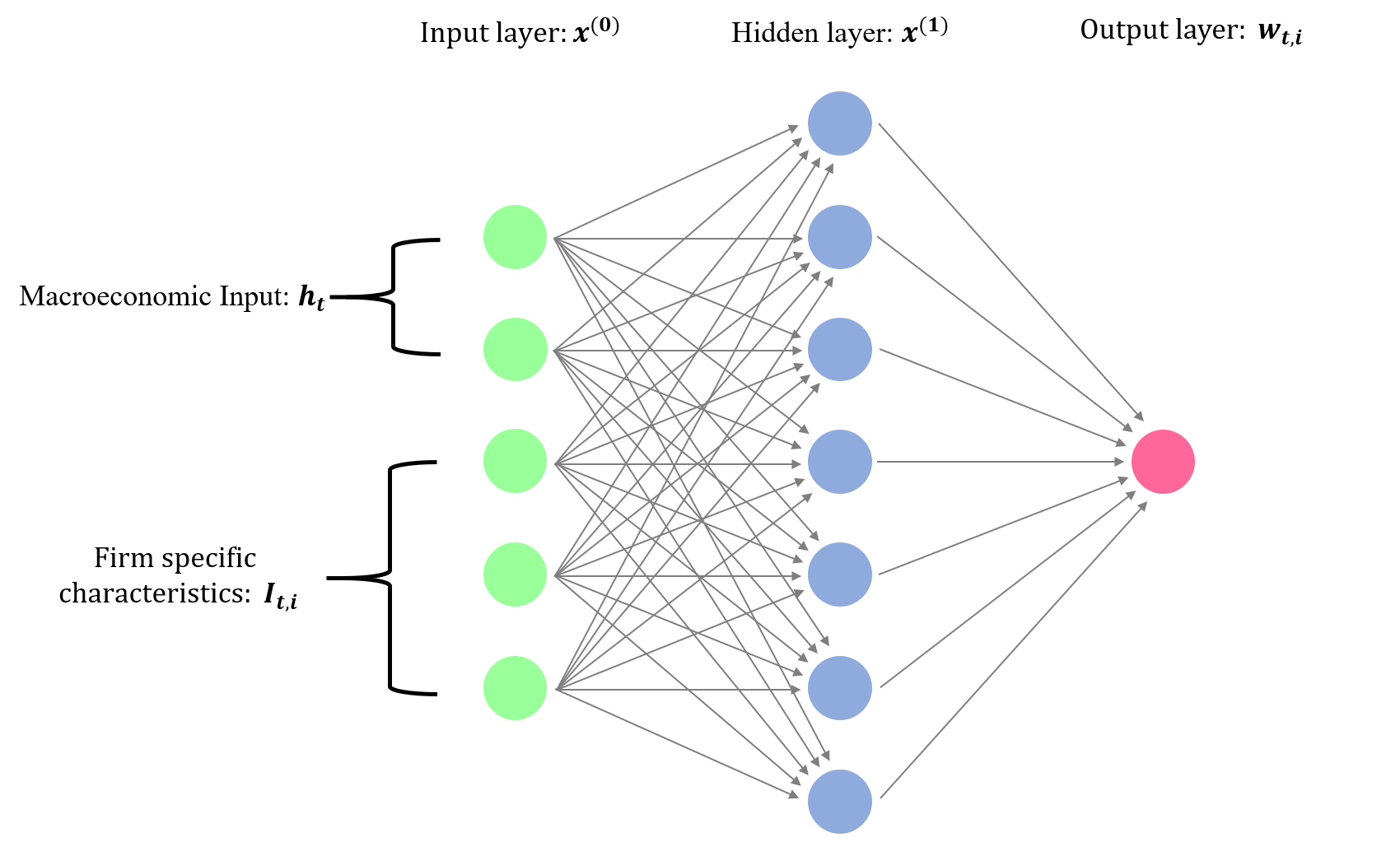}
\end{figure}

We start with a one-layer neural network. It combines the original covariates $x=x^{(0)} \in \mathbbm R^{K^{(0)}}$ linearly and applies a non-linear transformation. This non-linear transformation is based on an element-wise operating activation function. We choose the popular function known as the rectified linear unit (ReLU)\footnote{Other activation functions include sigmoid, hyperbolic tangent function and leaky ReLU. ReLU activation functions have a number of advantages including the non-saturation of its gradient, which greatly accelerates the convergence of stochastic gradient descent compared to the sigmoid/hyperbolic functions (Krizhevsky et al. (2012) and fast calculations of expensive operations.}, which component-wise thresholds the inputs and is defined as 
\begin{align*}
\text{ReLU}(x_k)=\max(x_k,0). 
\end{align*}
The result is the hidden layer $x^{(1)}=(x_1^{(1)},...,x_{K^{(1)}}^{(1)})$ of dimension $K^{(1)}$ which depends on the parameters $W^{(0)}=(w_1^{(0)},...,w_{K^{(0)}}^{(0)})$ and the bias term $w_0^{(0)}$. 
The output layer is simply a linear transformation of the output from the hidden layer. 
\begin{align*}
x^{(1)} &= \text{ReLU}({W^{(0)\top}} x^{(0)} + w_0^{(0)})= \text{ReLU}\left( w_0^{(0)} + \sum_{k=1}^{K^{(0)}} w_k^{(0)} x^{(0)}_k \right)   \\
y&= W^{(1)\top} x^{(1)} + w_0^{(1)} \qquad \text{with } x^{(1)} \in \mathbbm R^{K^{(1)}}, W^{(0)} \in \mathbbm R^{K^{(1)} \times K^{(0)}} ,W^{(1)} \in \mathbbm R^{K^{(1)}}.
\end{align*}
Note that without the non-linearity in the hidden layer, the one-layer network would reduce to a generalized linear model. A deep neural network combines several layers by using the output of one hidden layer as an input to the next hidden layer. The details are explained in Appendix \ref{app:ffn}. The multiple layers allow the network to capture non-linearities and interaction effects in a more parsimonious way.

%
%
%


\subsection{Recurrent Neural Network (RNN) with LSTM}

A Recurrent Neural Network (RNN) with Long-Short-Term-Memory (LSTM) estimates the hidden macroeconomic state variables. Instead of directly passing macroeconomic variables $I_t$ as covariates to the feedforward network, we extract their dynamic patterns with a specific RNN and only pass on a small number of hidden states capturing these dynamics.

Many macroeconomic variables themselves are not stationary. Hence, we need to first perform transformations as suggested in \cite{mccracken2016fred}, which typically take the form of some difference of the time-series. There is no reason to assume that the pricing kernel has a Markovian structure with respect to the macroeconomic information, in particular after transforming them into stationary increments. For example, business cycles can affect pricing but the GDP growth of the last period is insufficient to learn if the model is in a boom or a recession. Hence, we need to include lagged values of the macroeconomic variables and find a way to extract the relevant information from a potentially large number of lagged values.

\begin{figure}[h!]
\centering
\tcapfig{Examples of Macroeconomic Variables}\label{fig:3macro}
\includegraphics[width=\textwidth]{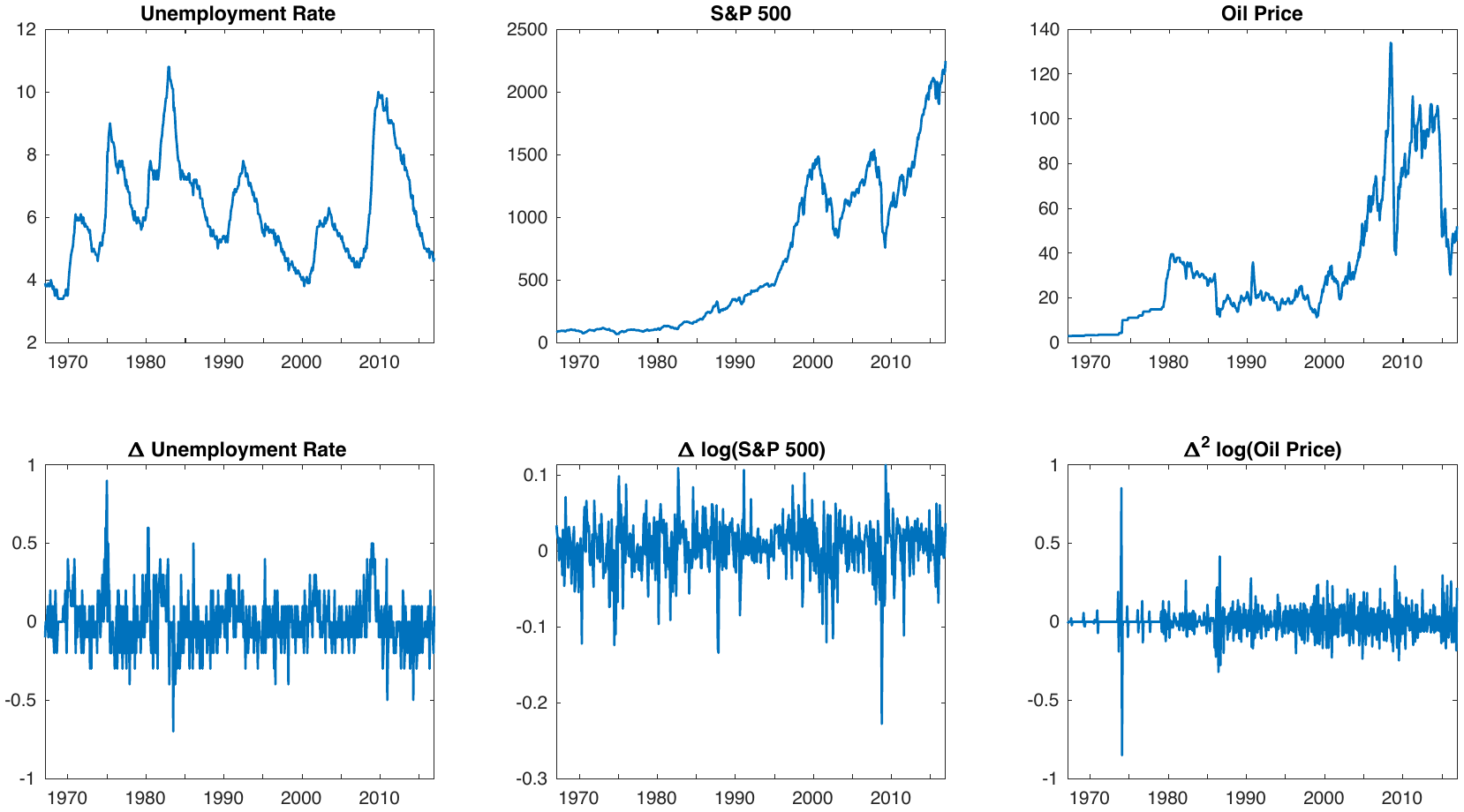}
\bnotefig{This figure shows examples of macroeconomic time series with standard transformations proposed by \cite{mccracken2016fred}.\par}
\end{figure}

As an illustration, we show in Figure \ref{fig:3macro} three examples of the complex dynamics in the macroeconomic time series that we include in our empirical analysis. We plot the time series of the U.S. unemployment rate, the S\&P 500 price and the oil price together with the standard transformations proposed by \cite{mccracken2016fred} to remove the obvious non-stationarities. Using only the last observation of the differenced data obviously results in a loss of information and cannot identify the cyclical dynamic patterns. 

Formally, we have a sequence of stationary vector valued processes $\{x_0,...,x_t\}$ where we set $x_t$ to the stationary transformation of $I_t$ at time $t$, i.e. typically an increment. Our goal is to estimate a functional mapping $h$ that transforms the time-series $x_t$ into ``state processes'' $h_t=h(x_0,...,x_t)$ for $t=1,...,T$. The simplest transformation is to simply take the last increment, that is $h^{\Delta}_t=h^{\Delta}(x_0,...,x_t)=x_t$. This approach is used in most papers including \cite{gu2018} and neglects the serial dependency structure in $x_t$.

Macroeconomic time series variables are strongly cross-sectionally dependent, that is, there is redundant information which could be captured by some form of factor model. A cross-sectional dimension reduction is necessary as the number of time-series observations in our macroeconomic panel is of a similar magnitude as the number of cross-sectional observations. \cite{ludvigson2007} advocate the use of PCA to extract a small number $K_h$ of factors which is a special case of the function $h^{\text{PCA}}(x_0,...,x_t)=W_x x_t$ for $W_x \in \mathbbm R^{p \times K_h}$. This aggregates the time series to a small number of latent factors that explain the correlation in the innovations in the time series, but PCA cannot identify the current state of the economic system which depends on the dynamics.

RNNs are a family of neural networks for processing sequences of data. They estimate non-linear time-series dependencies for vector valued sequences in a recursive form. A vanilla RNN model takes the current input variable $x_t$ and the previous hidden state $h^{\text{RNN}}_{t-1}$ and performs a non-linear transformation to get the current state $h_t^{\text{RNN}}$.
\begin{align*}
h_t^{\text{RNN}}=h^{\text{RNN}}(x_0,...,x_t)=\sigma(W_h h^{\text{RNN}}_{t-1}+W_x x_t+w_0),
\end{align*}
where $\sigma$ is the non-linear activation function. Intuitively, a vanilla RNN combines two steps: First, it summarize cross-sectional information by linearly combining a large vector $x_t$ into a lower dimensional vector. Second, it is a non-linear generalization of an autoregressive process where the lagged variables are transformations of the lagged observed variables. This type of structure is powerful if only the immediate past is relevant, but it is not suitable if the time series dynamics are driven by events that are further back in the past. Conventional RNNs can encounter problems with exploding and vanishing gradients when considering longer time lags. This is why we use the more complex Long-Short-Term-Memory cells. The LSTM is designed to deal with lags of unknown and potentially long duration in the time series, which makes it well-suited to detect business cycles.

Our LSTM approach can deal with both the large dimensionality of the system and a very general functional form of the states while allowing for long-term dependencies. Appendix \ref{app:lstm} provides a detailed explanation of the estimation method. Intuitively, an LSTM uses different RNN structures to model short-term and long-term dependencies and combines them with a non-linear function. We can think of an LSTM as a flexible hidden state space model for a large dimensional system. On the one hand it provides a cross-sectional aggregation similar to a latent factor model. On the other hand, it extracts dynamics similar in spirit to state space models, like for example the simple linear Gaussian state space model estimated by a Kalman filter. The strength of the LSTM is that it combines both elements in a general non-linear model. In our simulation example in Section \ref{sec:simulation} we illustrate that an LSTM can successfully extract a business cycle pattern which essentially captures deviations of a local mean from a long-term mean. Similarly, the state processes in our empirical analysis also seem to based on the relationship between the short-term and long-term averages of the macroeconomic increments and hence represent a business cycle type behavior.

The output of the LSTM is a small number of state processes $h_t=h^{\text{LSTM}}(x_0,...,x_t)$ which we use instead of the macroeconomic variables $I_t$ as an input to our SDF network. Note, that each state $h_t$ depends only on current and past macroeconomic increments and has no look-ahead bias.

\subsection{Generative Adversarial Network (GAN)}

The conditioning function $g$ is the output of a second feedforward network. Inspired by Generative Adversarial Networks (GAN), we chose the moment conditions that lead to the largest pricing discrepancy by having two networks compete against each other. One network creates the SDF $M_{t+1}$, and the other network creates the conditioning function.

We take three steps to train the model. Our initial first step SDF minimizes the unconditional loss. Second, given this SDF we maximize the loss by optimizing the parameters in the conditional network. Finally, given the conditional network we update the SDF network to minimize the conditional loss.\footnote{A conventional GAN network iterates this procedure until convergence. We find that our algorithm converges already after the above three steps, i.e. the model does not improve further by repeating the adversarial game. Detailed results on the GAN iterations for the empirical analysis are in Figure IA.1 in the Internet Appendix.} The logic behind this idea is that by minimizing the largest conditional loss among all possible conditioning functions, the loss for any function is small. Note that both, the SDF network and the conditional network each use a FFN network combined with an LSTM that estimates the macroeconomic hidden state variables, i.e. instead of directly using $I_t$ as an input each network summarizes the whole macroeconomic time series information in the state process $h_t$ (respectively $h^g_t$ for the conditional network). The two LSTMs are based on the criteria function of the two networks, that is $h_t$ are the hidden states that can minimize the pricing errors, while $h_t^g$ generate the test assets with the largest mispricing:\footnote{We allow for potentially different macroeconomic states for the SDF and the conditional network as the unconditional moment conditions that identify the SDF can depend on different states than the SDF weights.}
\begin{align*}
\{ \hat \omega, \hat h_t, \hat g,\hat h_t^g \}=\min_{\omega,h_t} \max_{g,h_t^g}  L(\omega | \hat g, h_t^g, h_t, I_{t,i}).
\end{align*}


\subsection{Hyperparameters and Ensemble Learning}\label{sec:implementation}

Due to the high dimensionality and non-linearity of the problem, training a deep neural network is a complex task. Here, we summarize the implementation and provide additional details in Appendix \ref{app:implementation}.
We prevent the model from overfitting and deal with the large number of parameters by using ``Dropout'', which is a form of regularization that has generally better performance than conventional $l_1/l_2$ regularization. We optimize the objective function accurately and efficiently by employing an adaptive learning rate for a gradient-based optimization. 

We obtain robust and stable fits by ensemble averaging over several fits of the models. A distinguishing feature of neural networks is that the estimation results can depend on the starting value used in the optimization. The standard practice which has also been used by \cite{gu2018} is to train the models separately with different initial values chosen from an optimal distribution. Averaging over multiple fits achieves two goals: First, it diminishes the effect of a local suboptimal fit. Second, it reduces the estimation variance of the estimated model. All our neural networks including the forecasting approach are averaged over nine model fits.\footnote{An ensemble over nine models produces very robust and stable result and there is no effect of averaging over more models. The results are available upon request.} Let $\hat w^{(j)}$ and $\hat \beta^{(j)}$ be the optimal portfolio weights respectively SDF loadings given by the $j^{th}$ model fit. The ensemble model is an average of the outputs from models with the same architecture but different starting values for the optimization, that is $\hat \omega= \frac{1}{9}  \sum_{j=1}^9 \hat \omega^{(j)}$ and $\hat \beta= \frac{1}{9}  \sum_{j=1}^9 \hat \beta^{(j)}$. Note that for vector valued functions, for example the conditioning function $g$ and macroeconomic states $h$, it is not meaningful to report their model averages as different entries in the vectors are not necessarily reflecting the same object in each fit.

We split the data into a training, validation and testing sample. The validation set is used to tune the hyperparameters, which includes the depth of the network (number of layers), the number of basis functions in each layer (nodes), the number of macroeconomic states, the number of conditioning instruments and the structure of the conditioning network. We choose the best configuration among all possible combinations of hyperparameters by maximizing the Sharpe ratio of the SDF on the validation data.\footnote{We have used different criteria functions, including the error in minimizing the moment conditions, to select the hyperparameters. The results are virtually identical and available upon request.} The optimal model is evaluated on the test data. Our optimal model has two layers, four economic states and eight instruments for the test assets. Our results are robust to the tuning parameters as discussed in Section \ref{sec:robust}. In particular, our results do not depend on the structure of the network and the best performing networks on the validation data provide essentially an identical model with the same relative performance on the test data. The FFN for the forecasting approach uses the optimal hyperparameters selected by \cite{gu2018}. This has the additional advantage of making our results directly comparable to their results.

\subsection{Model Comparison}

We evaluate the performance of our model by calculating the Sharpe ratio of the SDF, the amount of explained variation and the pricing errors. We compare our GAN model with its linear special case, which is a linear factor model, and the deep-learning forecasting approach. The one factor representation yields three performance metrics to compare the different model formulations. First, the SDF is by construction on the globally efficient frontier and should have the highest conditional Sharpe ratio. We use the unconditional Sharpe ratio of the SDF portfolio $SR= \frac{\mathbbm E[F]}{\sqrt{Var[F]}}$ as a measure to assess the pricing performance of models. 
The second metric measures the variation explained by the SDF. The explained variation is defined as $1-\frac{\sum_{i=1}^N \mathbbm E[\epsilon_i^2]}{\sum_{i=1}^N \mathbbm E[R_i^e]}$ where $\epsilon_i$ is the residual of a cross-sectional regression on the loadings. As in \cite{kelly2018} we do not demean returns due to their non-stationarity and noise in the mean estimation. Our explained variation measure can be interpreted as a time series $R^2$.
The third performance measure is the average pricing error normalized by the average mean return to obtain a cross-sectional $R^2$ measure $1- \frac{{\frac{1}{N} \sum_{i=1}^N \mathbbm E[\epsilon_i]^2}}{{\frac{1}{N} \sum_{i=1}^N \mathbbm E[R_i]^2}}$.

The output for our GAN model are the SDF factor weights $\hat \omega_{GAN}$. We obtain the risk exposure $\hat \beta_{GAN}$ by fitting a feedforward network to predict $R_{t+1}^eF_{t+1}$ and hence estimate $\mathbb{E}_t[R_{t+1}^eF_{t+1}]$. Note, that this loading estimate $\hat \beta_{GAN}$ is only proportional to the population value $\beta$ but this is sufficient for projecting on the systematic and non-systematic component. The conventional forecasting approach, which we label FFN, yields the conditional mean $\hat \mu_{FFN}$, which is proportional to $\beta$ and hence is used as $\hat \beta_{FNN}$ in the projection. At the same time $\hat \mu_{FFN}$ is proportional to the SDF factor portfolio weights and hence also serves as $\hat \omega_{FFN}$. Hence, the fundamental difference between GAN and FFN is that GAN estimates a conditional second moment $\beta_{GAN}=\mathbb{E}_t[R_{t+1}^eF_{t+1}]$, while FFN estimates a conditional first moment $\beta_{FFN}=\mathbb{E}_t[R_{t+1}^e]$ for cross-sectional pricing.


Note that the linear model, labeled as LS, is a special case with an explicit solution 
\begin{align*}
\hat \theta_{LS} &= \left(\frac{1}{T}\sum_{t=1}^T \left(\frac{1}{N}\sum_{i=1}^N \Ri I_{t,i}  \right)\left(\frac{1}{N} \sum_{i=1}^N \Ri I_{t,i}\right)^{\top} \right)^{-1}  \left(\frac{1}{N T} \sum_{t=1}^T \sum_{i=1}^N \Ri I_{t,i}  \right) \\
&= \left( \frac{1}{T} \sum_{t=1}^T \tilde F_{t+1} \tilde F_{t+1}^{\top} \right)^{-1} \left(\frac{1}{T} \sum_{t=1}^T \tilde F_{t+1}^{\top}  \right)
\end{align*}
and SDF factor portfolio weights $\omega_{LS}=\hat \theta^{\top}_{LS}I_{t,i}$. The risk exposure $\hat \beta_{LS}$ is obtained by a linear regression of $R_{t+1}^eF_{t+1}$ on $I_{t,i}$. As the number of characteristics is very large in our setup, the linear model is likely to suffer from over-fitting. The non-linear models include a form of regularization to deal with the large number of characteristics. In order to make the model comparison valid, we add a regularization to the linear model as well. The regularized linear model EN adds an elastic net penalty to the regression to obtain $\hat \theta_{EN}$ and to the predictive regression for $\hat \beta_{EN}$:\footnote{The elastic net includes lasso and ridge regularization as a special case. We select the tuning parameters of the elastic net optimally on the validation data.} 
\begin{align*}
\hat \theta_{EN} = \arg \min_{\theta} \left(\frac{1}{T} \sum_{t=1}^T \tilde F_{t+1}- \frac{1}{T}\sum_{t=1}^T \tilde F_{t+1} \tilde F_{t+1}^{\top}\theta\right)^2 + \lambda_2 \|\theta\|_2^2 + \lambda_1 \| \theta \|_1 .
\end{align*}
The linear approach with elastic net is closely related to \cite{kozak2017} who perform mean-variance optimization with an elastic net penalty on characteristic based factors.\footnote{There are five differences to their paper. First, they use a modified ridge penalty based on a Bayesian prior. Second, they also include product terms of the characteristics. Third, their second moment matrix uses demeaned returns, i.e. the two approaches choose different mean-variance efficient portfolios on the globally efficient frontier. Fourth, we allow for different linear weights on the long and the short leg of the characteristic based factors. Fifth, they advocate to first apply PCA to the characteristics managed factors before solving the mean-variance optimization with elastic net penalty. \cite{lettaupelger2018} generalize the robust SDF recovery to the RP-PCA space. \cite{bryzgalova2019} also include additional mean shrinkage in the robust SDF recovery and propose decision trees as an alternative to PCA. \cite{bryzgalova2019} also show that mean-variance optimization with regularization can be interpreted as an adversarial approach with parameter uncertainty. We use conventional long-short factors as a benchmark as those are the most commonly used linear models in the literature. PCA based methods are deferred to Section \ref{sec:IPCA}.} 
In addition we also report the maximum Sharpe ratios for the tangency portfolios based on the Fama-French 3 and 5 factor models.\footnote{The tangency portfolio weights are obtained on the training data set and used on the validation and test data set.}

For the four models GAN, FFN, EN and LS we obtain estimates of $\omega$ for constructing the SDF and estimates of $\beta$ for calculating the residuals $\epsilon$. 
We obtain the systematic and non-systematic return components by projecting returns on the estimated risk exposure $\hat \beta$:
\begin{align*}
 \hat \epsilon_{t+1}=\left(I_N-\hat \beta_t(\hat \beta_t^{\top}\hat \beta_t)^{-1}\hat \beta_t^{\top}\right) R_{t+1}^e.
\end{align*}
For each model we report (1) the unconditional Sharpe ratio of the SDF factor, (2) the explained variation in individual stock returns and and (3) the cross-sectional mean\footnote{We weight the estimated means by their rate of convergence to account for the differences in precision.} $R^2$:
\begin{align*}
 {SR}= \frac{\hat {\mathbbm E}[F_t]}{\sqrt{\widehat {Var}(F_t)}}  , \; {EV}=1- \frac{\left(\frac{1}{T}\sum_{t=1}^T\frac{1}{N_t}\sum_{i=1}^{N_t}(\hat \epsilon_{t+1,i})^2 \right)}{\left(\frac{1}{T}\sum_{t=1}^T\frac{1}{N_t}\sum_{i=1}^{N_t}(R_{t+1,i}^e)^2  \right)}, \; \text{XS-$R^2$} = 1- \frac{\frac{1}{N}\sum_{i=1}^N\frac{T_i}{T}\left(\frac{1}{T_i}\sum_{t\in T_i}\hat \epsilon_{t+1,i}\right)^2}{\frac{1}{N}\sum_{i=1}^N\frac{T_i}{T}\left(\frac{1}{T_i}\sum_{t\in T_i}\hat R_{t+1,i}\right)^2} .
\end{align*}
These are generalization of the standard metrics used in linear asset pricing. 


We also evaluate our models on conventional characteristic sorted portfolios. The portfolio loadings are the average of the stock loadings weighted by the portfolio weights. 
In more detail, our models provide risk loadings $\beta_{t,i}$'s for each individual stock $i$. The risk loadings $\beta_t$'s for the portfolios are obtained by aggregating the corresponding stock specific loadings. 
We obtain the portfolio error from a cross-sectional regression of the portfolio returns on the portfolio $\beta_t$ at each point in time. This is similar to a standard cross-sectional Fama-MacBeth regression in a linear model with the main difference that the $\beta_t$'s are obtained from our SDF models on individual stocks. The measures $EV$ and XS-$R^2$ for portfolios follow the same procedure as for individual stocks but use portfolio instead of stock returns. For the individual quantiles we also report the pricing error $\hat \alpha_i$ normalized by the root-mean-squared average returns of all corresponding quantile sorted portfolios, that is, $\hat \alpha_i= \frac{\hat{\mathbbm E}[ \hat \epsilon_{t,i}]}{\sqrt{\frac{1}{N}\sum_{i=1}^N \hat{\mathbbm E}[ R_{t,i}]^2}}$.\footnote{Note, that {\it XS-}$R^2=1 - \sum_{i=1}^N \hat \alpha_i^2$.} 

Appendix \ref{sec:simulation} includes a simulation that illustrates that all three evaluation metrics (SR, EV and XS-$R^2$) are necessary to assess the quality of an SDF. A model like FFN can achieve high Sharpe ratios by loading on some extreme portfolios but it does not imply that it captures the loading structure correctly.\footnote{\cite{pelgerxiong2018} provide the theoretical arguments and show empirically in a linear setup why ``proximate'' factors that only capture the extreme factor weights correctly have similar time series properties as the population factors but their portfolio weights are not the correct loadings.} Similarly, linear factors can achieve high Sharpe ratios but by construction cannot capture non-linear and interaction effects in the SDF loadings which is reflected in lower EV and XS-$R^2$. It does not matter how flexible the model is (e.g. FFN), by conditioning only on the most recent macroeconomic observations, general macroeconomic dynamics are ruled out, which seems to be the most strongly reflected in the Sharpe ratio. The no-arbitrage condition in the GAN model helps to deal with a low signal-to-noise ratio and to correctly estimate the SDF loadings of stocks that have small risk premia which is reflected in the XS-$R^2$.

\section{Empirical Results for U.S. Equities}\label{sec:empirical}

\subsection{Data}


We collect monthly equity return data for all securities on CRSP. The sample period spans January 1967 to December 2016, totaling 50 years. We divide the full data into 20 years of training sample (1967 - 1986), 5 years of validation sample (1987 - 1991), and 25 years of out-of-sample testing sample (1992 - 2016). We use the one-month Treasury bill rates from the Kenneth French Data Library as the risk-free rate to calculate excess returns. 

In addition, we collect the 46 firm-specific characteristics listed either on Kenneth French Data Library or used by \cite{freyberger2017dissecting}.\footnote{We use the characteristics that \cite{freyberger2017dissecting} used in the 2017 version of their paper.} All these variables are constructed either from accounting variables from the CRSP/Compustat database or from past returns from CRSP. We follow the standard conventions in the variable definition, construction and their updating. Yearly updated variables are updated at the end of each June following the Fama-French convention, while monthly changing variables are updated at the end of each month for the use in the next month. The full details on the construction of these variables are in the Internet Appendix. In Table \ref{tab:category} we sort the characteristics into the six categories {\it past returns, investment, profitability, intangibles, value and trading frictions}.

The number of all available stocks from CRSP is around 31,000. As in \cite{kelly2018} or \cite{freyberger2017dissecting}, we are limited to the returns of stocks that have all firm characteristics information available in a certain month, which leaves us with around 10,000 stocks. This is the largest possible data set that can be used for this type of analysis.\footnote{Using stocks with missing characteristic information requires data imputation based on model assumptions. \cite{kelly2019} replace a missing characteristic with the cross-sectional median of that characteristic during that month. However, this approach introduces an additional source of error and ignores the dependency structure in the characteristic space and thus creates artificial time-series fluctuation in the characteristics, which we want to avoid. Hence, we follow the same approach as in \cite{freyberger2017dissecting} and \cite{kelly2018} to use only stocks that have all firm characteristics available in a given month, which has the additional benefit of removing predominantly stocks with a very small market capitalization. Note that we do not require for one stock to have the characteristics to exist throughout its entire time-series. We simply only include the returns at time $t$ for stock $i$ if for this point in time it has all characteristics, that is, stock $i$ does not necessarily have a complete time-series, which is allowed in our approach.} 

For each characteristic variable in each month, we rank them cross-sectionally and convert them into quantiles. This is a standard transformation to deal with the different scales and has also been used in \cite{kelly2018}, \cite{kozak2017} or \cite{freyberger2017dissecting} among others. In the linear model the projection $\tilde F_{t+1}= \frac{1}{N} \sum_{i=1}^N I_{t,i} R^{e}_{t+1,i}$ results in long-short factors with an increasing positive weight for stocks that have a characteristic value above the median and a decreasing negative weight for below median values.\footnote{\cite{kelly2018} and \cite{kozak2017} construct factors in this way.} We increase the flexibility of the linear model by including the positive and negative leg separately for each characteristic, i.e. we take the rank-weighted average of the stocks with above median characteristic values and similarly for the below median values. This results in two ``factors'' for each characteristic. Note, that our model includes the conventional long-short factors as a special case where the long and short legs receive the same absolute weight of opposite sign in the SDF. These factors are still zero cost portfolios as they are based on excess returns.\footnote{In the first version of this paper we used the conventional long-short factors. However, our empirical results suggest that the long and short leg have different weights in the SDF and this additional flexibility improves the performance of the linear model. These findings are also in line with \cite{lettaupelger2018} who extract linear factors from the extreme deciles of single sorted portfolios and show that they are not spanned by long-short factors that put equal weight on the extreme deciles of each characteristic.} 


We collect 178 macroeconomic time series from three sources. We take 124 macroeconomic predictors from the FRED-MD database as detailed in \cite{mccracken2016fred}. Next, we add the cross-sectional median time series for each of the 46 firm characteristics. The quantile distribution combined with the median level for each characteristics are close to representing the same information as the raw characteristic information but in a normalized form. Third, we supplement the time series with the 8 macroeconomic predictors from \cite{welch2007comprehensive} which have been suggested as predictors for the equity premium and are not already included in the FRED-MD database. 

We apply standard transformations to the time series data. We use the transformations suggested in \cite{mccracken2016fred}, and define transformations for the 46 median and the 8 time series from \cite{welch2007comprehensive} to obtain stationary time series. A detailed description of the macroeconomic variables as well as their corresponding transformations are collected in the Internet Appendix.

\subsection{An Illustrative Example of GAN}

We illustrate how GAN works with a simple example that uses only the three characteristics size (\texttt{LME}), book-to-market ratio (\texttt{BEME}) and investment (\texttt{Investment}) for all stocks in our sample but leaves out the macroeconomic information. We show that is not only crucial which characteristics are included in the SDF weights $\omega$ but also which are included in the test assets constructed by the conditioning function $g$. UNC denotes the model that is unconditional in the test assets, that is, the test assets are the individual stock returns and the objective function is based on the unconditional moments. We allow the SDF weights to depend on size and value information denoted by UNC (SV) or also include investment for UNC (SVI). The GAN model allows for test assets that depend on the characteristics, that is, $g$ is a non-trivial function. We first list the characteristics included in the SDF weights $\omega$ and then those included in the test assets modeled by $g$. For example, GAN (SVI-SV) uses size, value and investment in $\omega$, but only size and value in $g$. In order to keep this simple example interpretable, we restrict $g$ to a scalar function. The loadings $\beta$ depend on the same information as the SDF weights $\omega$. We also include the benchmark model that is estimated on all the data and discussed in more detail in the next subsection. We evaluate the asset pricing performance on two well-known sets of test assets: 25 portfolios double-sorted on size and book-to-market (SV 25) and 35 portfolios that include in addition 10 decile portfolios sorted on investment (SVI 35). We infer the portfolio's $\beta$ from the SDF loadings of the individual stocks and the portfolio weights. 

\begin{figure}[h!]
\tcapfig{GAN Illustration}\label{fig:GAN_toy}
\centering
\includegraphics[width=0.95\linewidth]{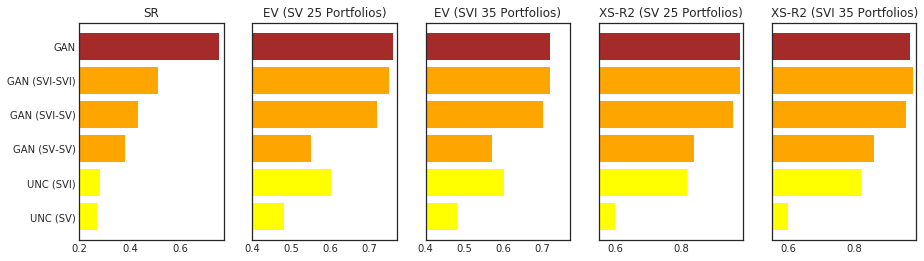}
\bnotefig{This figure shows the out-of-sample monthly Sharpe ratio (SR), explained time series variation (EV) and cross-sectional $R^2$ for different SDF models. UNC (SV) and UNC (SVI) are unconditional models with respect to the test assets, that is, they use only size and value respectively size, value and investment for the SDF weights, but set $g$ to a constant. The GAN models use a non-trivial $g$. GAN (SVI-SV) allows the SDF weight $\omega$ to depend on size, value and investment, but the test asset function $g$ to depend only on size and investment. The model labeled as GAN is our benchmark model estimated with all characteristics and macroeconomic information. We evaluate the model on 25 double-sorted size and book-to-market portfolios (SV 25) and we add another 10 decile portfolios sorted on investment (SVI 35). The portfolios are value-weighted.}
\end{figure}

\begin{figure}[h!]
  \tcapfig{GAN Conditioning Function $g$ and Portfolio Pricing}\label{fig:SDF_GAN_g}
  \begin{subfigure}[t]{.45\textwidth}
  \centering
    \includegraphics[width=1\textwidth]{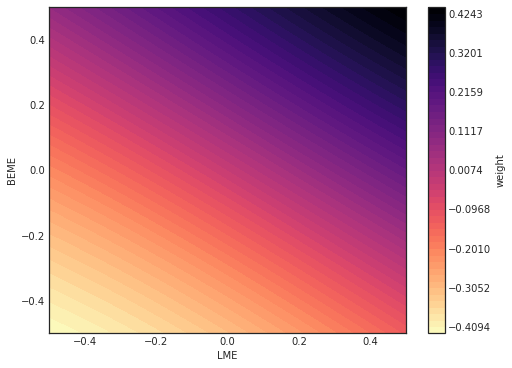}
    \caption{Conditioning function $g$ for GAN (SV-SV)}
  \end{subfigure}\hfill
  \begin{subfigure}[t]{.45\textwidth}
  \centering
    \includegraphics[width=1\textwidth]{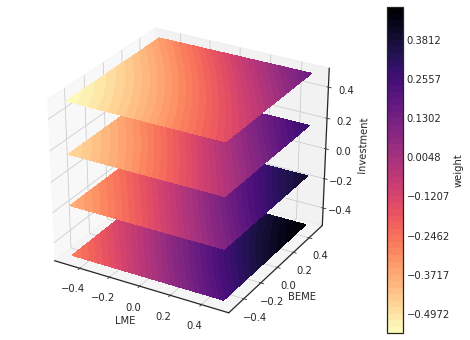}
    \caption{Conditioning function $g$ for GAN (SVI-SVI)}
  \end{subfigure}
  \begin{subfigure}[t]{.47\textwidth}
  \centering
    \includegraphics[width=1\textwidth]{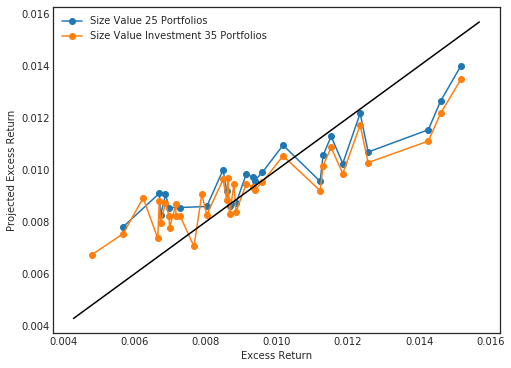}
    \caption{Portfolio pricing for GAN (SVI-SVI)}
  \end{subfigure}\hfill
  \begin{subfigure}[t]{.47\textwidth}
  \centering
    \includegraphics[width=1\textwidth]{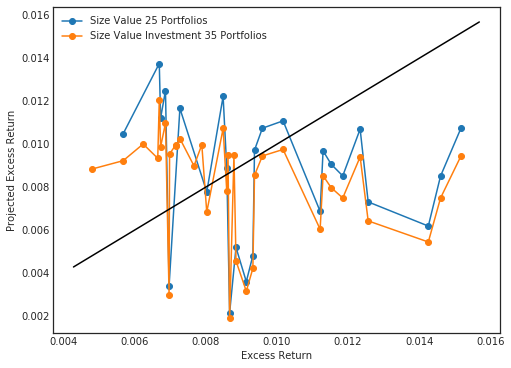}
    \caption{Portfolio pricing for UNC (SVI)}
  \end{subfigure}
  \bnotefig{This figures show the composition of the conditioning function $g$ and average vs. model implied average returns for the sorted portfolios. Subfigure (a) shows $g$ based only on size and value while subfigure (b) also includes investment. Subfigures (c) and (d) show the average excess returns and model-implied average excess returns from a cross-sectional regression for the 25 and 35 sorted portfolios. The tuning parameters are chosen optimally on the validation data and are different from the general benchmark GAN. All results are evaluated out-of-sample.}
\end{figure}

Figure \ref{fig:GAN_toy} shows the out-of-sample Sharpe ratio, explained variation and cross-sectional $R^2$. First, not surprisingly including more information in the SDF weights $\omega$ leads to a better asset pricing model. UNC (SVI) explains more variation and mean returns for portfolios sorted on investment but also for the size and book-to-market portfolios compared to UNC (SV). However, the key insight is that the information in the test assets matters crucially for the SDF recovery. The test assets of GAN (SVI-SVI) include investment information, which results in a better model than GAN (SVI -SV) that has only size and book-to-market information for the test assets. Depending on the metric, GAN (SVI-SVI) is roughly twice as good as UNC (SVI). The top bar is the full benchmark model. Not surprisingly, the higher SR confirms that there is substantially more information that can be extracted by including the other characteristics and macroeconomic time-series. However, if the goal is to simply explain the 25 Fama-French double-sorted portfolios, GAN (SVI -SVI) already provides a good model.

Figure \ref{fig:SDF_GAN_g} explains why we observe these findings. The top figures show heat maps for the scalar conditioning function $g$. For GAN (SV-SV) the test assets become long-short portfolios with extreme weights in small value stocks and large growth stocks.\footnote{Note that the sign of $g$ is not identified and we could multiply $g$ by $-1$ and obtain the same output.} When we add investment information, the test assets become essentially long-short portfolios with extreme weighs for small, conservative value stocks and large, aggressive growth stocks. GAN has in a data driven way discovered the structure of the Fama-French type test assets! Not surprisingly, an asset pricing model trained on these test assets will better explain portfolios sorted on these characteristics. The bottom figures show the model-implied average excess returns from a cross-sectional regression and average excess returns for the 25 and 35 sorted portfolios for GAN (SVI-SVI) and UNC (SVI). In an ideal model the points would line up on the 45 degree line. UNC (SVI) fails in explaining small value stocks while the GAN formulation captures mean returns very well for all quantiles. The Internet Appendix contains the detailed results for all the models and test assets.

In summary, the simple example illustrates that the problem of estimating an asset pricing model cannot be separated from the problem of choosing informative test assets. In the next section we move to our main analysis that includes all firm characteristics and macroeconomic information.

\subsection{Cross Section of Individual Stock Returns}

The GAN SDF has a higher out-of-sample Sharpe ratio while explaining more variation and pricing than the other benchmark models. Table \ref{tab:SDF-Comparison} reports the three main performance measures, Sharpe ratio, explained variation and cross-sectional $R^2$, for the four model specifications. The annual out-of-sample Sharpe ratio of GAN is around 2.6 and almost twice as high as with the simple forecasting approach FFN. The non-linear and interaction structure that GAN can capture results in a 50\% increase compared to the regularized linear model. Hence, the more flexible form matters, but an appropriately designed linear model can already achieve an impressive performance. The non-regularized linear model has the worst performance in terms of explained variation and pricing error. GAN explains 8\% of the variation of individual stock returns which is twice as large as the other models. Similarly, the cross-sectional $R^2$ of 23\% is substantially higher than for the other models. Interestingly, the regularized linear model based on the no-arbitrage objective function explains the time-series and cross-section of stock returns at least as good as the flexible neural network without the no-arbitrage condition. Each model here uses the optimal set of hyperparameters to maximize the validation Sharpe ratio. In case of the LS, EN and FFN this implies to leave out the macroeconomic variables.\footnote{The results are not affected by normalizing the SDF weights to have $\| \omega \|_1=1$. The explained variation and pricing results are based on a cross-sectional projection at each time step which is independent of any scaling.} 

The benchmark criteria differ on the out-of-sample test and in-sample training data. It is important to keep in mind that risk premia and risk exposure of individual stocks are time-varying.\footnote{\cite{pesaran1996} among others show the time variation in risk premia.} Hence, there is fundamentally no reason to expect the benchmark numbers on different time windows to be the same. Nevertheless, the higher benchmark numbers on the in-sample data suggests a certain degree of overfitting. Thus, the relevant metric is the relative out-of-sample performance between different models as also emphasized among others by \cite{martin2020} and \cite{gu2018}.

\begin{table}[h!]
\centering
\tcaptab{Performance of Different SDF Models}\label{tab:SDF-Comparison}
{\small
\begin{tabular}{cccc|ccc|ccc}
\toprule
& \multicolumn{3}{c}{SR} & \multicolumn{3}{c}{EV} & \multicolumn{3}{c}{XS-$R^2$}\\
\cmidrule(l){2-10}
Model & Train & Valid & Test & Train & Valid & Test & Train & Valid & Test \\
\midrule
LS & 1.80 & 0.58 & 0.42 & 0.09 & 0.03 & 0.03 & 0.15 & 0.00 & 0.14 \\
EN & 1.37 & 1.15 & 0.50 & 0.12 & 0.05 & 0.04 & 0.17 & 0.02 & 0.19 \\
FFN & 0.45 & 0.42 & 0.44 & 0.11 & 0.04 & 0.04 & 0.14 & -0.00 & 0.15 \\
\midrule
GAN & 2.68 & 1.43 & 0.75 & 0.20 & 0.09 & 0.08 & 0.12 & 0.01 & 0.23 \\
\bottomrule
\end{tabular}}
\bnotetab{This table shows the monthly Sharpe ratio (SR) of the SDF, explained time series variation (EV) and cross-sectional mean $R^2$ for the GAN, FFN, EN and LS model.}
\end{table}

 \begin{figure}[h!]
\centering
\tcapfig{Performance of Models with Different Macroeconomic Variables}\label{fig:no-macro}
\includegraphics[width=1\linewidth]{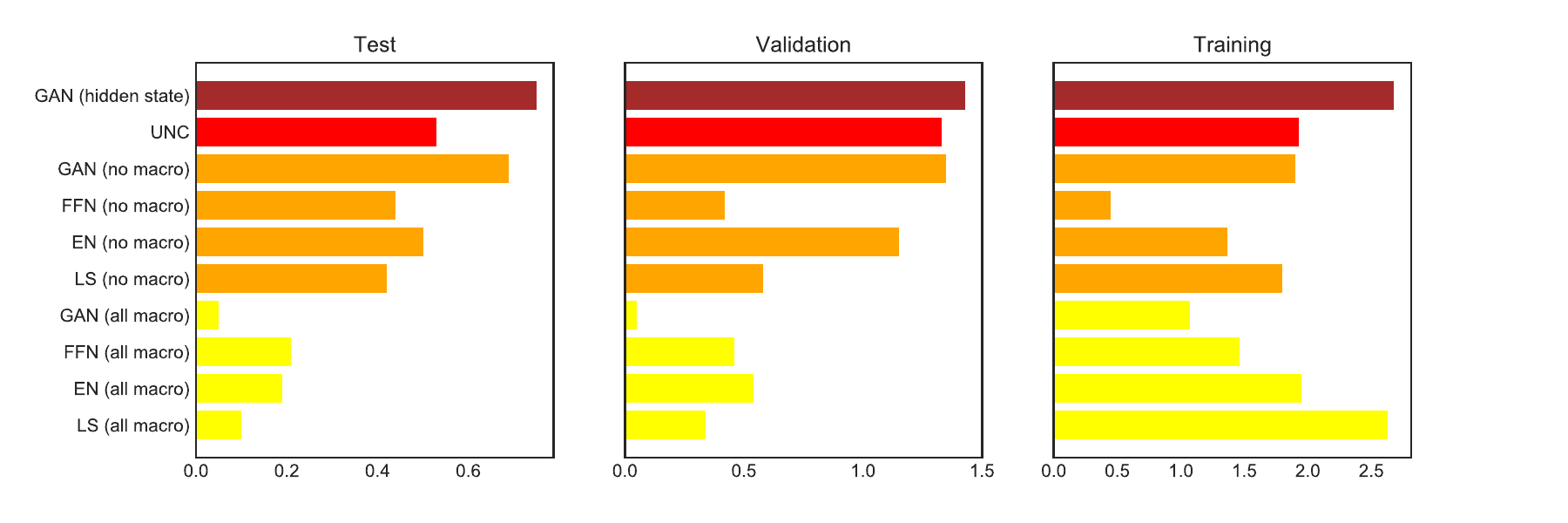}
\bnotefig{This figure shows the Sharpe ratio of SDFs for different inclusions of the macroeconomic information. The GAN (hidden states) is our reference model. UNC is a special version of our model that uses only unconditional moments (but includes LSTM macroeconomic states in the FFN network for the SDF weights). GAN (no macro), FFN (no macro), EN (no macro) and LS (no macro) use only firm specific information as conditioning variables but no macroeconomic variables. GAN (all macro), FFN (all macro), EN (all macro) and LS (all macro) include all 178 macro variables as predictors (respectively conditioning variables) without using LSTM to transform them into macroeconomic states.} 
\end{figure}

Figure \ref{fig:no-macro} summarizes the effect of conditioning on the hidden macroeconomic state variables. First, we add the 178 macroeconomic variables as predictors to all networks without reducing them to the hidden state variables. The performance for the out-of-sample Sharpe ratio of the LS, EN, FFN and GAN model completely collapses. First, conditioning only on the last normalized observation of the macroeconomic variables, which is usually an increment, does not allow to detect a dynamic structure, e.g. a business cycle. The decay in the Sharpe ratio indicates that using only the past macroeconomic information results in a loss of valuable information. Even worse, including the large number of irrelevant variables actually lowers the performance compared to a model without macroeconomic information. Although the models use a form of regularization, a too large number of irrelevant variables makes it harder to select those that are actually relevant. The results for the in-sample training data illustrate the complete overfitting when the large number of macroeconomic variables is included. FFN, EN and LS without macroeconomic information perform better and that is why we choose them as the comparison benchmark models. GAN without the macroeconomic but only firm-specific variables has an out-of-sample Sharpe ratio that is around 10\% lower than with the macroeconomic hidden states. This is another indication that it is relevant to include the dynamics of the time series. The UNC model uses only unconditional moments as the objective function, that is, we use a constant conditioning function $g$, but include the LSTM hidden states in the factor weights. The Sharpe ratio is around 20\% lower than the GAN with hidden states. These results confirm the insights from the last subsection. Hence, it is not only important to include all characteristics and the hidden states in the weights and loadings of SDF but also in the conditioning function $g$ to identify the assets and times that matter for pricing.

\subsection{Predictive Performance}

The no-arbitrage factor representation implies a connection between average returns of stocks and their risk exposure to the SDF measured by $\beta$. The fundamental equation
\begin{align*}
\E[\Ri]  &= \bi \E[F_{t+1}]
\end{align*}
implies that as long as the conditional risk premium $\E[F_{t+1}]$ is positive, which is required by no-arbitrage, assets with a higher risk exposure $\bi$ should have higher expected returns. We test the predictive power of our model by sorting stocks into decile portfolios based on their risk loadings.

\begin{figure}[h!]
\centering
\tcapfig{Cumulative Excess Return of Decile Sorted Portfolios with GAN}\label{fig:beta-sorting}
\includegraphics[width=0.75\linewidth]{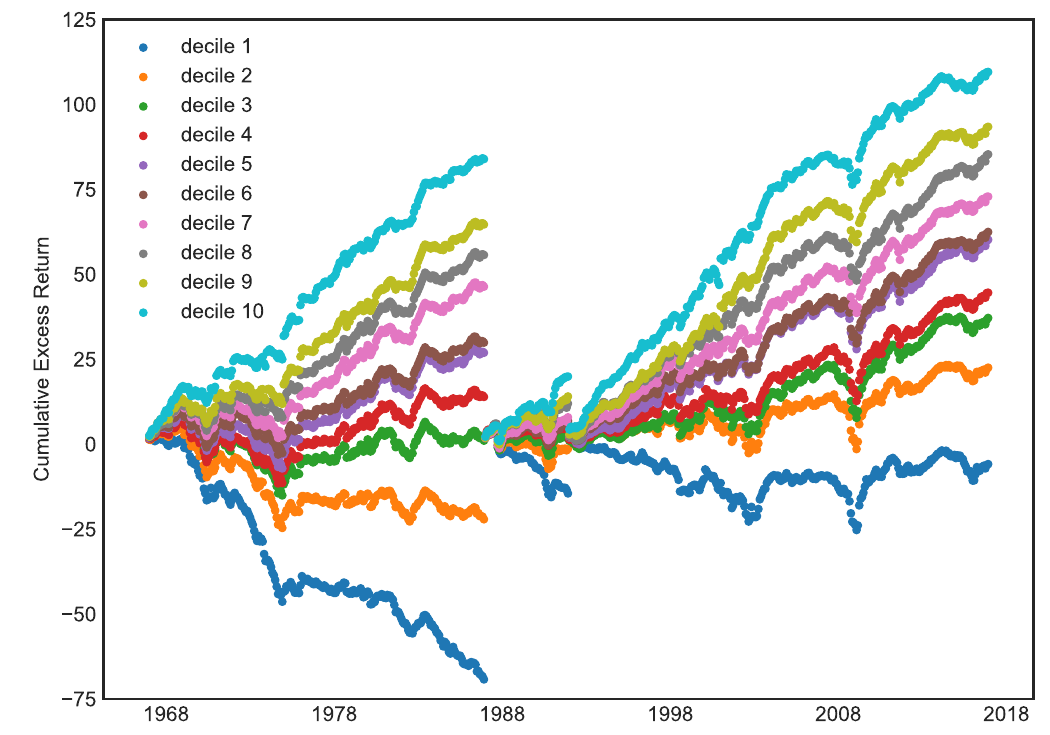}
\bnotefig{This figure shows the cumulative excess return of decile sorted portfolios based on the risk loadings $\beta$. The first portfolio is based on the smallest decile of risk loadings, while the last decile portfolio is constructed with the largest loading decile. Within each decile the stocks are equally weighted.}
\end{figure}

\begin{figure}[h!]
  \tcapfig{Expected Excess Returns of $\beta$-Sorted Portfolios as a Function of $\beta$}\label{fig:betaline}
  \begin{subfigure}[t]{.33\textwidth}
  \centering
    \includegraphics[width=1.0\textwidth]{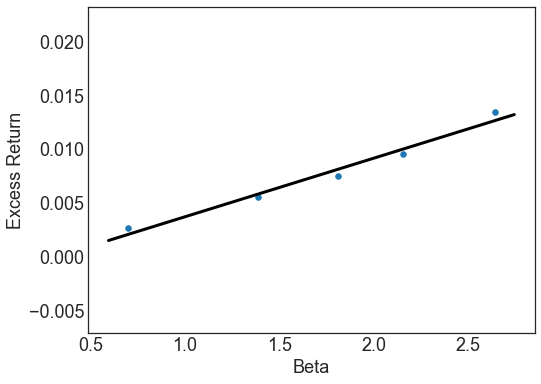}
    \caption{\small 5 $\beta$-sorted quintiles}
  \end{subfigure}\hfill
  \begin{subfigure}[t]{.33\textwidth}
  \centering
    \includegraphics[width=1.0\textwidth]{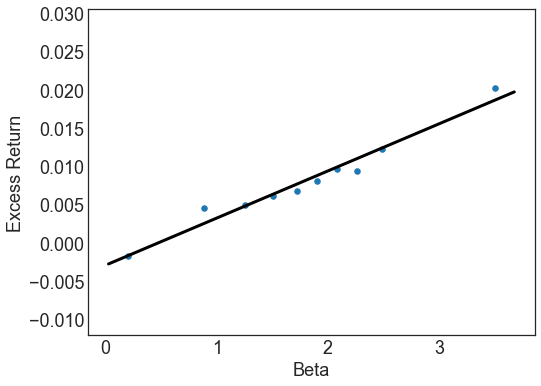}
    \caption{\small 10 $\beta$-sorted deciles}
  \end{subfigure}
    \begin{subfigure}[t]{.33\textwidth}
  \centering
    \includegraphics[width=1.0\textwidth]{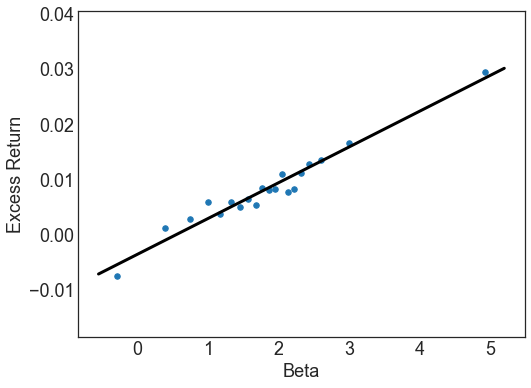}
    \caption{\small 20 $\beta$-sorted quantiles}
  \end{subfigure}\hfill
\bnotefig{This figure shows expected excess returns of $\beta$-sorted portfolios for GAN on the test sample. Stocks are sorted into 5, 10 or 20 quantiles every month. We plot them against the $\beta$ of each portfolio that averages the individual stock $\beta_{t,i}$ over stocks and time within each portfolio. The linear line denotes a linear regression with intercept, which yields an $R^2$ of 0.98 (quintiles), 0.97 (deciles) and 0.95 (20 quantiles). The SDF itself has a $\beta$ equal to one.}
\end{figure}

In Figure \ref{fig:beta-sorting} we plot the cumulative excess return of decile sorted portfolios based on risk loadings $\beta$'s. Portfolios based on higher $\beta$'s have higher subsequent returns. This clearly indicates that risk loadings predict future stock returns. In particular, the highest and lowest deciles clearly separate. The Internet Appendix collects the corresponding results for the other estimation approaches with qualitatively similar findings, i.e. the risk loadings predict future returns.

The no-arbitrage condition does not only apply a monotonic but a linear relationship between stock $\beta$'s and conditional expected returns. An unconditional one-factor model such as the CAPM can be tested by evaluating the fit of the security market line, i.e. the relationship between expected returns and the $\beta$'s of the assets. In our case the stock specific $\beta_{t,i}$ are time-varying, but by construction the $\beta$-sorted portfolios should have a close to constant slope with respect to the SDF.\footnote{Our estimated loadings are only proportional to the SDF $\beta_t$. At each time $t$ we scale the loadings such that the SDF portfolio has a loading of one. This is equivalent to obtaining the correctly scaled $\beta_t$ without explicitly calculating the second conditional moment of the SDF.} In Figure \ref{fig:betaline} we plot the expected excess returns of the 10 $\beta$-sorted deciles as well as for 5 and 20 $\beta$-sorted quantile portfolios against their average $\beta$'s. No-arbitrage imposes a linear relationship and a zero intercept. Indeed, for all three plots the relationship is almost perfectly linear with a $R^2$ of 0.98, 0.97 and 0.95 respectively. However, the intercept seems to be slightly below zero. This indicates a very good but not perfect fit.

\begin{table}[h!]
{\footnotesize
\tcaptab{~Time Series Pricing Errors for $\beta$-Sorted Portfolios}\label{tab:betaalpha}
\scalebox{0.92}{
\begin{tabular}{ccc||cccc||cccc||cccc}
\toprule
 & \multicolumn{2}{c||}{Average Returns} & \multicolumn{4}{c||}{Market-Rf} & \multicolumn{4}{c||}{Fama-French 3} & \multicolumn{4}{c}{Fama-French 5}\\
\cmidrule{2-15}
 & Full & Test & \multicolumn{2}{c}{Full} & \multicolumn{2}{c||}{Test} & \multicolumn{2}{c}{Full} & \multicolumn{2}{c||}{Test} & \multicolumn{2}{c}{Full} & \multicolumn{2}{c}{Test} \\
 Decile & & & $\alpha$ & t & $\alpha$ & t & $\alpha$ & t & $\alpha$ & t & $\alpha$ & t & $\alpha$ & t \\
\midrule
1 & -0.12 & -0.02 & -0.19 & -8.92 & -0.11 & -3.43 & -0.21 & -12.77 & -0.13 & -5.01 & -0.20 & -11.99 & -0.12 & -4.35 \\
2 & -0.00 & 0.05 & -0.07 & -4.99 & -0.04 & -1.56 & -0.09 & -8.79 & -0.05 & -3.22 & -0.09 & -8.29 & -0.05 & -2.68 \\
3 & 0.04 & 0.08 & -0.02 & -2.01 & -0.00 & -0.16 & -0.04 & -5.18 & -0.02 & -1.40 & -0.04 & -4.87 & -0.01 & -1.05 \\
4 & 0.07 & 0.09 & -0.00 & -0.03 & 0.01 & 0.68 & -0.02 & -2.30 & -0.00 & -0.35 & -0.02 & -2.86 & -0.01 & -0.54 \\
5 & 0.10 & 0.12 & 0.03 & 2.75 & 0.04 & 2.50 & 0.01 & 2.08 & 0.03 & 2.46 & 0.01 & 1.36 & 0.03 & 2.17 \\
6 & 0.11 & 0.12 & 0.04 & 3.16 & 0.05 & 2.77 & 0.02 & 2.75 & 0.03 & 2.85 & 0.01 & 1.51 & 0.02 & 2.20 \\
7 & 0.14 & 0.15 & 0.07 & 5.62 & 0.07 & 3.92 & 0.05 & 6.61 & 0.05 & 4.39 & 0.04 & 5.16 & 0.04 & 3.41 \\
8 & 0.18 & 0.18 & 0.11 & 7.41 & 0.10 & 5.12 & 0.08 & 9.32 & 0.08 & 5.83 & 0.07 & 8.05 & 0.07 & 4.86 \\
9 & 0.22 & 0.21 & 0.15 & 7.83 & 0.13 & 5.37 & 0.11 & 9.16 & 0.11 & 5.71 & 0.11 & 8.58 & 0.11 & 5.39 \\
10 & 0.37 & 0.37 & 0.29 & 9.22 & 0.27 & 6.05 & 0.24 & 10.03 & 0.25 & 6.27 & 0.25 & 10.43 & 0.27 & 6.59 \\
\midrule
10-1 & 0.48 & 0.39 & 0.47 & 18.93 & 0.38 & 10.29 & 0.45 & 18.50 & 0.38 & 10.14 & 0.46 & 18.13 & 0.39 & 9.96 \\

\midrule
  \multicolumn{3}{c||}{GRS Asset Pricing Test} & GRS & p & GRS & p & GRS & p & GRS & p & GRS & p & GRS & p \\
  & & & 42.23 & 0.00 & 11.58 & 0.00 & 39.72 & 0.00 & 11.25 & 0.00 & 37.64 & 0.00 & 10.75 & 0.00 \\
\bottomrule
\end{tabular}}
}
\bnotetab{This table shows the average returns, time series pricing errors and corresponding t-statistics for $\beta$-sorted decile portfolios based on GAN. The pricing errors are based on the CAPM and Fama-French 3 and 5 factors models. Returns are annualized. The GRS-test is under the null hypothesis of correctly pricing all decile portfolios and includes the p-values. We consider the full time period and the test period. Within each decile the stocks are equally weighted.}
\end{table}

The systematic return difference of the $\beta$-sorted portfolios is not explained by the market or Fama-French factors. Table \ref{tab:betaalpha} reports the time series pricing errors with corresponding t-statistics for the 10 decile-sorted portfolios for the three factor models. Obviously, the pricing errors are highly significant and expected returns of almost all decile portfolios are not explained by the Fama-French factors. The GRS test clearly rejects the null-hypothesis that either of the factor models prices this cross-section. These $\beta$-sorted portfolios equally weight the stocks within each decile. The Internet Appendix shows that the findings extend to value weighted $\beta$-sorted portfolios.

\subsection{Pricing of Characteristic Sorted Portfolios}

Our approach achieves an unprecedented pricing performance on standard test portfolios. Asset pricing testing is usually conducted on characteristic sorted portfolios that isolate the pricing effect of a small number of characteristics. We sort the stocks into value weighted decile and double-sorted 25 portfolios based on the characteristics.\footnote{The Internet Appendix collects the results for additional characteristic sorts with similar findings. Here we report only the results for value weighted portfolios. The results for equally weighted portfolios are similar. The results for the unregularized linear model are the worst and available upon request.} 
Table \ref{tab:decile4} starts with four sets of decile sorted portfolios. We choose short-term reversal and momentum as these are among the two most important variables as discussed in the next sections and size and book-to-market sorted portfolios which are well-studied characteristics. GAN can substantially better capture the variation and mean return for short-term reversal and momentum sorted decile portfolios. EN and FFN have a very similar performance. The better GAN results are driven by explaining the extreme decile portfolios (the 10th decile for short-term reversal and the first decile for momentum). All approaches perform very similarly for the middle portfolios. It turns out that book-to-market and size sorted portfolios are very ``easy'' to price. All models have time series $R^2$ above 70\% and cross-sectional $R^2$ close to 1. Hence, all models seems to capture this pricing information very well, although the GAN results are still slightly better than for the other models.


\begin{table}[h!]
{\footnotesize
\tcaptab{~Explained Variation and Pricing Errors for Decile Sorted Portfolios}\label{tab:decile}
\centering
\begin{tabular}{lccc|ccc||lccc|ccc}
\toprule
& \multicolumn{3}{c|}{Explained Variation} & \multicolumn{3}{c||}{Cross-Sectional $R^2$} &  &\multicolumn{3}{c|}{Explained Variation} & \multicolumn{3}{c}{Cross-Sectional $R^2$}\\
\midrule
 Charact. & EN& FFN & GAN & EN& FFN & GAN  &Charact. & EN & FFN & GAN & EN& FFN & GAN \\
 \midrule
ST\_REV & 0.43 & 0.58 & 0.70 & 0.45 & 0.79 & 0.94 & Q & 0.68 & 0.70 & 0.78 & 0.97 & 0.92 & 0.96 \\
SUV & 0.42 & 0.75 & 0.83 & 0.64 & 0.97 & 0.99 & Investment & 0.54 & 0.65 & 0.75 & 0.91 & 0.94 & 0.98 \\
r12\_2 & 0.26 & 0.27 & 0.54 & 0.66 & 0.71 & 0.93 & PM & 0.52 & 0.42 & 0.68 & 0.90 & 0.86 & 0.93 \\
NOA & 0.58 & 0.69 & 0.78 & 0.94 & 0.96 & 0.95 & DPI2A & 0.57 & 0.70 & 0.78 & 0.90 & 0.95 & 0.97 \\
SGA2S & 0.52 & 0.63 & 0.73 & 0.93 & 0.95 & 0.96 & ROE & 0.59 & 0.56 & 0.76 & 0.91 & 0.86 & 0.97 \\
LME & 0.83 & 0.78 & 0.86 & 0.96 & 0.95 & 0.97 & S2P & 0.69 & 0.79 & 0.82 & 0.98 & 0.98 & 0.97 \\
RNA & 0.50 & 0.48 & 0.69 & 0.93 & 0.87 & 0.96 & FC2Y & 0.56 & 0.71 & 0.76 & 0.91 & 0.94 & 0.95 \\
LTurnover & 0.52 & 0.57 & 0.68 & 0.88 & 0.89 & 0.96 & AC & 0.63 & 0.79 & 0.82 & 0.96 & 0.98 & 0.98 \\
Lev & 0.52 & 0.63 & 0.73 & 0.90 & 0.92 & 0.95 & CTO & 0.59 & 0.73 & 0.79 & 0.92 & 0.96 & 0.97 \\
Resid\_Var & 0.52 & 0.27 & 0.65 & 0.84 & 0.73 & 0.97 & LT\_Rev & 0.60 & 0.59 & 0.72 & 0.93 & 0.85 & 0.94 \\
ROA & 0.51 & 0.44 & 0.70 & 0.92 & 0.93 & 0.98 & OP & 0.56 & 0.48 & 0.74 & 0.97 & 0.88 & 0.98 \\
E2P & 0.48 & 0.44 & 0.67 & 0.86 & 0.80 & 0.95 & PROF & 0.58 & 0.62 & 0.76 & 0.91 & 0.98 & 0.95 \\
D2P & 0.47 & 0.51 & 0.72 & 0.82 & 0.85 & 0.94 & IdioVol & 0.43 & 0.27 & 0.66 & 0.79 & 0.72 & 0.97 \\
Spread & 0.49 & 0.32 & 0.60 & 0.76 & 0.71 & 0.92 & r12\_7 & 0.37 & 0.42 & 0.66 & 0.84 & 0.86 & 0.93 \\
CF2P & 0.46 & 0.47 & 0.66 & 0.90 & 0.89 & 0.99 & Beta & 0.45 & 0.46 & 0.62 & 0.83 & 0.87 & 0.97 \\
BEME & 0.70 & 0.75 & 0.82 & 0.97 & 0.94 & 0.98 & OA & 0.65 & 0.78 & 0.83 & 0.88 & 0.92 & 0.93 \\
Variance & 0.48 & 0.27 & 0.61 & 0.74 & 0.72 & 0.90 & ATO & 0.58 & 0.70 & 0.77 & 0.96 & 0.98 & 0.99 \\
D2A & 0.57 & 0.71 & 0.78 & 0.96 & 0.96 & 0.97 & MktBeta & 0.44 & 0.44 & 0.64 & 0.81 & 0.85 & 0.97 \\
PCM & 0.66 & 0.79 & 0.82 & 0.97 & 0.98 & 0.99 & OL & 0.60 & 0.73 & 0.78 & 0.95 & 0.97 & 0.97 \\
A2ME & 0.72 & 0.79 & 0.83 & 0.97 & 0.96 & 0.98 & C & 0.51 & 0.65 & 0.73 & 0.90 & 0.93 & 0.95 \\
AT & 0.77 & 0.70 & 0.83 & 0.77 & 0.89 & 0.92 & r36\_13 & 0.54 & 0.53 & 0.69 & 0.92 & 0.82 & 0.93 \\
Rel2High & 0.46 & 0.33 & 0.60 & 0.90 & 0.83 & 0.97 & NI & 0.51 & 0.60 & 0.75 & 0.88 & 0.96 & 0.99 \\
CF & 0.61 & 0.64 & 0.78 & 0.89 & 0.85 & 0.96 & r2\_1 & 0.51 & 0.52 & 0.69 & 0.87 & 0.90 & 0.95 \\
\bottomrule
\end{tabular}
\bnotetab{This table shows the out-of-sample explained variation and cross-section $R^2$ for 46 decile-sorted and value weighted portfolios.}
}
\end{table}

\begin{figure}[h!]
\tcapfig{Predicted returns for value weighted characteristic sorted portfolios}\label{fig:PredictedReturnVw}
  \begin{subfigure}[t]{.48\textwidth}
  \centering
    \includegraphics[width=1.0\textwidth]{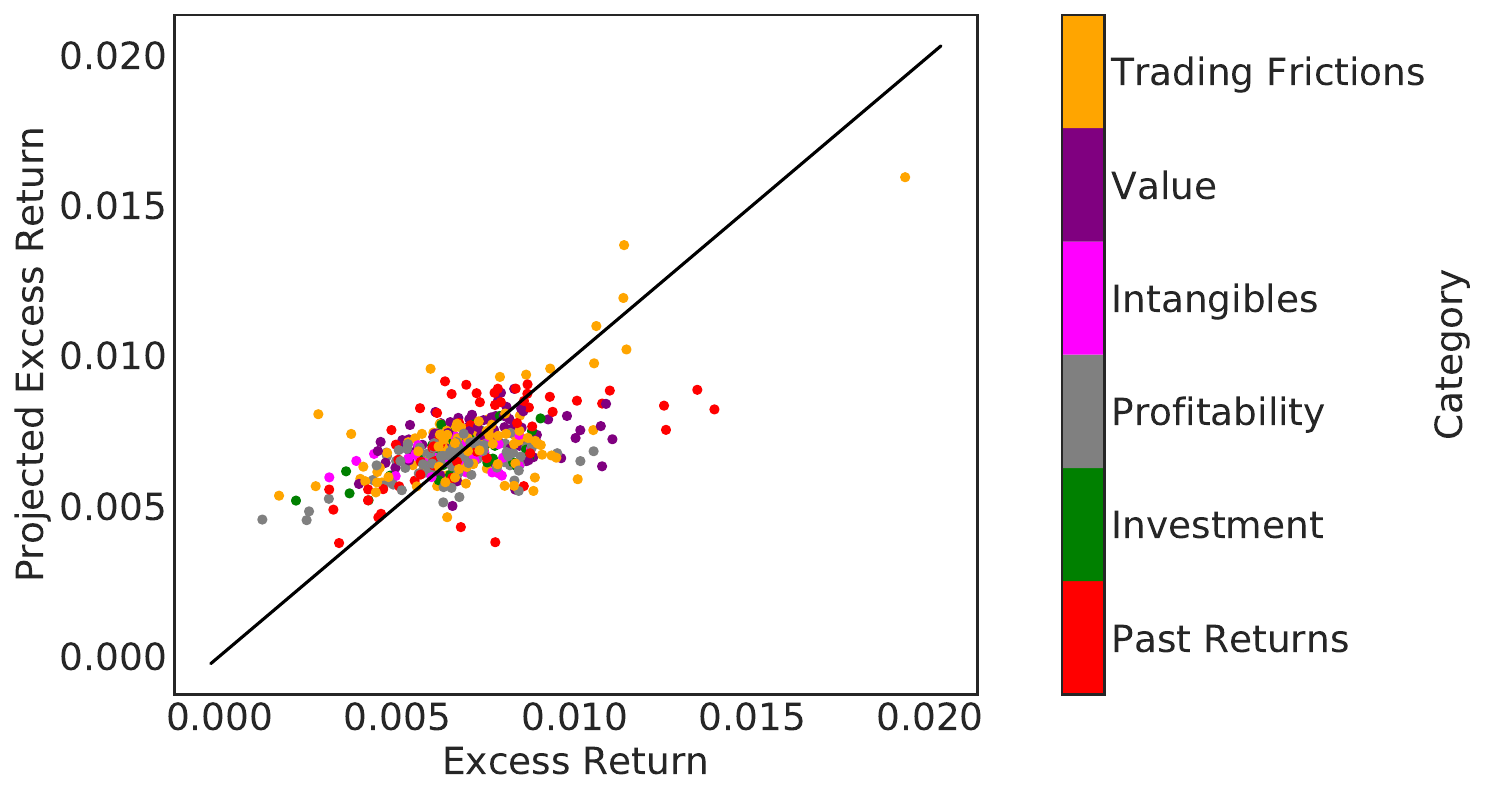}
    \caption{GAN}
  \end{subfigure}\hfill
  \begin{subfigure}[t]{.48\textwidth}
  \centering
    \includegraphics[width=1.0\textwidth]{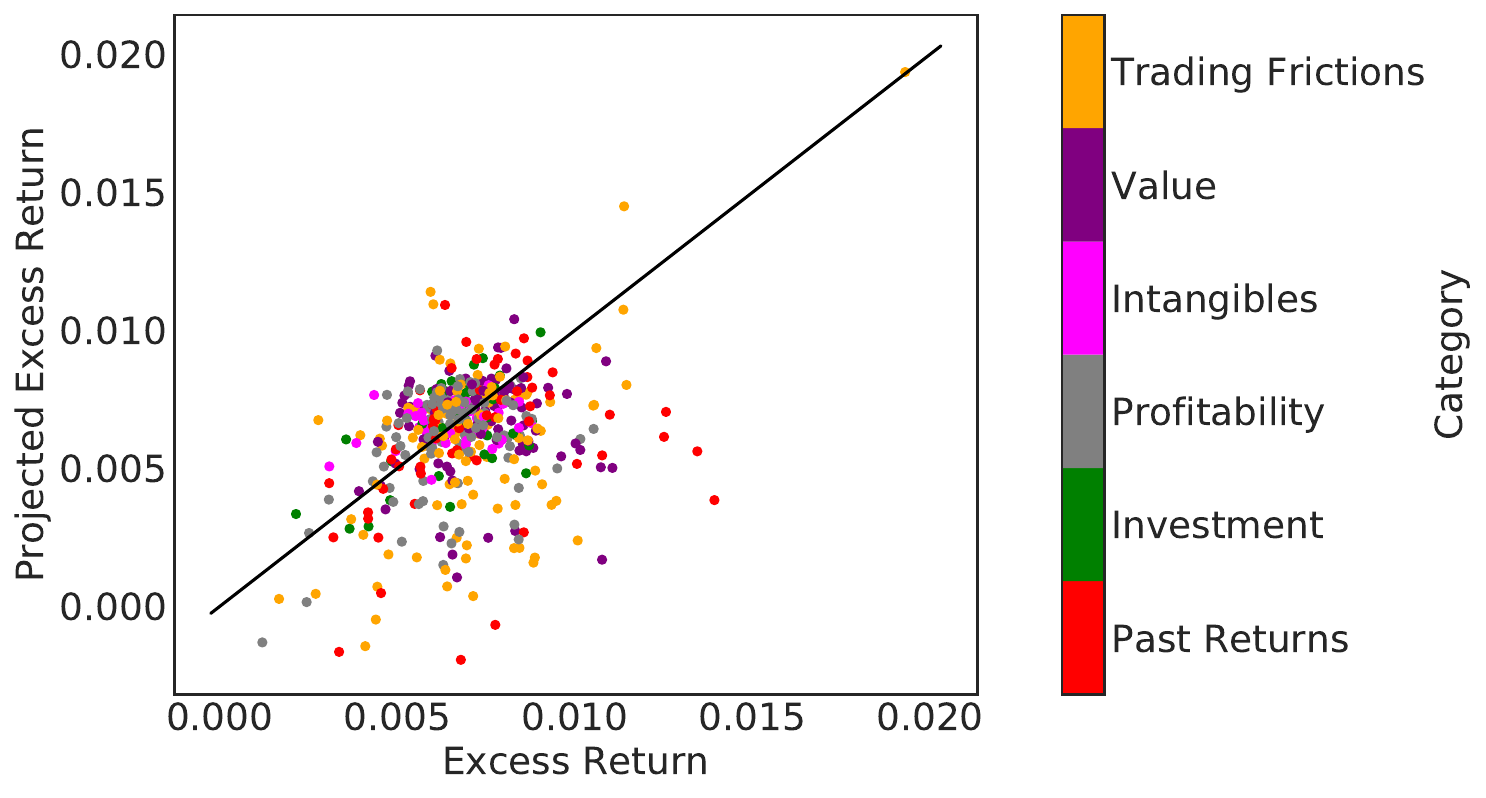}
    \caption{FFN}
  \end{subfigure}
    \begin{subfigure}[t]{.48\textwidth}
  \centering
    \includegraphics[width=1.0\textwidth]{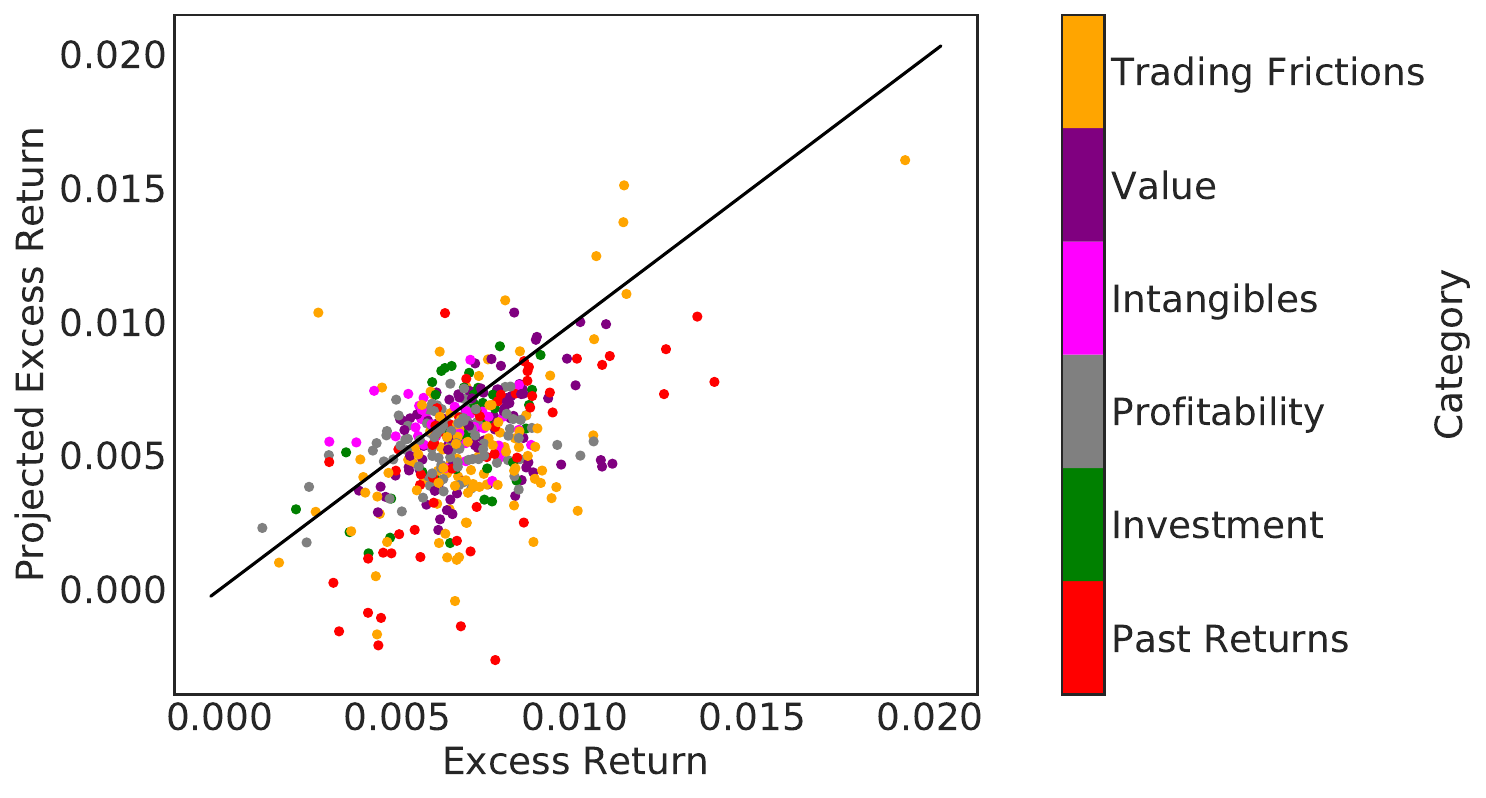}
    \caption{EN}
  \end{subfigure}\hfill
  \begin{subfigure}[t]{.48\textwidth}
  \centering
    \includegraphics[width=1.0\textwidth]{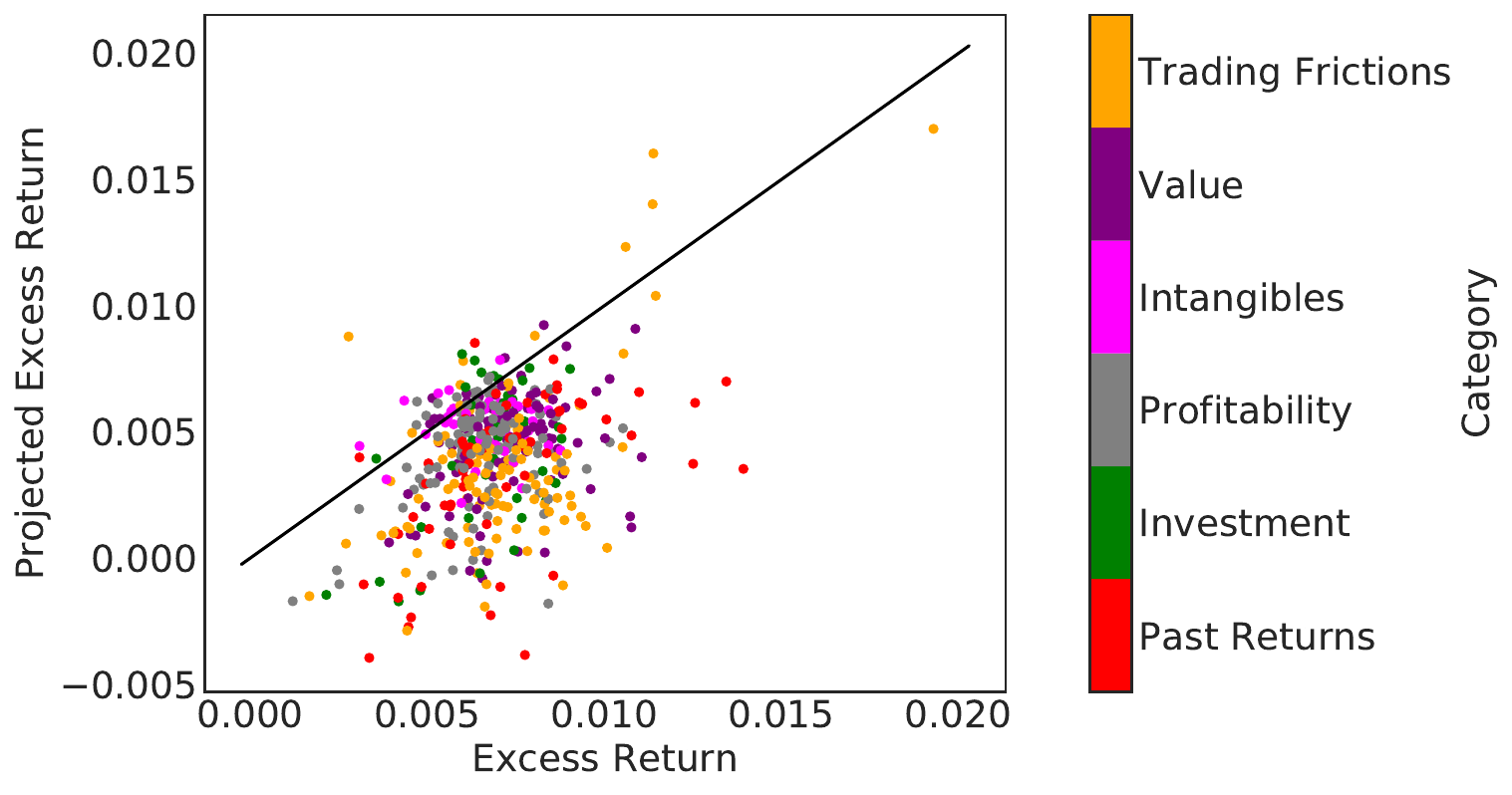}
    \caption{LS}
  \end{subfigure}
\bnotefig{This figure shows the predicted and average excess returns for value weighted characteristic sorted decile portfolios for the four SDF models. We have in total 460 decile portfolios in six different anomaly categories.}
\end{figure}

Table \ref{tab:double_sort_mom} repeats the same analysis on short-term reversal and momentum double-sorted and size and book-to-market double-sorted portfolios. The takeaways are similar to the decile sorted portfolios. GAN outperforms FFN and EN on the momentum related portfolios, while all three models are able to explain the size and value double-sorted portfolios. Importantly, the linear EN becomes worse on the double-sorted reversal and momentum portfolios. This is due to the extreme corner portfolios, which are in particular low momentum and high short-term reversal stocks. This implies that the linear model cannot capture the interaction between characteristics, while the GAN model successfully identifies the potentially non-linear interaction effects.

Our findings generalize to other decile sorted portfolios. Table \ref{tab:decile} collects the explained variation and cross-sectional $R^2$ for all decile-sorted portfolios. It is striking that GAN is always better than the other two models in explaining variation. At the same time GAN achieves a cross-sectional $R^2$ higher than 90\% for all characteristics. In the few cases where the other models have a slightly higher cross-sectional $R^2$, this number is very close to 1, i.e. all models can essentially perfectly explain the pricing information in the deciles. In summary GAN strongly dominates the other methods in explaining sorted portfolios. The results show (1) that the non-linearities and interactions matter as GAN is better than EN and (2) the no-arbitrage condition extracts additional information as GAN is better than FFN.

\begin{figure}[h!]
\tcapfig{Predicted returns for equally weighted characteristic sorted portfolios}\label{fig:PredictedReturnEw}
  \begin{subfigure}[t]{.48\textwidth}
  \centering
    \includegraphics[width=1.0\textwidth]{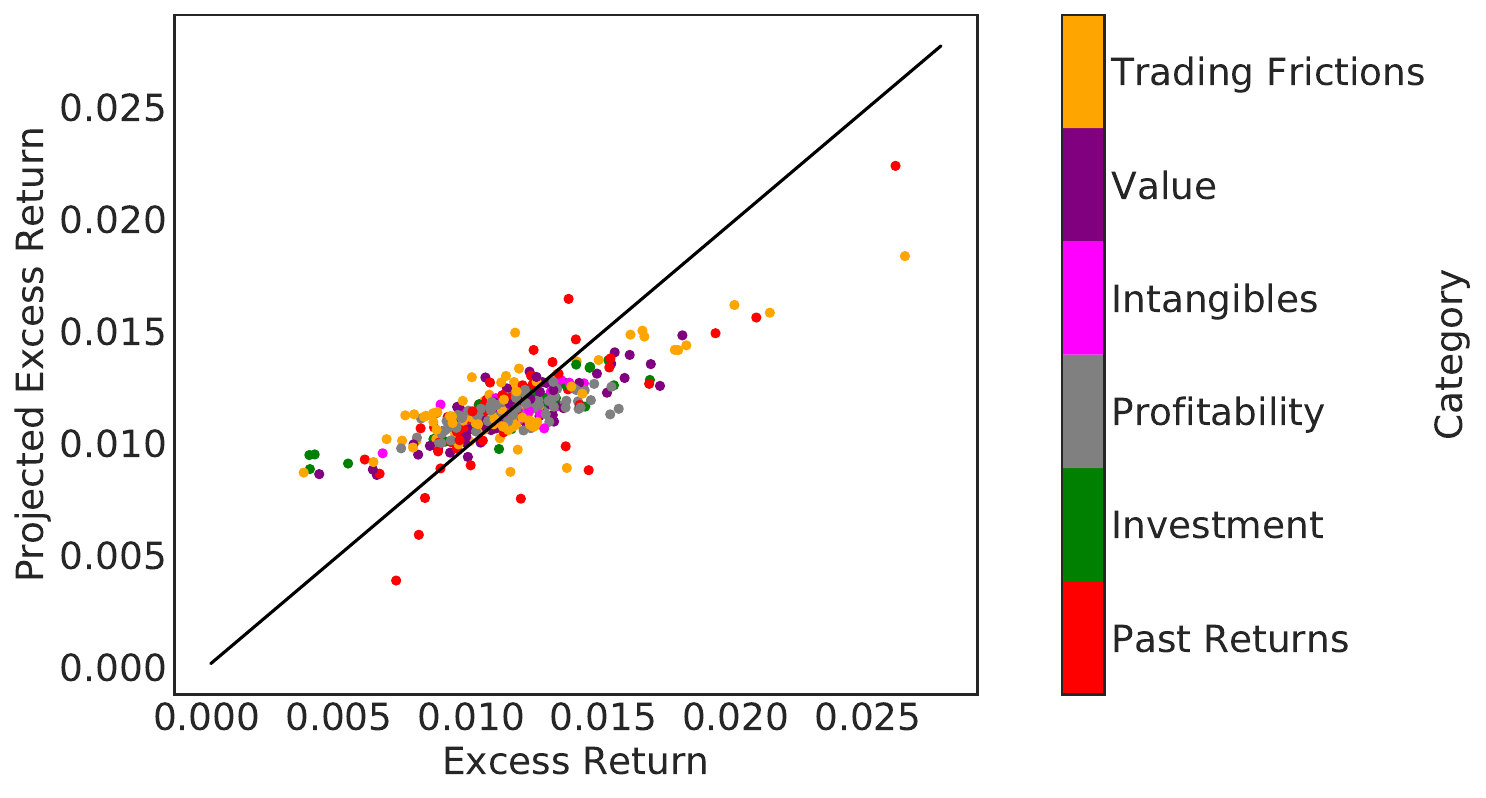}
    \caption{GAN}
  \end{subfigure}\hfill
  \begin{subfigure}[t]{.48\textwidth}
  \centering
    \includegraphics[width=1.0\textwidth]{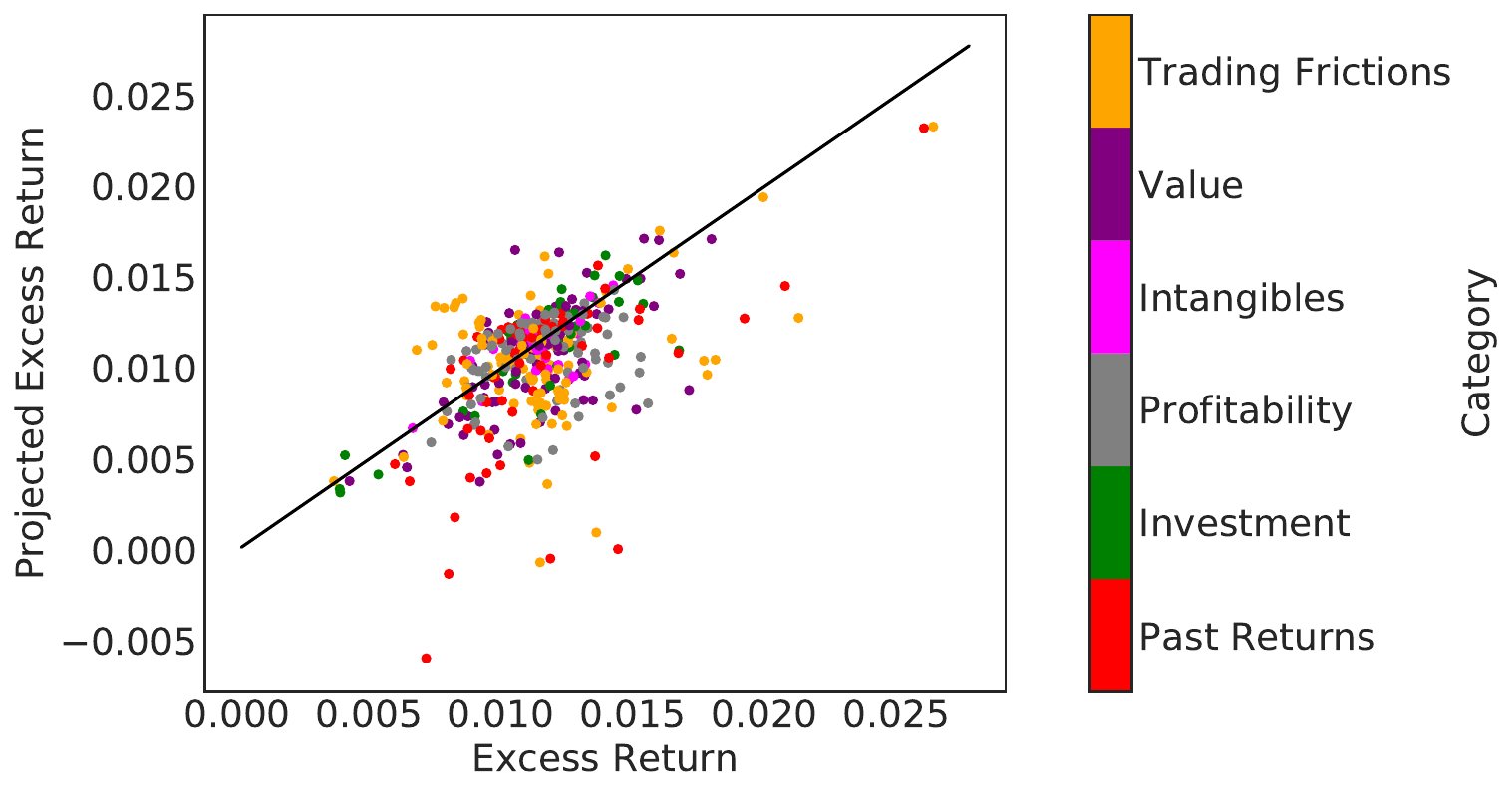}
    \caption{FFN}
  \end{subfigure}
    \begin{subfigure}[t]{.48\textwidth}
  \centering
    \includegraphics[width=1.0\textwidth]{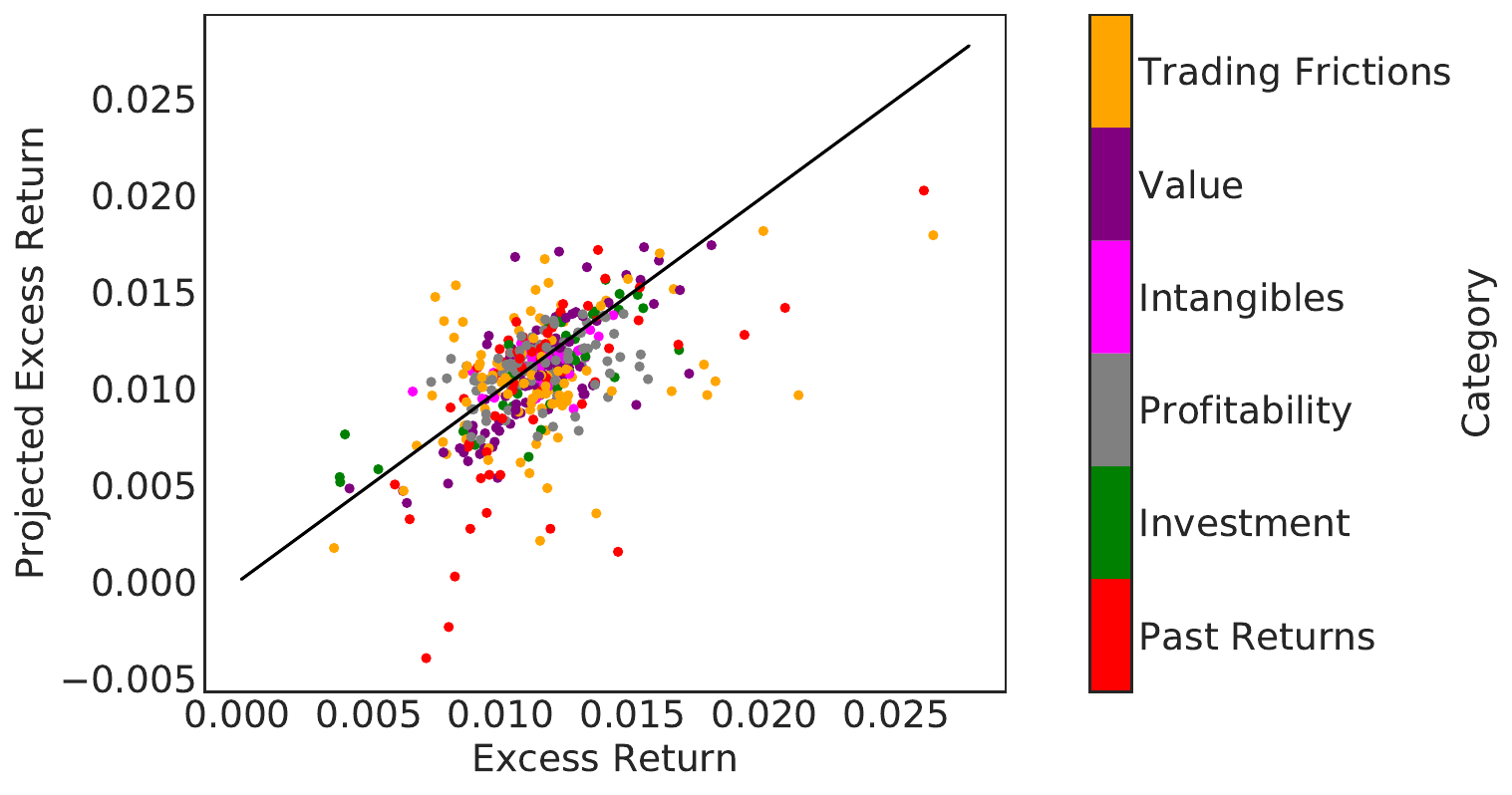}
    \caption{EN}
  \end{subfigure}\hfill
  \begin{subfigure}[t]{.48\textwidth}
  \centering
    \includegraphics[width=1.0\textwidth]{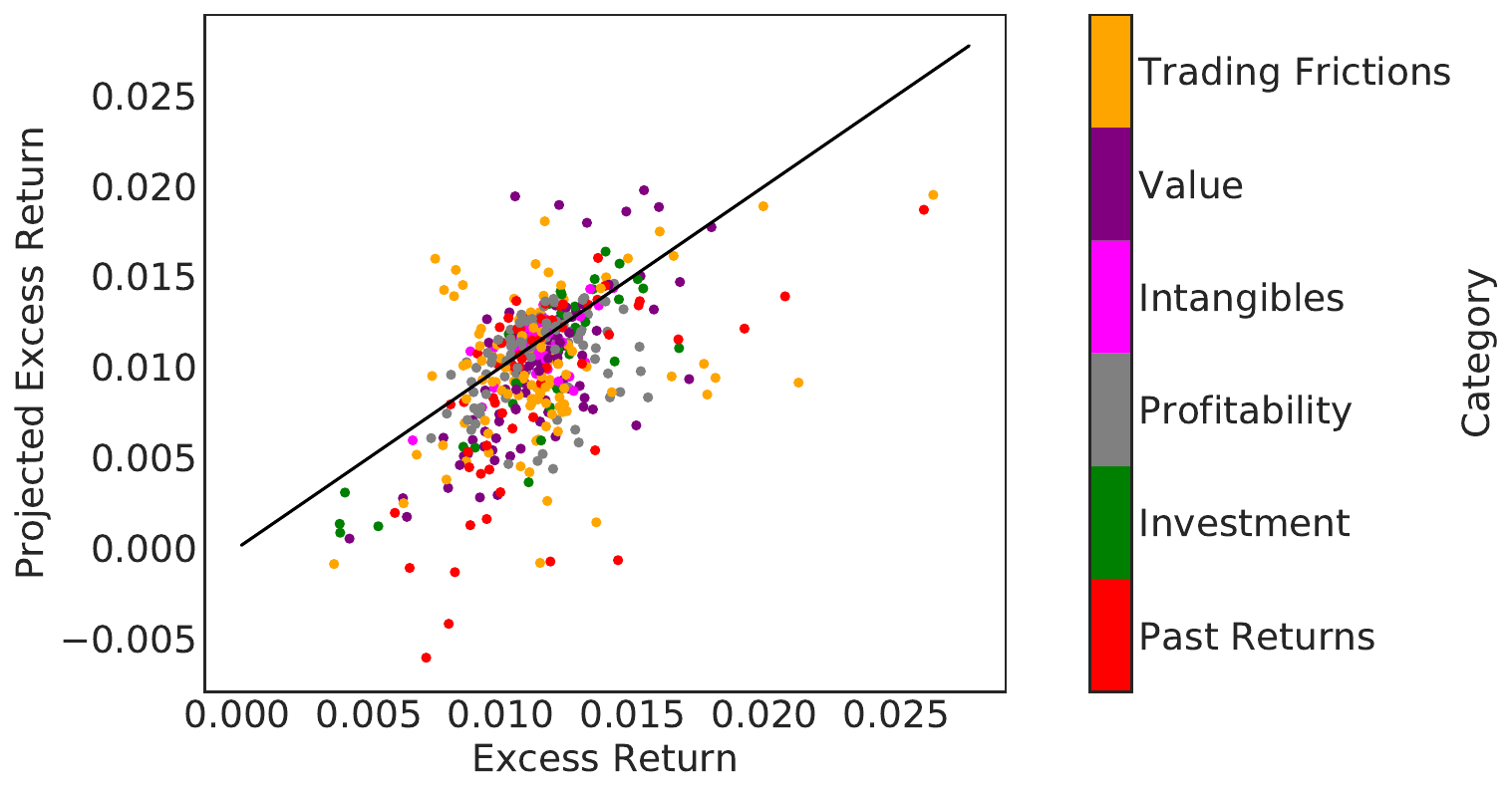}
    \caption{LS}
  \end{subfigure}
\bnotefig{This figure shows the predicted and average excess returns for equally weighted characteristic sorted decile portfolios for the four SDF models. We have in total 460 decile portfolios in six different anomaly categories.}
\end{figure}

Figure \ref{fig:PredictedReturnVw} visualizes the ability of GAN to explain the cross-section of expected returns for all value weighted characteristic sorted deciles. We plot the average excess return and the model implied average excess return. The GAN SDF captures the correct monotonic behavior, but its prediction is biased towards the mean. In contrast, the prediction of the other three models show a larger discrepancy which holds for characteristics of all groups. Figure \ref{fig:PredictedReturnEw} shows the prediction results for equally weighted decile portfolios. All models seem to perform slightly better, but the general findings are the same.

\subsection{Variable Importance}

What is the structure of the SDF factor? As a first step in Table \ref{tab:corr} we compare the GAN factor with the Fama-French 5 factor model. None of the five factors has a high correlation with our factor with the profitability factor having the highest correlation with 17\%. The market factor has only a correlation of 10\%. Next, we run a time series regression to explain the GAN factor portfolio with the Fama-French 5 factors. Only the profitability factor is significant. The strongly significant pricing error indicates that these factors fail to capture the pricing information in our SDF portfolio.

We rank the importance of firm-specific and macroeconomic variables for the pricing kernel based on the sensitivity of the SDF weight $\omega$ with respect to these variables. Our sensitivity analysis is similar to \cite{sirignano2016deep} and \cite{horel2019} and based on the average absolute gradient. More specifically, we define the sensitivity of a particular variable as the average absolute derivative of the weight $w$ with respect to this variable: 
\begin{align*}
\text{Sensitivity}(x_j)=\frac{1}{C}\sum_{i=1}^N\sum_{t=1}^T \Big|\frac{\partial w(I_t, I_{t,i})}{\partial x_j}\Big|,
\end{align*}
where $C$ a normalization constant. This simplifies to the standard slope coefficient in the special case of a linear regression framework. A larger sensitivity means that a variable has a larger effect on the SDF weight $\omega$.


\begin{figure}[h!]
\centering
\tcapfig{Characteristic Importance for GAN SDF}
\includegraphics[width=0.78\linewidth]{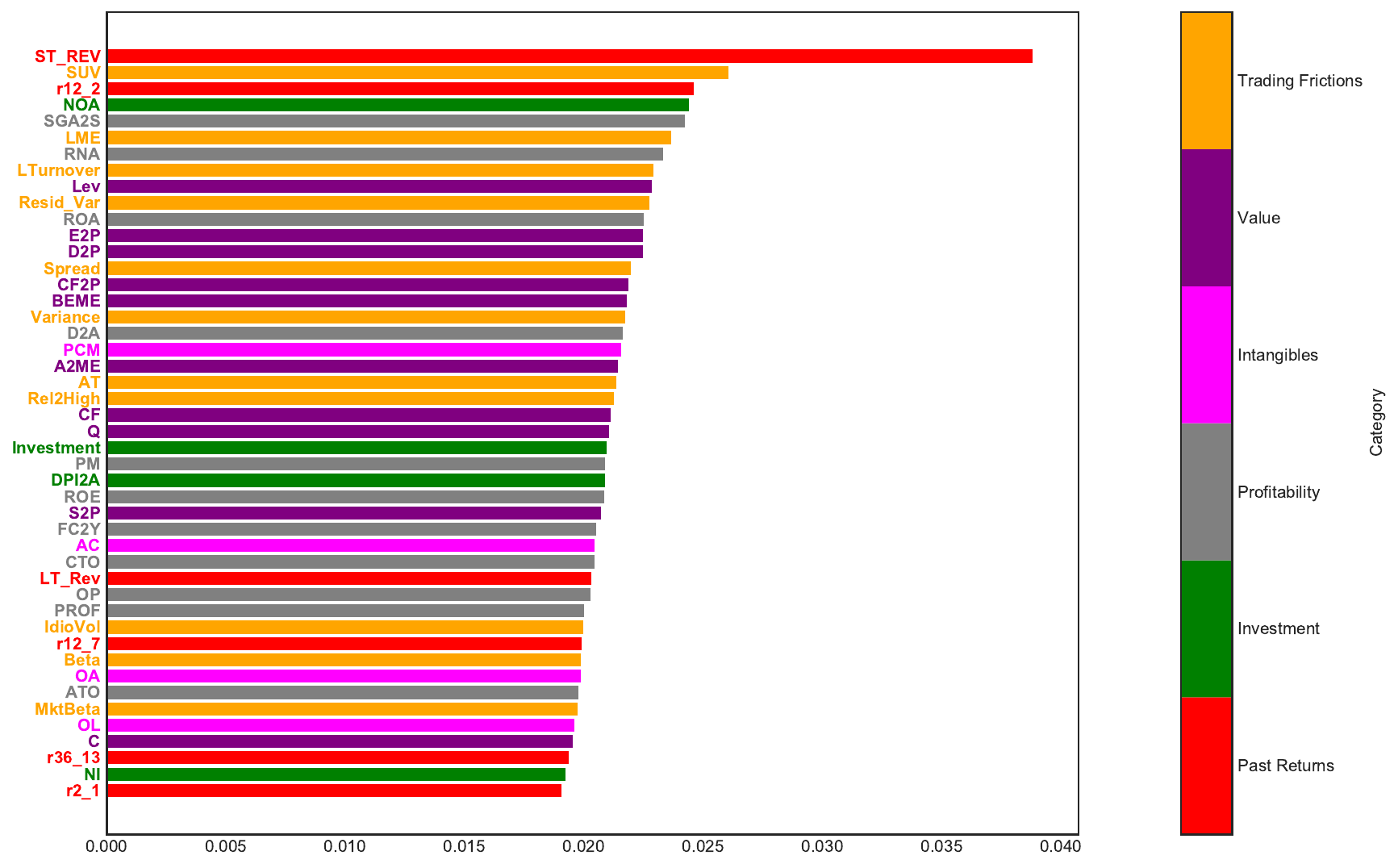}\label{fig:VI-GAN}
\bnotefig{The figure shows the GAN variable importance ranking of the 46 firm-specific characteristics in terms of average absolute gradient (VI) on the test data. The values are normalized to sum up to one. }
\end{figure}

\begin{figure}[h!]
\centering
\tcapfig{Characteristic Importance for FFN SDF}\label{fig:VI-FFN}
\includegraphics[width=0.78\linewidth]{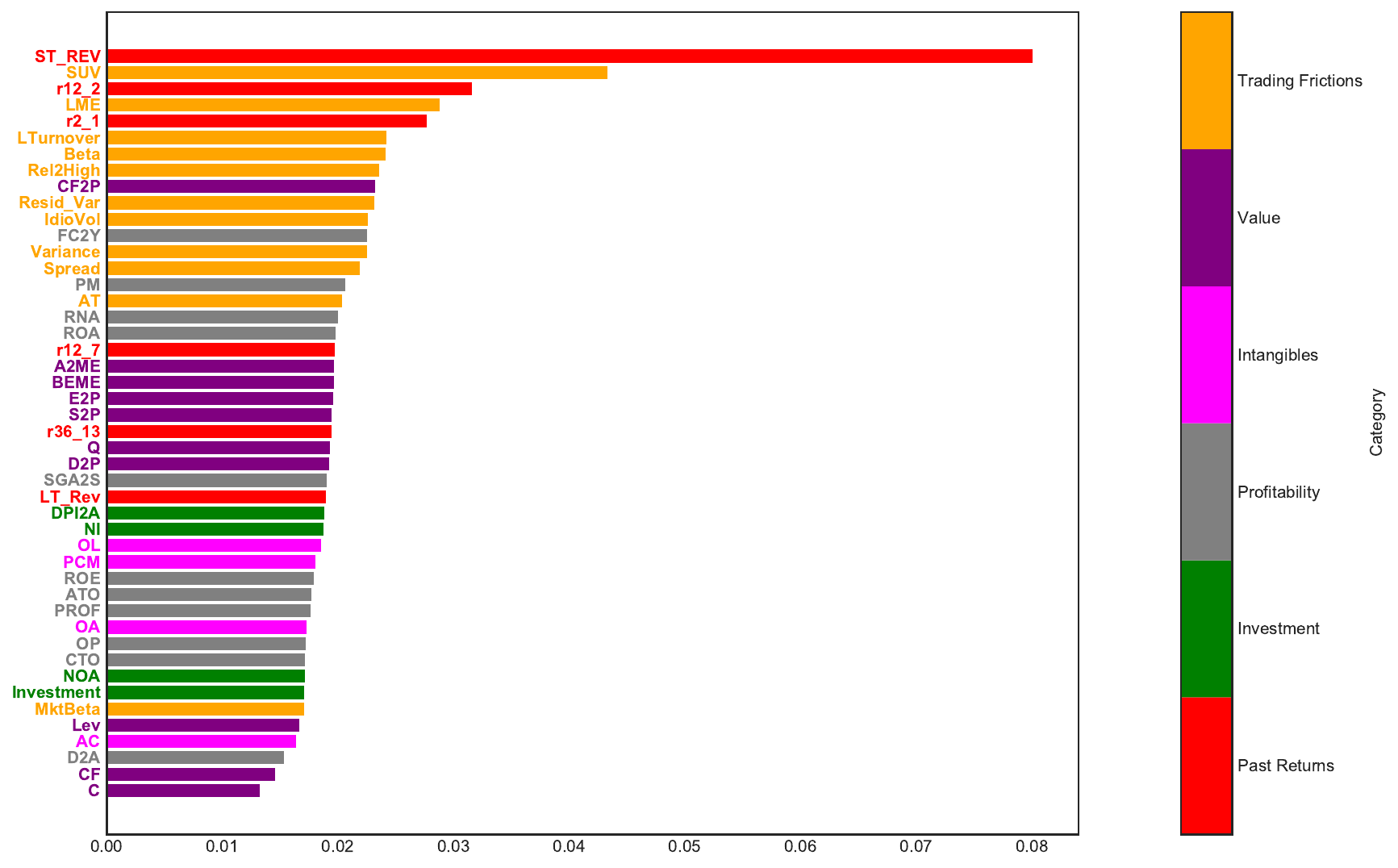}
\bnotefig{The figure shows the FFN variable importance ranking of the 46 firm-specific characteristics in terms of average absolute gradient (VI) on the test data. The values are normalized to sum up to one.}
\end{figure}

Figure \ref{fig:VI-GAN} ranks the variable importance of the 46 firm-specific characteristics for GAN. The sum of all sensitivities is normalized to one. Figures \ref{fig:VI-FFN}, \ref{fig:VI-EN} and \ref{fig:VI-LS} collect the corresponding results for FFN, EN and LS. All three models GAN, FFN and EN select trading frictions and past returns as being the most relevant categories. The most important variables for GAN are Short-Term Reversal (\texttt{ST\_REV}), Standard Unexplained Volume (\texttt{SUV}) and Momentum (\texttt{r12\_2}). Importantly, for GAN all 6 categories are represented among the first 20 variables, which includes value, intangibles, investment and profitability characteristics. The SDF composition is different for FNN, where the first 14 characteristics are almost only in the trading friction and past return category. More specifically, this SDF loads heavily on short-term reversal, illiquidity measured by unexplained volume and size, which raises the suspicion that a simple forecasting approach might focus mainly on illiquid penny stocks. The no-arbitrage condition with informative test assets seems to be necessary to discipline the model to capture the pricing information in other characteristics. Figure \ref{fig:VI-GAN-g} shows the variable importance ranking for the conditioning vector $g$. The GAN test assets depend on all six major anomaly categories. These test assets ensure that the GAN SDF also reflects this information. The linear model with regularization also selects variables from all six categories among the first 9 variables. Note, that the elastic net penalty removes characteristics that are close substitutes, for example, as the dividend-price ratio (\texttt{D2P}) and book-to-market ratio (\texttt{BEME}) capture similar information, the regularized model only selects one of them. The linear model without regularization cannot handle the large number of variables and not surprisingly results in a different ranking.

Figure \ref{fig:VI-GAN-Macro} shows the importance of the macroeconomic variables for the GAN model. These variables are first summarized into the four hidden states processes before they enter the weights of the SDF. First, it is apparent that most macroeconomic variables have a very similar importance. This is in line with a model where there is a strong dependency between the macroeconomic time series which is driven by a low dimensional non-linear factor structure. A simple example would be the factor model in \cite{ludvigson2009} where the information in a macroeconomic data set very similar to ours is summarized by a small number of PCA factors. As the first PCA factor is likely to pick up a general economic market trend, it would affect all variables. If the SDF structure depends on this PCA factor, all macroeconomic variables will appear to be important (of potentially similar magnitude). It is important to keep in mind that a simple PCA analysis of the macroeconomic variables does not work in our asset pricing context. The reason is that the PCA factors would mainly be based on increments of the macroeconomic time series and hence would not capture the dynamic pattern.\footnote{The results for PCA based macroeconomic factors are available upon request. We also want to clarify that for other applications PCA based factors based on macroeconomic time series might actually capture the relevant information.} The two most relevant variables that stand out in our importance ranking are the median bid-ask spread (\texttt{Spread}) and the federal fund rate (\texttt{FEDFUNDS}). These can be interpreted as capturing the overall economic activity level and overall market volatility.

\begin{figure}[h!]
\centering
\tcapfig{Macroeconomic Hidden State Processes (LSTM Outputs)}\label{fig:macro-state}
\includegraphics[width=0.85\linewidth]{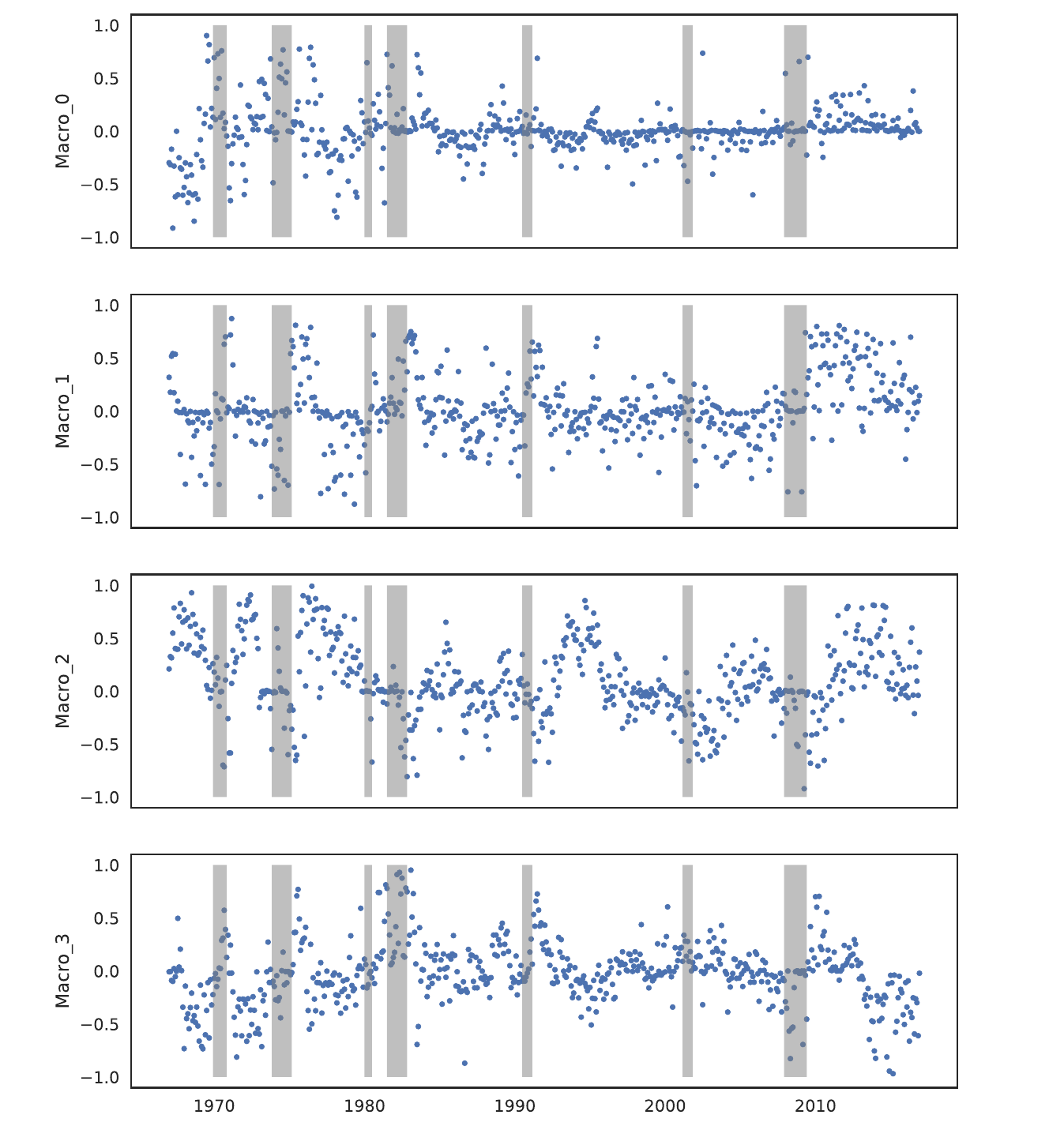}
\bnotefig{The figure shows the four macroeconomic hidden state processes extracted with LSTM in the GAN model. The gray areas mark NBER recession periods. The time-series are one representative estimation in the multiple ensemble estimations.}
\end{figure}

We show that the hidden macroeconomic states are closely linked to business cycles and overall economic activity. Figure \ref{fig:macro-state} plots the time series of the four hidden macroeconomic state variables. These variables are the outputs from the LSTM that encodes the history of macroeconomic information. Here we report one representative fit of the LSTM from the nine ensemble estimates.\footnote{Each ensemble fit returns a four dimensional vector of the state processes. However, it is not meaningful to average these vectors as the first state process in one fit does not need to correspond to the same process in another fit. It is only meaningful to report model averages of scalar output variables.} The grey shaded areas indicate NBER recessions.\footnote{NBER based Recession Indicators for the United States from the Peak through the Trough are taken from https://fred.stlouisfed.org/series/USRECM.} First, it is apparent that the state variables, in particular for the third and fourth state, peak during times of recessions. Second, the state processes seem to have a cyclical behavior which confirms our intuition that the relevant macroeconomic information is likely to be related to business cycles. The cycles and peaks of the different state variables do not coincide at all times indicating that they capture different macroeconomic risks.

\subsection{SDF Structure}

We study the structure of the SDF weights and betas as a function of the characteristics. Our main findings are two-fold: Surprisingly, individual characteristics have an almost linear effect on the pricing kernel and the risk loadings, i.e. non-linearities matter less than expected for individual characteristics. Second, the better performance of GAN is explained by non-linear interaction effects, i.e. the general functional form of our model is necessary for capturing the dependency between multiple characteristics.

\begin{figure}[h!]
  \tcapfig{SDF weight $\omega$ as a Function of Characteristics for GAN}\label{fig:lineinteraction}
  \begin{subfigure}[t]{.45\textwidth}
  \centering
    \includegraphics[width=0.8\textwidth]{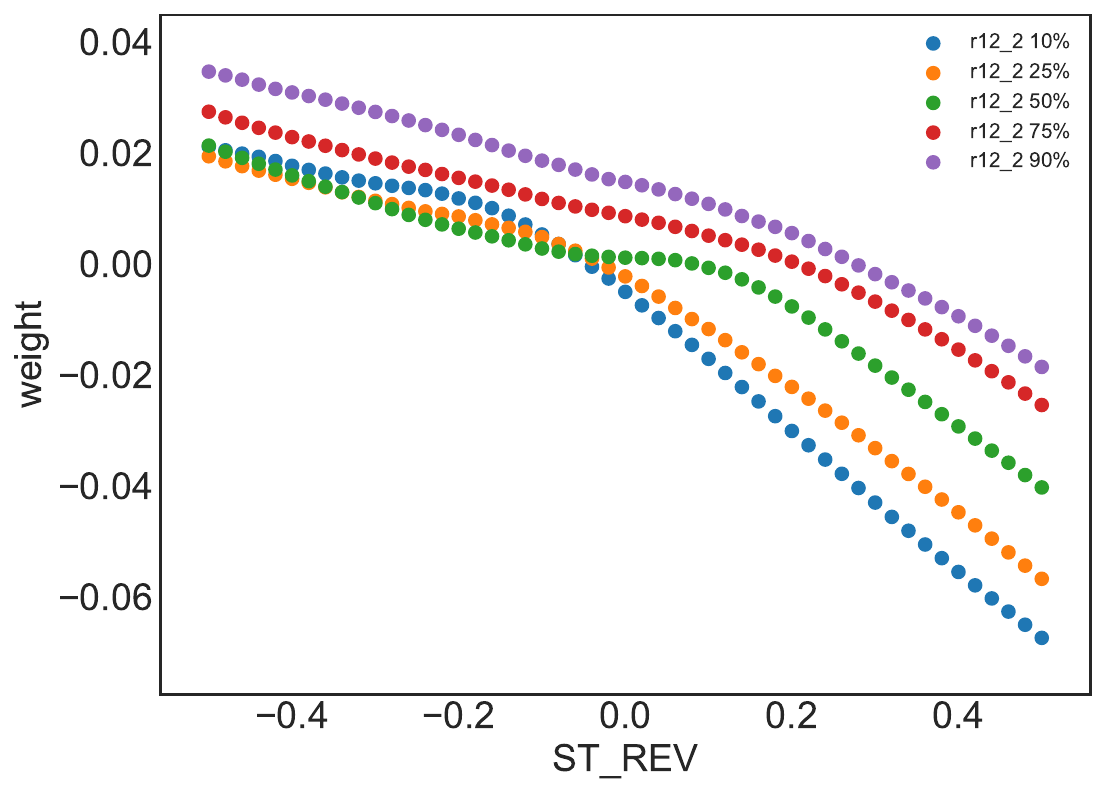}
    \includegraphics[width=0.8\textwidth]{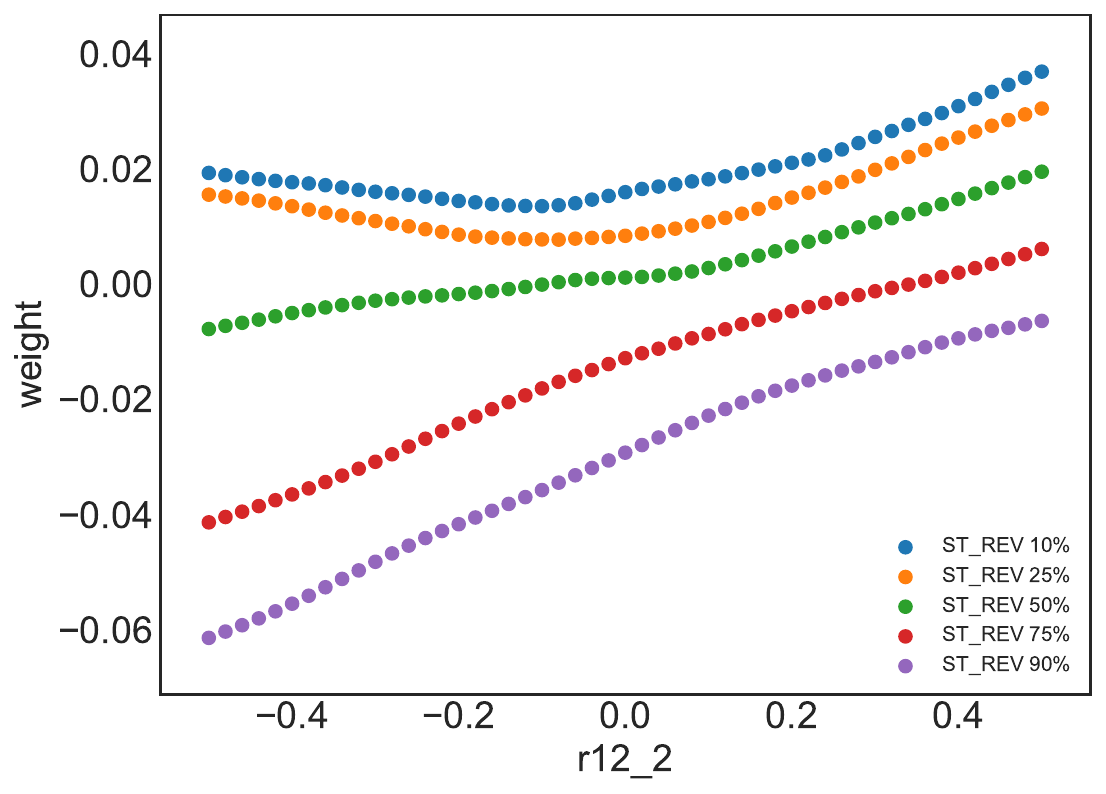}
    \caption{Interaction between Short-Term Reversal (\texttt{ST\_REV}) and Momentum (\texttt{r12\_2})}
  \end{subfigure}\hfill
  \begin{subfigure}[t]{.45\textwidth}
  \centering
    \includegraphics[width=0.8\textwidth]{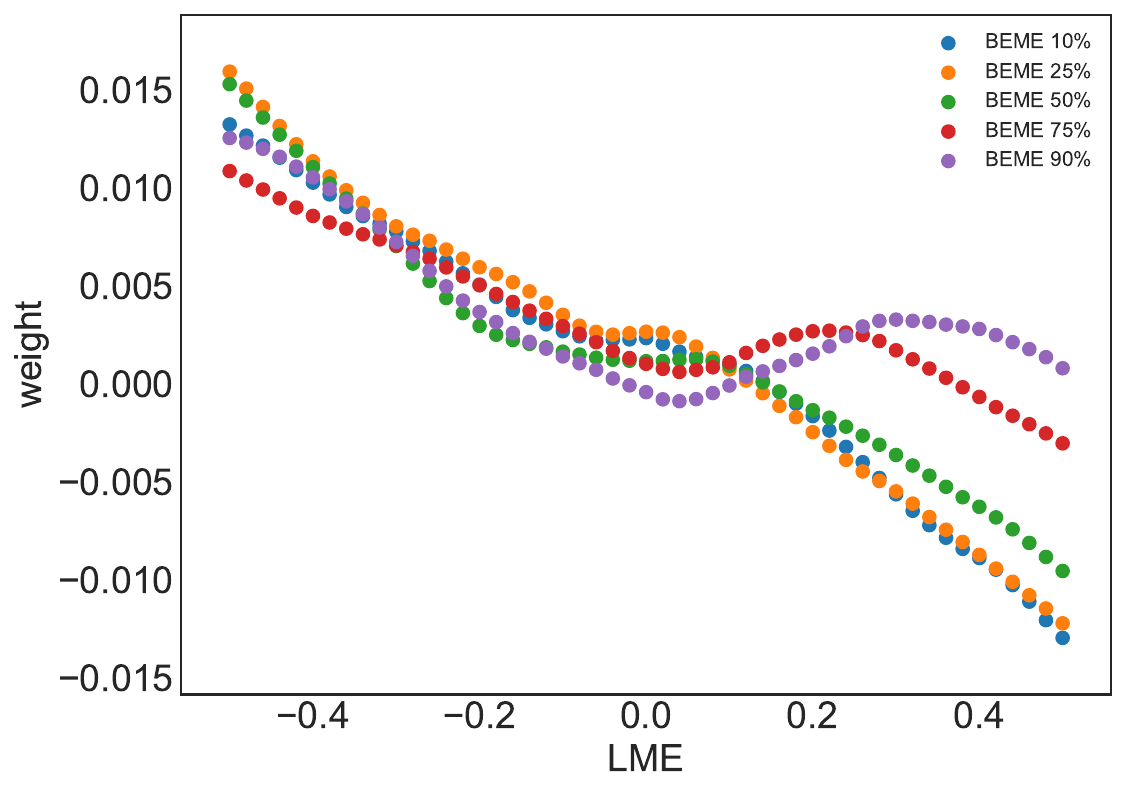}
    \includegraphics[width=0.8\textwidth]{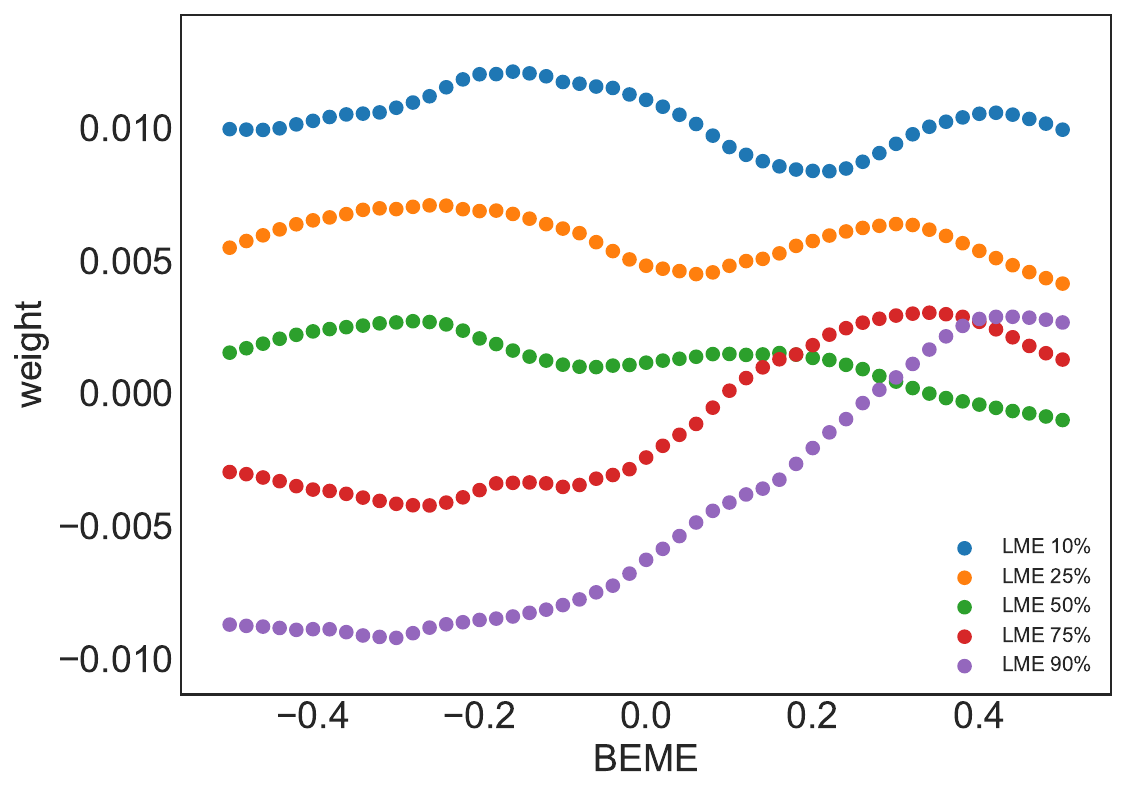}
    \caption{Interaction between Size (\texttt{LME}) and Book to Market Ratio (\texttt{BEME})}
  \end{subfigure}
  \bnotefig{These figures show the SDF weight $\omega$ as function of short-term reversal, momentum, size and book-to-market ratio for different quantiles of the second variable while keeping the remaining variables at their mean level.}
\end{figure}

Figure \ref{fig:SDF_weight_1D} plots the one-dimensional relationship between the SDF weights $\omega$ and one specific characteristic. The other variables are fixed at their mean values.\footnote{As the characteristics are normalized to quantiles their mean is equal to their median value.} In the case of a linear model these plots simply show the slope of a linear regression coefficient. As we include a separate long and short leg for the linear model, we allow for a kink at the median value. Otherwise the linear model would simply be a straight line. For the non-linear GAN and FFN the one-dimensional relationship can take any functional form. We show the univariate functional form for the three most relevant characteristics in Figure \ref{fig:SDF_weight_1D}, while the Internet Appendix collects the results for the other characteristics. It is striking how close the functional form of the SDF for GAN and FFN is to a linear function. This explains why linear models are actually so successful in explaining single-sorted characteristics. For a small number of characteristics, for example short-term reversal, GAN has some non-linearities around the median. These are exactly the decile sorted portfolios for which GAN performs better than FFN and EN. However, for most characteristics the pricing kernel depends almost linearly on the characteristics as long as we consider a one-dimensional relationship. However, it seems to be relevant to allow the low and high quantiles to have different linear slopes. The linear model without regularization obtains a relationship for some characteristics that is completely out of line with the other models. Given the worse overall performance of LS, this suggests that LS suffers from severe over-fitting.

\begin{figure}[h!]
  \tcapfig{SDF weight $\omega$ as a Function of Characteristics for GAN}\label{fig:SDF_GAN_23D}
  \begin{subfigure}[t]{.45\textwidth}
  \centering
    \includegraphics[width=1.1\textwidth]{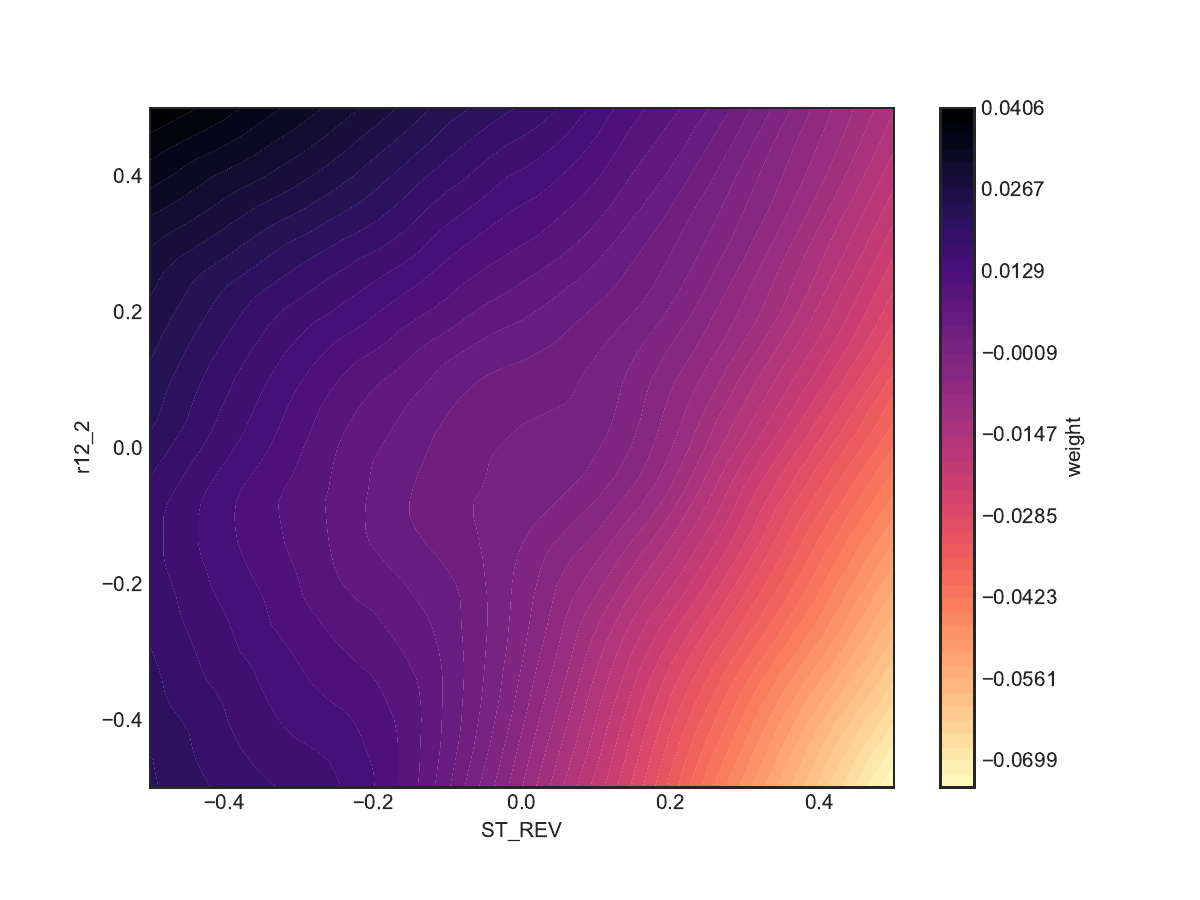}
    \subcap{Interaction between Short-Term Reversal (\texttt{ST\_REV}) and Momentum (\texttt{r12\_2})}
  \end{subfigure}\hfill
  \begin{subfigure}[t]{.45\textwidth}
  \centering
    \includegraphics[width=1.1\textwidth]{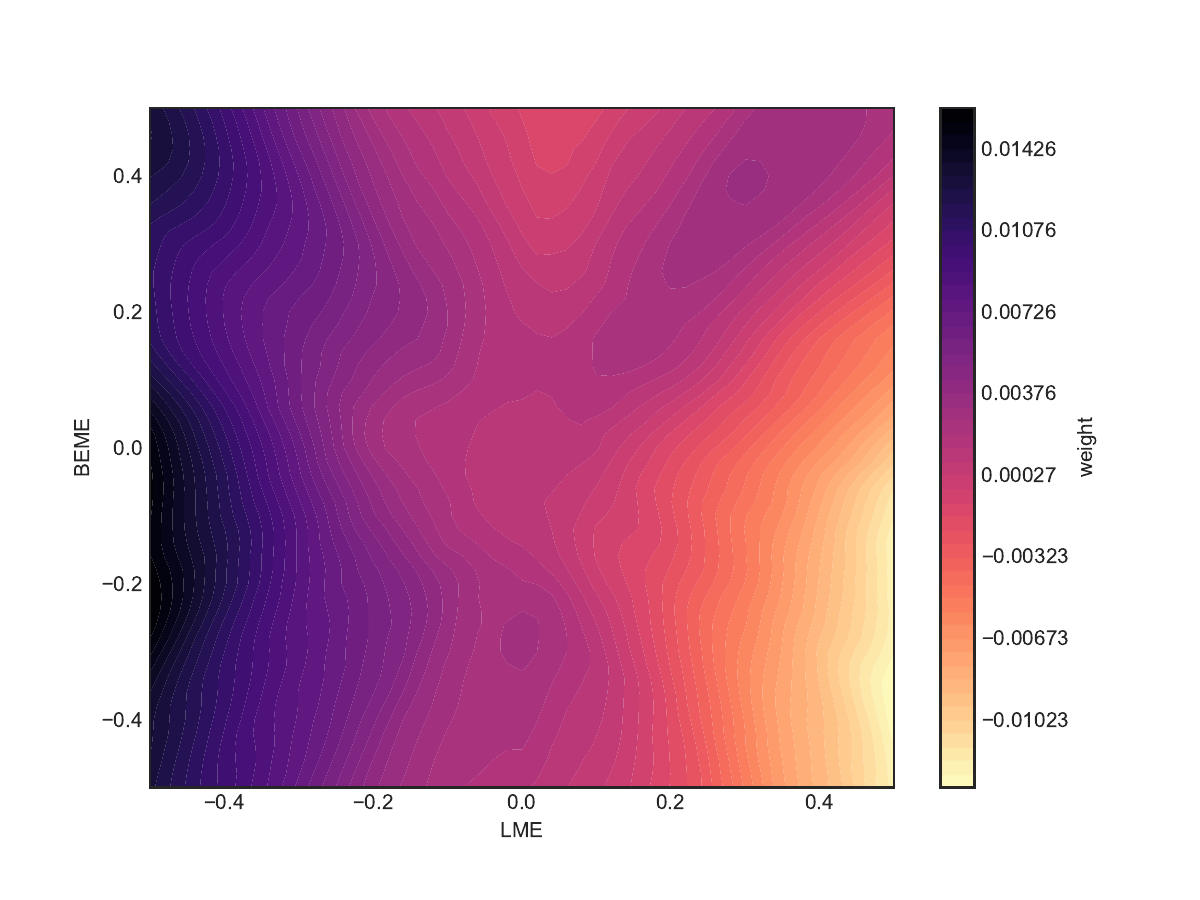}
    \subcap{Interaction between Size (\texttt{LME}) and Book to Market Ratio (\texttt{BEME})}
  \end{subfigure}
  \begin{subfigure}[t]{.45\textwidth}
  \centering
    \includegraphics[width=1.1\textwidth]{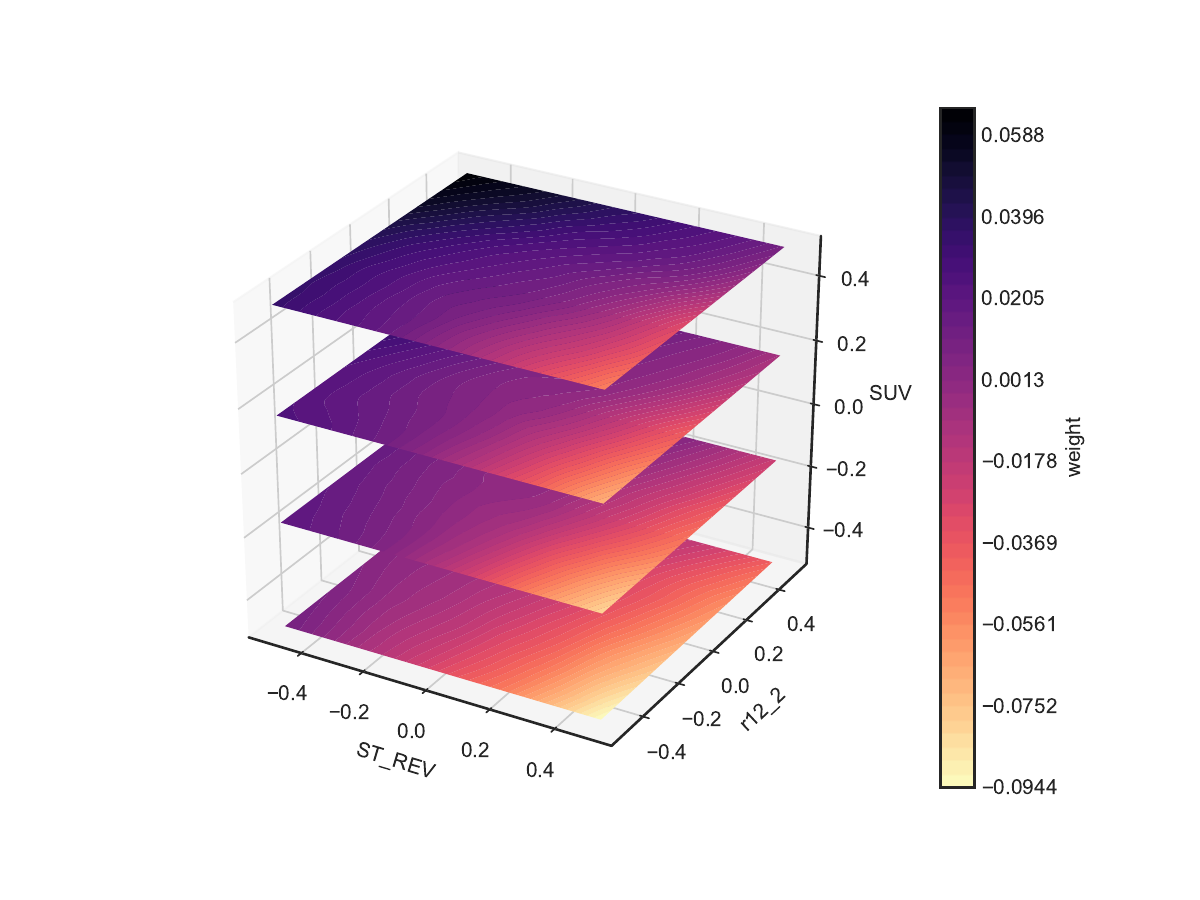}
    \subcap{Interaction between Short-Term Reversal (\texttt{ST\_REV}), Momentum (\texttt{r12\_2}) and Standard Unexplained Volume (\texttt{SUV}) }
  \end{subfigure}\hfill
  \begin{subfigure}[t]{.45\textwidth}
  \centering
    \includegraphics[width=1.1\textwidth]{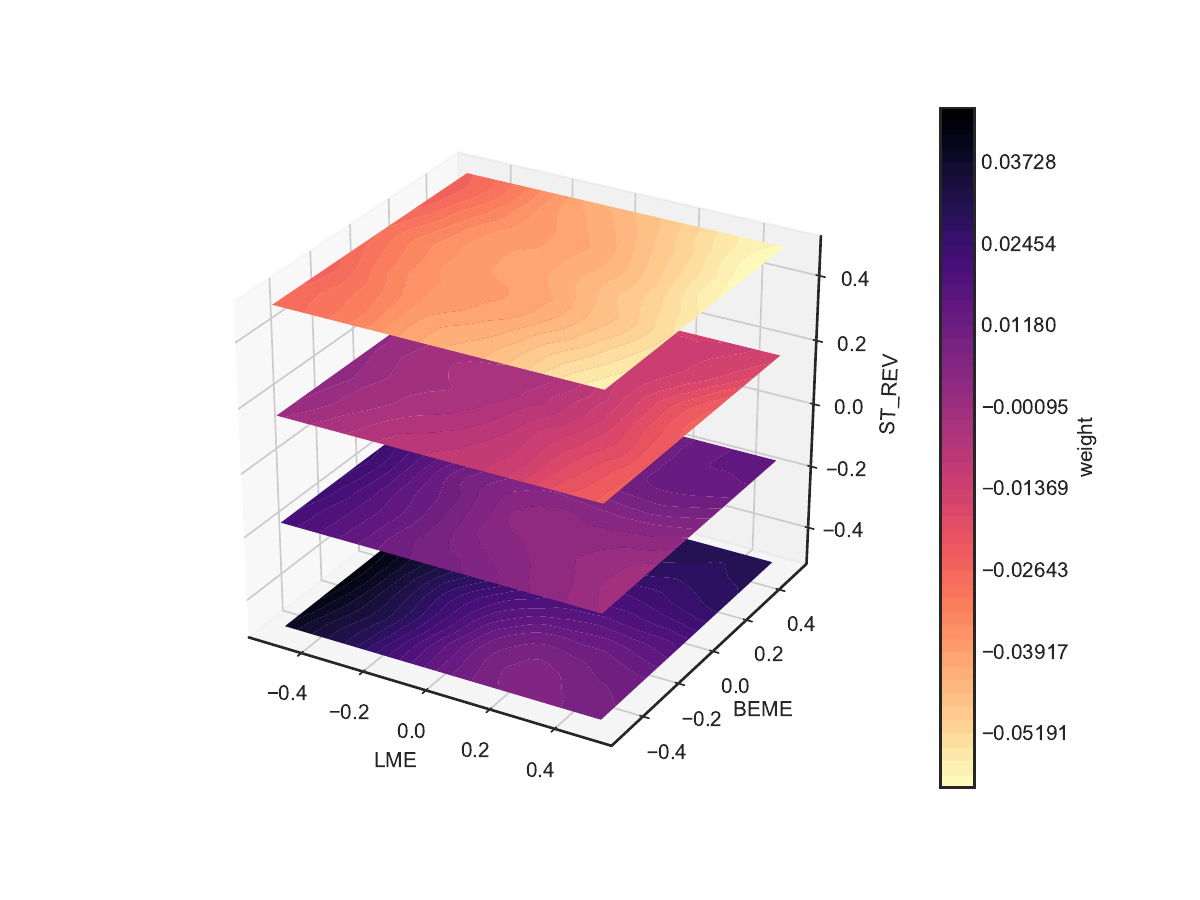}
    \subcap{Interaction between Size (\texttt{LME}), Book to Market Ratio (\texttt{BEME}) and Short-Term Reversal (\texttt{ST\_REV})}
  \end{subfigure}
    \bnotefig{These figures show the SDF weight $\omega$ as two- and three-dimensional function of characteristics keeping the remaining variables at their mean level.}
\end{figure}

Figures \ref{fig:lineinteraction} and \ref{fig:SDF_GAN_23D} show the crucial finding for this section: Non-linearities matter for interactions. 
Figure \ref{fig:lineinteraction} plots the SDF weight of one characteristic conditioned on a quantile of a second characteristic that can be different than the median. In an additive model without interaction all lines would be parallel shifts. This is exactly what we see for the two linear models.\footnote{As the linear model with regularization removes variables, it is possible that the SDF weights for one characteristic conditioned on different quantiles of the second characteristic collapse to one line.} Interestingly, for size and value, the FFN model also has almost parallel shifts in the SDF weights, implying that it does not capture interactions. However, for GAN small stocks have a very different exposure to value than large cap stocks. The line plots for GAN reveal more complex interaction patterns than for the other models. In general, the shape seems to become more non-linear when conditioning the second characteristic on an extreme quantile.

Instead of conditioning on only five quantiles for the second characteristic, we plot the two-dimensional pricing kernel for GAN in Figure \ref{fig:SDF_GAN_23D}. It confirms that the combined size and book-to-market characteristics have a highly non-linear effect on the GAN pricing kernel. The triple interaction in Figure \ref{fig:SDF_GAN_23D} shows that low short-term reversal, high momentum and high explained volume has the highest positive weight while high reversal, low momentum and low unexplained volume has the largest negative weight in the kernel when conditioning on these three characteristics. Low reversal and low momentum or high reversal and high momentum have an almost neutral effect independent of unexplained volume. The interaction effect for size, book-to-market and short-term reversal is even more complicated.

\subsection{Robustness Results}\label{sec:robust}


Our findings are robust to small cap stocks, the choice of the tuning parameters, the time period under consideration and are not exploiting limits to arbitrage. In this subsection, we evaluate and refit the model without small cap stocks, compare the performance and structure of the SDF for different tuning parameters and time periods and control the information used to construct the test assets.

The qualitative findings are robust to small cap stocks. It is well-known that penny stocks can achieve high Sharpe ratios and are hard to price by conventional asset pricing models. However, trading in these small cap stocks is limited due to low liquidity and high spreads. Hence, the high Sharpe ratios or large alphas of small cap stocks can potentially not be exploited. Here, we compare the model performance restricted to medium and large cap stocks. Our cross-section of stocks in the test data  is composed of 2,000 to 3,000 individual stocks per month. The Internet Appendix shows that the restriction to the stocks with a market capitalization larger than $0.001\%$ of the total market capitalization leaves us on average with the largest 1,500 stocks. Restricting the sample to stocks with market cap above $0.01\%$ of the total market cap yields on average the largest 550 stocks, that is, the sample is close to the S\&P 500 index.

\begin{table}[h!]
\centering
\tcaptab{~Different SDF Models Evaluated on Large Market Cap Stocks}\label{tab:smallcap}
{\small
\begin{tabular}{cccc|ccc|ccc}
\toprule
& \multicolumn{3}{c}{SR} & \multicolumn{3}{c}{EV} & \multicolumn{3}{c}{Cross-Sectional $R^2$}\\
\cmidrule(l){2-10}
Model & Train & Valid & Test & Train & Valid & Test & Train & Valid & Test \\
\midrule
& \multicolumn{9}{c}{Evaluated for size $\geq0.001\%$ of total market cap} \\
\midrule
LS & 1.44 & 0.31 & 0.13 & 0.07 & 0.05 & 0.03 & 0.14 & 0.03 & 0.10 \\
EN & 0.93 & 0.56 & 0.15 & 0.11 & 0.09 & 0.06 & 0.17 & 0.05 & 0.14 \\
FFN & 0.42 & 0.20 & 0.30 & 0.11 & 0.10 & 0.05 & 0.19 & 0.08 & 0.18 \\
\midrule
GAN & 2.32 & 1.09 & 0.41 & 0.23 & 0.22 & 0.14 & 0.20 & 0.13 & 0.26 \\
\midrule
& \multicolumn{9}{c}{Evaluated for $\geq0.01\%$ of total market cap} \\
\midrule
LS & 0.32 & -0.11 & -0.06 & 0.05 & 0.07 & 0.04 & 0.13 & 0.05 & 0.09 \\
EN & 0.37 & 0.26 & 0.23 & 0.09 & 0.12 & 0.07 & 0.17 & 0.08 & 0.14\\
FFN & 0.32 & 0.17 & 0.24 & 0.13 & 0.22 & 0.09 & 0.22 & 0.15 & 0.26 \\
\midrule
GAN & 0.97 & 0.54 & 0.26 & 0.28 & 0.34 & 0.18 & 0.27 & 0.23 & 0.32 \\
\midrule
& \multicolumn{9}{c}{Estimated and evaluated for size $\geq0.001\%$ of total market cap} \\
\midrule
LS & 1.91 & 0.40 & 0.19 & 0.08 & 0.06 & 0.04 & 0.18 & 0.05 & 0.12 \\
EN & 1.34 & 0.92 & 0.42 & 0.13 & 0.13 & 0.07 & 0.23 & 0.09 & 0.19 \\
FFN & 0.37 & 0.19 & 0.28 & 0.13 & 0.13 & 0.07 & 0.21 & 0.10 & 0.21 \\
\midrule
GAN & 3.57 & 1.18 & 0.42 & 0.24 & 0.23 & 0.14 & 0.23 & 0.13 & 0.26 \\
\bottomrule
\end{tabular}
}
\bnotetab{The table shows monthly Sharpe ratios (SR) of the SDF factors, explained time series variation (EV) and cross-sectional $R^2$ for the GAN, FFN, EN and LS models. 
In the first two subtables the model is estimated on all stocks but evaluated on stocks with market capitalization larger than $0.01\%$ or $0.001\%$ of the total market capitalization. In the last subtable the model is estimated and evaluated on stocks market with capitalization larger than $0.001\%$.}
\end{table}

Table \ref{tab:smallcap} reports the model performance for these two subsets of the data. The SDF weights are obtained on all individual stocks, but the Sharpe-ratio and the explained time series and cross-sectional variation is calculated on stocks with market cap larger than $0.001\%$ respectively $0.01\%$ of the total market capitalization. As expected the Sharpe ratios decline, but GAN still achieves an annual out-of-sample Sharpe ratio of 1.4 using only the 1,500 largest stocks. In contrast, the linear models collapse. Based on the 550 largest stocks the annual Sharpe ratio of GAN falls to 0.9, but is still larger than for the other models. Most importantly the explained variation of GAN is two to three times higher than for the linear or deep learning prediction model. Similarly, the gap in the cross-sectional $R^2$ is substantially wider on the larger stocks than on the whole sample. This suggests that FFN and the linear models are mainly fitting small stocks, while GAN also finds the systematic structure in the large cap stocks.

Table \ref{tab:smallcap} also estimates and evaluates the different models on stocks with market capitalization larger than $0.001\%$ of the total market capitalization.\footnote{We estimate the optimal tuning parameters for the model restricted to the large cap stocks. Using the same tuning parameters as for the total sample yields identical results.} The performance of GAN is essentially identical, suggesting that our approach finds the same SDF structure conditioned on large cap stocks if it is trained on all stocks or only the large stocks. In this sense our model is robust to the size of the companies. In contrast, the elastic net approach performs significantly better on large cap stocks when estimated on this sample. This is evidence that it overfits small stocks when applied to the full sample in contrast to our approach. The prediction approach has a very similar performance on the large cap stocks when estimated on this subset or on the full data set. This is indicative that it cannot capture the structure in large cap stocks. Even when optimally trained on the subset of large cap stocks the linear and prediction approach explain substantially less time-series and cross-sectional variation than GAN.

We re-estimate the GAN model independently of our benchmark fit and list the tuning parameters of the best four models on the validation data, labeled GAN 1, 2, 3 and 4 in Table \ref{tab:tuning}. All models have four macroeconomic states, but differ in terms of the depth of the network and the number of instruments that construct the test assets. The tuning parameters of our benchmark model would only be the second best model in this independent fit. Table \ref{tab:SDF-Comparison} reports the performance for the different fits and tuning parameters. The asset pricing performance is essentially identical for all models. Moreover, Table \ref{tab:corr} shows that the SDFs for the various models all have a correlation higher than 80\%. The Internet Appendix collects the variable importance results and functional form of the SDF weights $\omega$ for the alternative fits. In summary, we conclude that not only the pricing performance is extremely robust to the tuning parameters, but we are actually discovering the same economic model for different tuning parameters and our results are replicable.     

\begin{table}[h!]
\centering
\tcaptab{Performance of Alternative GAN Models}\label{tab:SDF-Comparison}
{\small
\begin{tabular}{cccc|ccc|ccc}
\toprule
& \multicolumn{3}{c}{SR} & \multicolumn{3}{c}{EV} & \multicolumn{3}{c}{Cross-Sectional $R^2$}\\
\cmidrule(l){2-10}
Model & Train & Valid & Test & Train & Valid & Test & Train & Valid & Test \\
\midrule
GAN 1 & 2.78 & 1.47 & 0.72 & 0.18 & 0.08 & 0.07 & 0.12 & 0.01 & 0.21 \\
GAN 2 & 3.02 & 1.39   & 0.77 & 0.18 & 0.08 & 0.07 & 0.12 & 0.00 & 0.22 \\
GAN 3 & 2.55 & 1.38 & 0.74 & 0.22 & 0.11 & 0.09 & 0.17 & 0.04 & 0.25 \\
GAN 4  & 2.44 & 1.38 & 0.77 & 0.19 & 0.08 & 0.07 & 0.11 & 0.01 & 0.22 \\
\midrule
GAN Rolling & N/A & N/A & 0.88 & N/A & N/A & 0.08 & N/A & N/A & 0.24 \\
\midrule
GAN No Frict & 2.94 & 1.37   & 0.77 & 0.20 & 0.10 & 0.08 & 0.14 & 0.01 & 0.23 \\
\bottomrule
\end{tabular}}
\bnotetab{This table shows the monthly Sharpe ratio (SR) of the SDF, explained time series variation (EV) and cross-sectional $R^2$ for alternative GAN models. GAN 1, 2, 3 and 4 are the four best GAN models on the validation data from an independent re-estimation of the model. GAN Rolling is re-estimated every year on a rolling window of 240 months. GAN No Frict is estimated without trading frictions and past returns for the conditioning function $g$.}
\end{table}

As another robustness test we estimate the GAN model on a rolling window of 240 months. In more detail, every year we move the training data window by one year to re-estimate the SDF weights and loadings.\footnote{The estimation of the GAN models is computationally very expensive and for this reason we are not re-estimating it every month.} Not surprisingly, we obtain slightly better asset pricing results, in particular for the Sharpe ratio, as reported in Table \ref{tab:SDF-Comparison}. Overall, the results are very close and the rolling window GAN SDF has a correlation of 70\% with our benchmark SDF. Figures \ref{fig:VI_window} and \ref{fig:SDF_window} show the variable importance and functional form of the rolling window SDF which share the same general patterns. We conclude that a time-varying estimate of GAN does not lead to major improvements and fits a similar economic structure. In particular, this confirms that our results are robust to the time window chosen for estimation.

Our GAN is not simply capturing pricing information that is subject to limits to arbitrage. A potential concern is that GAN constructs test assets that do not represent a risk-premium but anomalies of illiquid stocks that cannot be exploited. In that case the GAN SDF would not represent the economic risk that we want to capture for asset pricing. We want to avoid that GAN explicitly targets stocks that have high trading frictions. We exclude characteristics from the trading frictions and past return category from the conditional network and re-estimate the model labeled as GAN No Frict. The resulting variable importance for $g$ is shown in Figure \ref{fig:VI-GAN-NF}. This SDF has the same pricing performance and a correlation of 78\% with our benchmark SDF suggesting that our results are robust to this change.

\subsection{Machine Learning Investment}

\begin{figure}[h!]
\tcapfig{Cumulative Excess Returns of SDF}\label{fig:cumulated-return}
\centering
\includegraphics[width=0.6\linewidth]{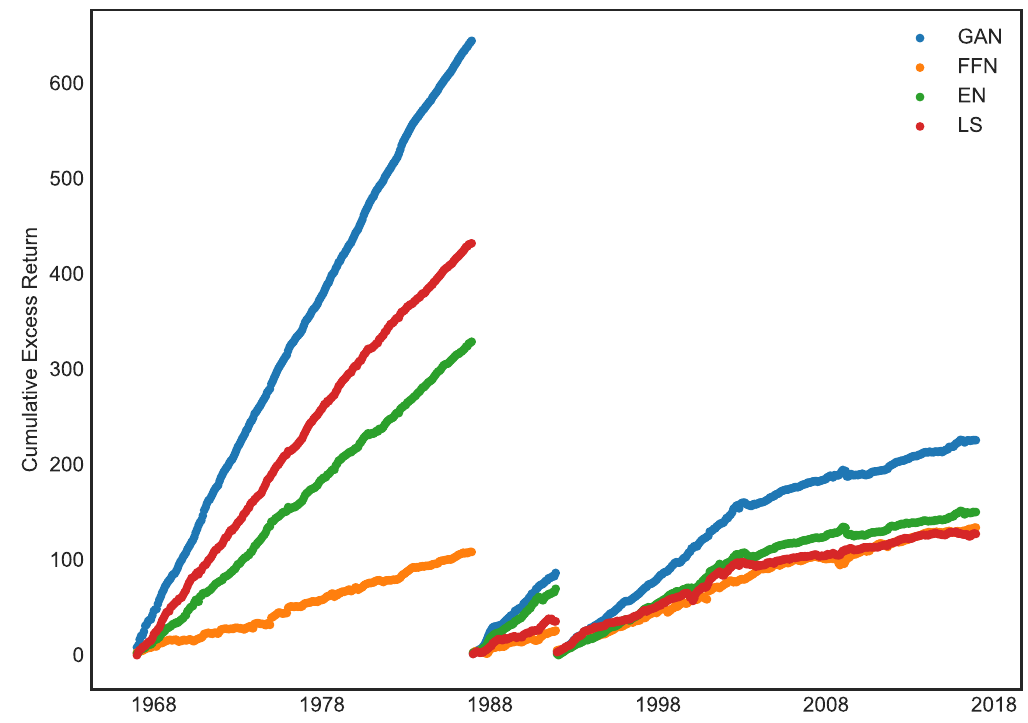}
\bnotefig{The figure shows the cumulative excess returns for the SDF for GAN, FFN, EN and LS. Each factor is normalized by its standard deviation for the time interval under consideration.}
\end{figure}

The GAN SDF is a tradeable portfolio with an attractive risk-return trade-off. Table \ref{tab:SDF-Factor} reports the monthly Sharpe ratios, maximum 1-month loss and maximum drawdown of the four benchmark models and also the Fama-French 3 and 5 factor models.\footnote{Max Drawdown is defined as the maximum number of consecutive months with negative returns. The maximum 1-month loss is normalized by the standard deviation of the asset.} 
The number of consecutive losses as measured by drawdown and the maximum loss for the GAN model is comparable to the other models, while the Sharpe ratio is by far the highest. Figure \ref{fig:cumulated-return} plots the cumulative return for each model normalized by the standard deviation. As suggested by the risk-measures the GAN return exceeds the other models while it avoids fluctuations and large losses. Table \ref{tab:turnover} lists the turnover for the different approaches. The GAN factor has a comparable or even lower turnover than the other SDF portfolios. This suggests that all approaches are exposed to similar transaction costs and it is valid to directly compare their risk-adjusted returns.\footnote{\cite{gu2018} report high out-of-sample Sharpe ratios for long-short portfolios based on the extreme quantiles of returns predicted by FFN. The Internet Appendix compares the Sharpe ratios for different extreme quantiles for equally and value-weighted long-short portfolios with FFN. We can replicate the high out-of-sample Sharpe ratios when using extreme deciles of 10\% or less and equally weighted portfolios. However, for value weighted portfolios the Sharpe ratio drops by around 50\%. This is a clear indication that the performance of these portfolios heavily depends on small stocks.} 

\begin{figure}[h!]
  \tcapfig{Trading Friction Cutoffs }\label{fig:tradingfriction}
  \begin{subfigure}[t]{.33\textwidth}
  \centering
\includegraphics[width=1\linewidth]{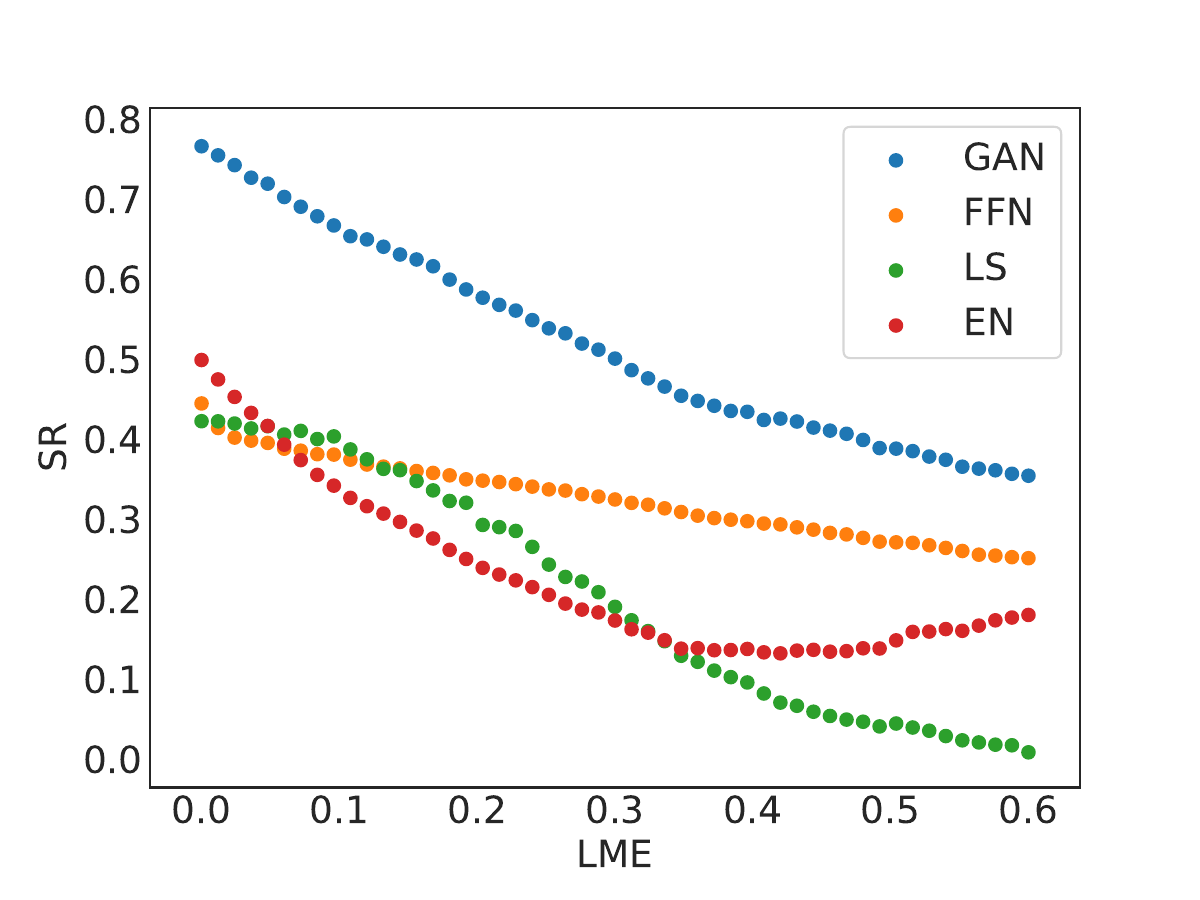}
    \caption{Size cutoff}
  \end{subfigure}
  \begin{subfigure}[t]{.33\textwidth}
  \centering
\includegraphics[width=1\linewidth]{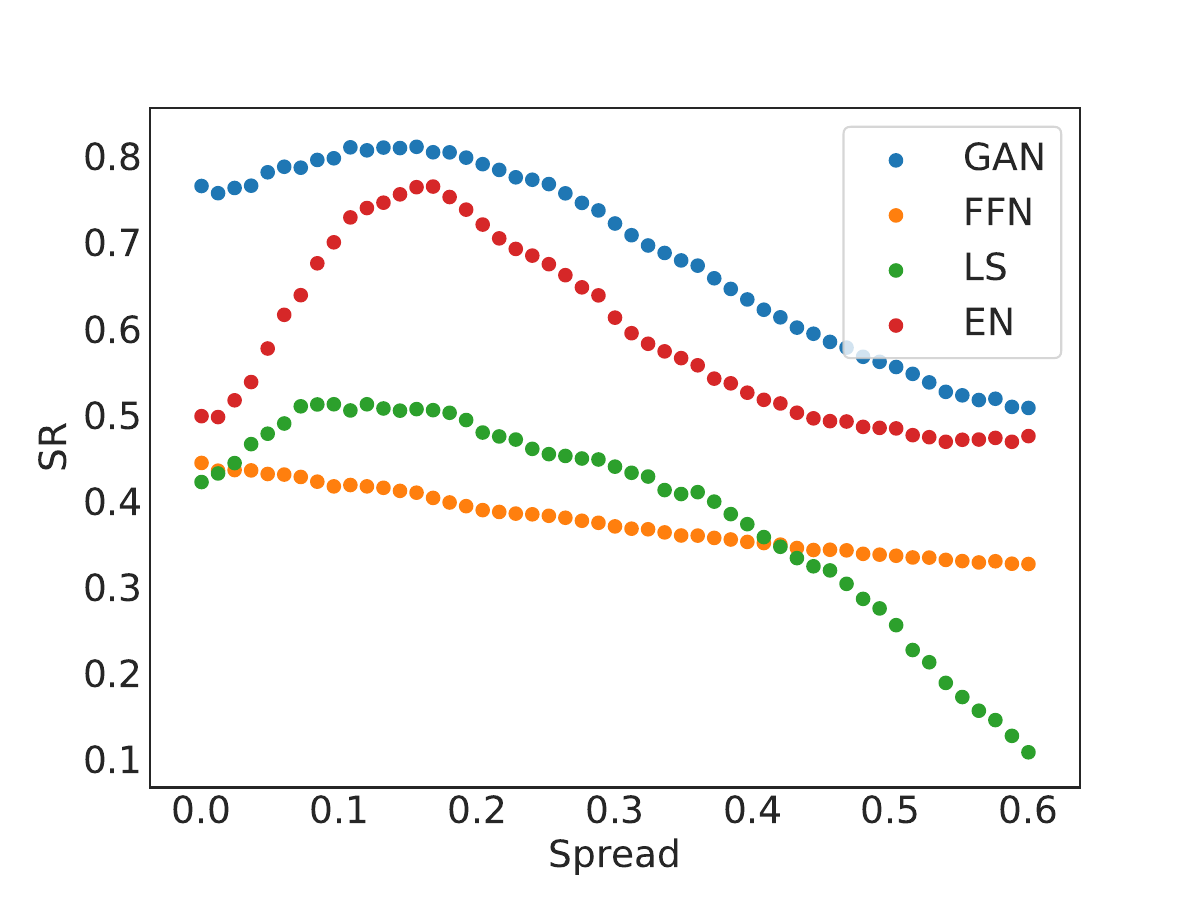}
    \caption{Spread cutoff}
  \end{subfigure}
    \begin{subfigure}[t]{.33\textwidth}
  \centering
\includegraphics[width=1\linewidth]{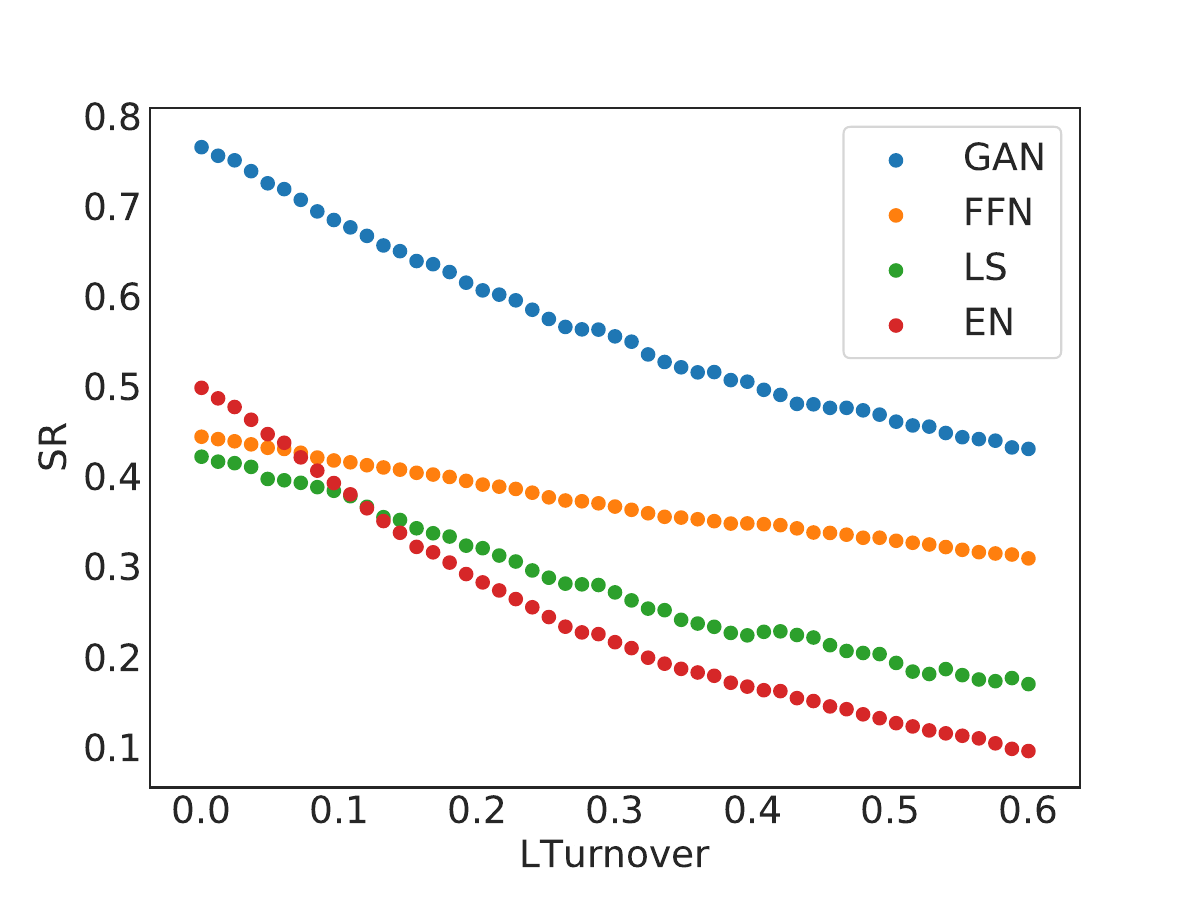}
    \caption{Turnover cutoff}
  \end{subfigure}
    \bnotefig{This figure shows the out-of-sample Sharpe ratios of the SDF for GAN, FFN, EN and LS after we have set the portfolio weights $\omega$ to zero if either the market capitalization (\texttt{LME}), bid-ask spread (\texttt{Spread}) or turnover (\texttt{Lturnover}) at the time of investment are below a specified cross-sectional quantile.}
\end{figure}

\cite{avramov2020} raise the concern that the performance of machine learning portfolios could deteriorate in the presence of trading costs due to high turnover or extreme positions.\footnote{In their comparison study \cite{avramov2020} also include a portfolio derived from GAN. However, they do not consider our SDF portfolio based on $\omega$ but use the SDF loadings $\beta$ to construct a long-short portfolio based on prediction quantiles.} This important insight can be taken into account when constructing machine learning investment portfolios. Figure \ref{fig:tradingfriction} shows the out-of-sample Sharpe ratios of the SDF portfolios after we have set the SDF weights $\omega$ to zero for stocks with either market capitalization, bid-ask spread or turnover below a specified cross-sectional quantile at the time of portfolio construction. The idea is to remove stocks that are more prone to trading frictions. There is a clear trade-off between trading-frictions and achievable Sharpe ratios. However, this indicates that a machine learning portfolio can be estimated to optimally trade-off the trading frictions and a high risk-adjusted return. For example, GAN without 40\% of the smallest stocks still has an annual SR of 1.73, without 40\% of the highest bid-ask spreads the SR is still 2.07, and without 40\% of the stocks with the least trading activity measured by turnover the SR is 1.87. Note that these are all lower bounds as GAN has not been re-estimated without these stocks, but we have just set the portfolio weights of the stocks below the cutoffs to zero. 

So far, most paper have separated the construction of profitable machine learning portfolios into two steps. In the first step, machine learning methods extract signals for predicting future returns. In a second step, these signals are used to form profitable portfolios, which are typically long-short investments based on prediction. However, we argue that these two steps should be merged together, that is machine learning techniques should extract the signals that are the most relevant for the overall portfolio design. This is exactly what we achieve when the objective is the estimation of the SDF, which is the conditionally mean-variance efficient portfolio. A step further is to include trading-frictions directly in this estimation, that is, machine learning techniques should extract the signals that are the most relevant for portfolio design under constraints. A promising step into this direction is presented in \cite{bryzgalova2019} who estimate mean-variance efficient portfolios with decision trees that can easily incorporate constraints, \cite{cong2020} who use a reinforcement learning approach and \cite{pelger2021} who estimate optimal statistical arbitrage strategies under trading friction constraints with neural networks specialized for time-series data.

\subsection{SDF of Multi-Factor Models}\label{sec:IPCA}

%
%
%
%
%
%
%
%
%
%

Our GAN framework is {\it complementary} to conditional and unconditional multi-factor models. Multi-factor models are based on the assumption that the SDF is a linear combination of the multiple factors. In an unconditional multi-factor model with constant factor loadings and pricing errors, the SDF can easily be constructed as the unconditional mean-variance efficient combination of the factors. The time series regression pricing error in such a multi-factor model is identical to that of a one-factor regression on the SDF. However, this relationship does not hold out-of-sample for unconditional models and breaks down in-sample and out-of-sample for conditional multi-factor models. So far, the factor literature has mainly focused on extracting the factors and their loadings, but has been largely silent on the construction of a coherent conditional SDF framework based on a conditional factor structure. Our GAN framework can help to close this gap. Formula \ref{eqn:unconditional} is the fundamental condition to construct the SDF and can incorporate the restriction that the SDF is a linear combination of factors. 

We use one of the most important conditional multi-factor models and combine it with the GAN framework to estimate its SDF. Instrumented Principal Component Analysis (IPCA) developed by \cite{kelly2018} allows for latent factors and time-varying loadings. Both elements are key as we need a conditional factor model with time-varying loadings to explain individual stock returns and want to estimate the best performing factors without taking an a prior stand on what the factors are. Note that this model includes simple unconditional factor models as a special case. As most of the literature \cite{kelly2018} use the multi-factor framework to calculate pricing errors and report Sharpe ratios for the unconditional mean-variance efficient combination of the factors, but they do not estimate pricing errors and Sharpe ratios for the same one-factor model of the SDF. Here we show that using the additional economic structure of spanning the SDF with IPCA factors and combining it with the GAN framework can lead to an even better asset pricing model. We condition only on firm characteristics to make the results more comparable with the original IPCA framework.

IPCA assumes a $K$-factor model where the loadings are a linear function of the characteristics:\footnote{\cite{kelly2018} also allow for an error term in the loading equation. However, this does not affect the estimation procedure and our discussion, but only the confidence intervals. In order to simplify notation we leave it out.}
\begin{align*}
\Ri=a_{t,i} + b_{t,i}^{\top} f_{t+1}^{\text{IPCA}} + \ei \qquad b_{t,i}= I_{i,t}^{\top} \Gamma_b.
\end{align*} 
Any multi-factor model assumes that the SDF is spanned by the factors, that is, 
\begin{align}\label{eqn:IPCA}
F= \sum_{k=1}^K \omega^{f}(I_{k,t}, I_t) f^{\text{IPCA}}_{t+1,k}. 
\end{align}
Appendix \ref{sec:appcondSDF} provides further details. Under weak assumptions, the SDF weights are given by the conditional tangency portfolio based on the factors: $\omega^{f}(I_{k,t}, I_t)= \text{Cov}_t \left(f_{t+1}^{\text{IPCA}},f_{t+1}^{\text{IPCA}\top} \right)^{-1} \mathbbm E_t \left[f_{t+1}^{\text{IPCA}} \right]$. However, the usual approach in most papers is to use constant weights, which are set to the unconditional mean-variance efficient portfolio weights, that is, $\omega^{\text{I-SR}}=\text{Cov} \left(f_{t+1}^{\text{IPCA}},f_{t+1}^{\text{IPCA}\top} \right)^{-1} \mathbbm E \left[f_{t+1}^{\text{IPCA}} \right]$. An alternative approach to obtain a one-factor representation with constant weights is to find a linear combination of the IPCA factors and loadings that either minimizes the cross-sectional pricing errors or the amount of unexplained variation. This means that we obtain the SDF weights $\omega^{\text{I-XS}} \in \mathbbm R^{K}$ and the corresponding loading weights $v^{\text{I-XS}} \in \mathbbm R^{K}$ such that the residuals $\Ri -  b_{t,i}^{\top} \left(v^{\text{I-XS}} \omega^{\text{I-XS}\top} \right)  f_{t+1}^{\text{IPCA}} = \hat{e}_{t+1,i}^{\text{I-XS}}$ maximize the $\text{XS-}R^2$. Similarly, we obtain $\omega^{\text{I-EV}}$ and  $v^{\text{I-EV}}$ to maximize $EV$. If the correct one-factor representation has constant weights on the IPCA factors and their loadings, all three criteria would represent valid identification conditions to recover those weights. Note that in contrast to \cite{kelly2018} we estimate the total pricing errors and not only the component that is spanned by the characteristics. In the case of $\omega^{\text{I-EV}}$ all the weight will be put on the first IPCA factor which by construction maximizes the amount of explained variation.

The SDF loadings of the one-factor combination that maximizes the Sharpe ratio are not simply the same combination of the IPCA loadings, that is, in the above notation $\omega^{\text{I-SR}}$ does not need to equal $v^{\text{I-SR}}$. 
The internally consistent way to estimate the loadings of $F^{\text{I-SR}}=\omega^{\text{I-SR}}f^{\text{IPCA}}$ is to run the IPCA loading regression:
\begin{align*}
\Gamma_{\beta}= \left(  \sum_{t=1}^{T-1} I_t I_t^{\top} \otimes \left(F^{\text{I-SR}}_{t+1}\right)^2 \right)^{-1} \left( \sum_{t=1}^{T-1} \left( I_t \otimes F_{t+1}^{\text{I-SR}} \right)^{\top}  \R \right),  \qquad \beta_{t,i}^{\text{I-SR}}=I_{i,t}^{\top}\Gamma_{\beta},  
\end{align*}
where by abuse of notation $I_t \in \mathbbm R^{q \times N_t}$ denotes here the firm-specific characteristics. This IPCA regression would simply return the IPCA loadings in a multi-factor model. In order to assess the effect of linearity between the characteristics and loadings, we also estimate the IPCA SDF loading $\beta_{i,t}^{\text{I-FFN}}$ with a feedforward neural network, that is, we estimate $\E \left[ \Ri F_{t+1}^{\text{I-SR}} \right]$.
The combination of GAN and IPCA estimates $\omega^{\text{I-GAN}}$ using Formula \ref{eqn:unconditional} but restricting the SDF to a linear combination of IPCA factors based on Equation \ref{eqn:IPCA} and then estimates the loadings with FFN in the prediction $\beta_{i,t}^{\text{I-GAN}}= \E \left[ \Ri F_{t+1}^{\text{I-GAN}} \right]$.

\begin{table}[h!]
\centering
\tcaptab{IPCA Asset Pricing with Different SDFs}
\label{tab-IPCA}
{\small
\begin{tabular}{l|l|llllllll}
\toprule
Model & Benchmark & 3     & 4     & 5     & 6     & 7     & 8     & 9     & 10    \\ 
\hline
& SR                        & 0.61  & 0.71  & 0.77  & 0.70  & 0.79  & 0.82  & 0.72  & 0.81  \\
IPCA GAN & EV               & 0.05  & 0.04  & 0.04  & 0.05  & 0.05  & 0.05  & 0.04  & 0.05  \\
$(\omega^{\text{I-GAN}}, \beta^{\text{I-GAN}})$& XS-$R^2$                  & 0.20  & 0.19  & 0.17  & 0.20  & 0.18  & 0.20  & 0.17  & 0.21  \\ 
\hline
& SR                                     & 0.69 & 0.79 & 0.82 & 0.84 & 0.83 & 0.86 & 0.86 & 0.94  \\
IPCA Max SR FFN Beta & EV               & 0.04  & 0.03  & 0.03  & 0.04  & 0.04  & 0.04  & 0.06 & 0.03  \\
$(\omega^{\text{I-SR}}, \beta^{\text{I-FFN}})$& XS-$R^2$                              & 0.14  & 0.13  & 0.11  & 0.14  & 0.14  & 0.15  & 0.19 & 0.14  \\ 
\hline
& SR                            & 0.69  & 0.79  & 0.82  & 0.84  & 0.83  & 0.86  & 0.86  & 0.94  \\
IPCA Max SR & EV                & 0.01  & 0.01  & 0.01  & 0.01  & 0.01  & 0.01  & 0.01  & 0.01  \\
$(\omega^{\text{I-SR}}, \beta^{\text{I-SR}})$& XS-$R^2$                      & -0.05 & -0.04 & -0.04 & -0.04 & -0.04 & -0.04 & -0.04 & -0.04  \\
\hline
& SR                            & 0.11  & 0.11  & 0.15  & 0.17  & 0.15  & 0.15  & 0.14  & 0.16  \\
IPCA Max EV & EV               & 0.04  & 0.04  & 0.04  & 0.04  & 0.04  & 0.04  & 0.04  & 0.04  \\
$(\omega^{\text{I-EV}}, \beta^{\text{I-EV}})$& XS-$R^2$                      & -0.02 & -0.03 & -0.03 & -0.03 & -0.03 & -0.03 & -0.03 & -0.03  \\ 
\hline
& SR                                 & -0.06 & 0.15  & 0.12  & 0.41  & 0.33  & 0.37  & 0.34  & 0.41  \\
IPCA Max XS-$R^2$ & EV               & -0.02 & -0.01 & -0.02 & -0.02 & -0.02 & -0.01 & -0.02 & -0.02  \\
$(\omega^{\text{I-XS}}, \beta^{\text{I-XS}})$& XS-$R^2$                          & -0.03 & 0.07  & 0.06  & 0.12  & 0.12  & 0.13  & 0.13  & 0.14  \\ 
\hline
& SR                                & 0.69  & 0.79  & 0.82  & 0.84  & 0.83  & 0.86  & 0.86  & 0.94  \\
IPCA Multifactor & EV                 & 0.05  & 0.05  & 0.06  & 0.06  & 0.06  & 0.06  & 0.06  & 0.07  \\
$(b_{t,i}\in \mathbbm R^{K} )$& XS-$R^2$                            & -0.04 & -0.03 & -0.02 & -0.01 & -0.02 & -0.01 & -0.02 & -0.02  \\ 
\bottomrule
\end{tabular}
}
\bnotetab{This table shows the out-of-sample asset pricing results for different SDFs based on IPCA. We consider $K=3$ to $10$ IPCA factors. $(\omega^{\text{I-GAN}}, \beta^{\text{I-GAN}})$ uses the GAN framework to estimate the SDF weights of IPCA factors and the SDF loading. $(\omega^{\text{I-SR}}, \beta^{\text{I-FFN}})$ is the unconditional mean-variance efficient combination of IPCA factors with flexible SDF weights. $(\omega^{\text{I-SR}}, \beta^{\text{I-SR}})$ restricts those weights to be linear. $(\omega^{\text{I-XS}}, \beta^{\text{I-EV}})$ combines the IPCA factors to maximize $EV$ while $(\omega^{\text{I-XS}}, \beta^{\text{I-XS}})$ maximizes $\text{XS-}R^2$. The multi-factor representation obtains the residuals with a cross-sectional regression on the multiple loadings. The SDF weights and loadings are estimated on the training data and tuning parameters are chosen optimally on the validation data set.}
\end{table}

Table \ref{tab-IPCA} summarizes the out-of-sample asset pricing results.\footnote{Additional results are in the Internet Appendix.} We can replicate the high Sharpe ratios of \cite{kelly2018} by forming the unconditional mean variance efficient combination of the IPCA factors. Note that IPCA factors use more information than our models in the main text which might explain the high Sharpe ratios. The IPCA regressions use the dependency structure in the characteristic variables over time which makes the factor weights and loadings at each point in time a function of all past characteristics. The last subtable shows the IPCA results in a multi-factor framework, that is, we use a cross-sectional regression with multiple loadings to obtain the residual. Similar to PCA methods, IPCA captures a large amount of the variation in returns, but not their means. In fact, the cross-sectional $R^2$ can become negative. If we construct a one-factor model that maximizes the Sharpe ratio, the explained variation of IPCA drops to around 1\% without explaining more of the mean returns. The one factor that explains the most variation is simply the first IPCA factor. Similar to PCA methods, this first factor does not explain mean returns. Note that IPCA factors are not necessarily orthogonal to each other and hence estimating the factor that maximizes $EV$ can lead to slightly different results when including more factors. As expected, the best combination for explaining mean returns leads to substantially higher $\text{XS-}R^2$, however at the cost of SR and $EV$. Estimating nonlinear loadings $\beta^{\text{I-FFN}}$ explains more variation and leads to smaller pricing errors. However, the best performing model among all three dimensions is IPCA combined with GAN. The Sharpe ratios are close to those of the mean-variance efficient combination and actually higher than our benchmark GAN for $K \geq 7$ , while the $\text{XS-}R^2$ almost reaches the level of our benchmark GAN. The explained variation is among the highest of all models and only better for our fully flexible benchmark GAN. In summary, the GAN framework is complementary to multi-factor models and can optimally make use of the additional structure imposed on the SDF and the additional information incorporated in factors.



\section{Conclusion}\label{sec:conclusion}

We propose a new way to estimate asset pricing models for individual stock returns that can take advantage of the vast amount of conditioning information, while keeping a fully flexible form and accounting for time-variation. For this purpose, we combine three different deep neural network structures in a novel way: A feedforward network to capture non-linearities, a recurrent (LSTM) network to find a small set of economic state processes, and a generative adversarial network to identify the portfolio strategies with the most unexplained pricing information. Our crucial innovation is the use of the no-arbitrage condition as part of the neural network algorithm. We estimate the stochastic discount factor that explains all stock returns from the conditional moment constraints implied by no-arbitrage. Our SDF is a portfolio of all traded assets with time-varying portfolio weights which are general functions of the observable firm-specific and macroeconomic variables. Our model allows us to understand what are the key factors that drive asset prices, identify mispricing of stocks and generate the conditional mean-variance efficient portfolio.  

Our primary conclusions are four-fold. First, we demonstrate the potential of machine learning methods in asset pricing. We are able to identify the key factors that drive asset prices and the functional form of this relationship on a level of generality and with an accuracy that was not possible with traditional econometric methods. Second, we show and quantify the importance of including a no-arbitrage condition in the estimation of machine learning asset pricing models. The ``kitchen-sink'' prediction approach with deep learning does not outperform a linear model with no-arbitrage constraints. This illustrates that a successful use of machine learning methods in finance requires both subject specific domain knowledge and a state-of-the-art technical implementation. Third, financial data have a time dimension which has to be taken into account accordingly. Even the most flexible model cannot compensate for the problem that macroeconomic data seems to be uninformative for asset pricing if only the last increments are used as input. We show that macroeconomic conditions matter for asset pricing and can be summarized by a small number of economic state variables, which depend on the complete dynamics of all time series. Fourth, asset pricing is actually surprisingly ``linear''. As long as we consider anomalies in isolation the linear factor models provide a good approximation. However, the multi-dimensional challenge of asset pricing cannot be solved with linear models and requires a different set of tools.

Our results have direct practical benefits for asset pricing researchers that go beyond our empirical findings. First, we provide a new set of benchmark test assets. New asset pricing models can be tested on explaining our SDF portfolio respectively the portfolios sorted according to the risk exposure in our model. These test assets incorporate the information of all characteristics and macroeconomic information in a small number of assets. Explaining portfolios sorted on a single characteristic is not a high hurdle to pass. Second, we provide a set of macroeconomic time series of hidden states that encapsulate the relevant macroeconomic information for asset pricing. These time series can also be used as an input for new asset pricing models.\footnote{The data are available on https://mpelger.people.stanford.edu/research.} 

Last but not least, our model is directly valuable for investors and portfolio managers. The main output of our model is the risk measure $\beta$ and the SDF weight $\omega$ as a function of characteristics and macroeconomic variables. Given our estimates, the user of our model can assign a risk measure and its portfolio weight to an asset even if it does not have a long time series available.

\singlespacing
\bibliographystyle{econometrica}
{\small
\bibliography{main}
}
\onehalfspacing

%


\appendix

 \setcounter{equation}{0}
 \renewcommand{\theequation}{\thesection.\arabic{equation}}




\renewcommand{\theequation}{A.\arabic{equation}}%
\renewcommand{\thefigure}{A.\arabic{figure}} \setcounter{figure}{0}
\renewcommand{\thetable}{A.\Roman{table}} \setcounter{table}{0}

\section{Estimation Method}

\subsection{Feedforward Network (FFN)}\label{app:ffn}

\begin{figure}[H]
\centering
\caption{Feedforward Network with 3 Hidden Layers}\label{fig:dnn}
\includegraphics[width=0.75\textwidth]{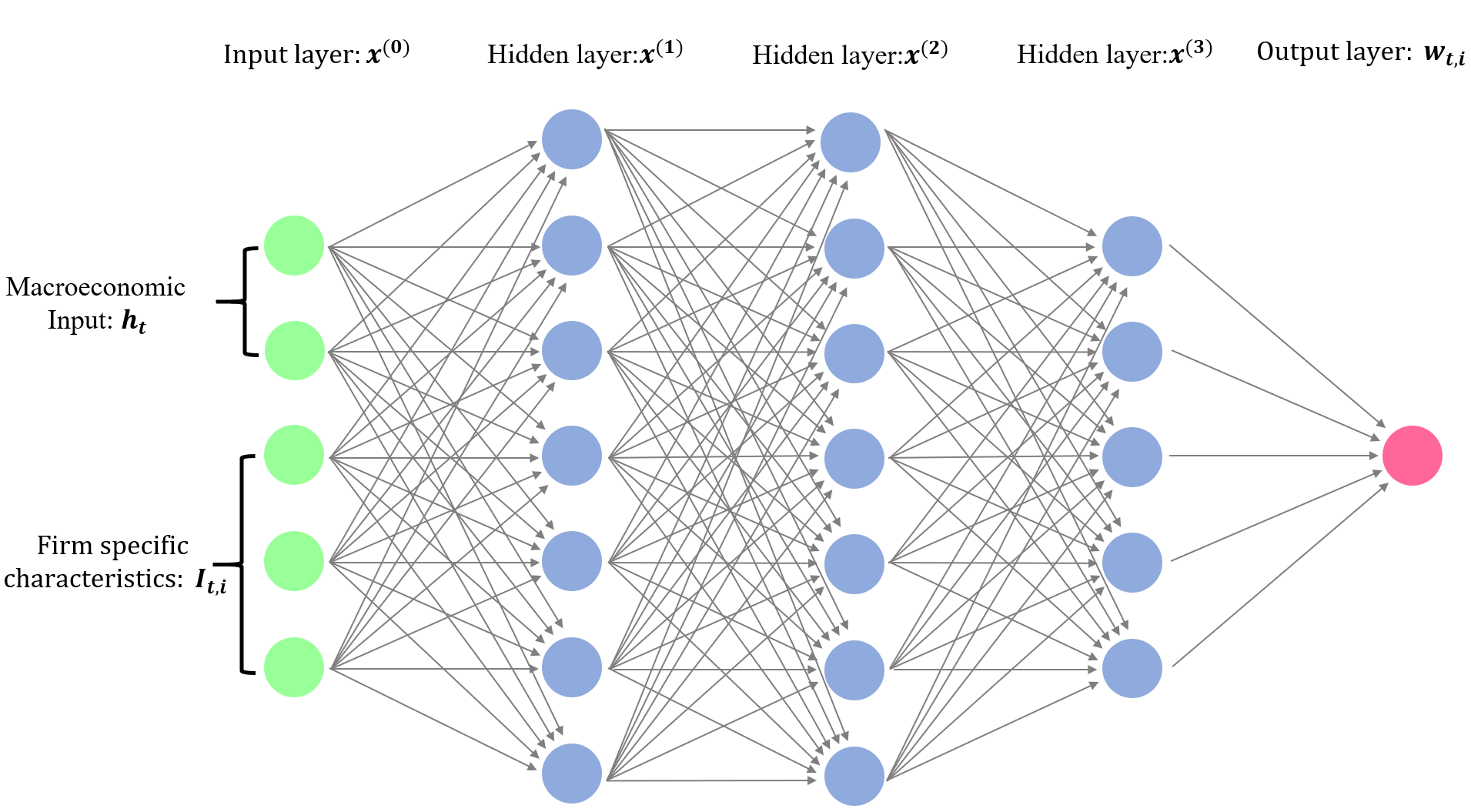}
\end{figure}

The deep neural network considers $L$ layers as illustrated in Figure \ref{fig:dnn}. Each hidden layer takes the output from the previous layer and transforms it into an output as 
\begin{align*}
x^{(l)} &= \text{ReLU}\left( {W^{(l-1)\top}} x^{(l-1)} + w_0^{(l-1)} \right) = \text{ReLU}\left( w_0^{(l-1)} + \sum_{k=1}^{K^{(l-1)}} w_k^{(l-1)} x^{(l-1)}_k \right)\\
y&= W^{(L)\top} x^{(L)} + w_0^{(L)}
\end{align*}
with hidden layer outputs $x^{(l)}=(x_1^{(l)},...,x_{K^{(l)}}^{(l)}) \in \mathbbm R^{K^{(l)}}$ and parameters $W^{(l)}=(w_1^{(l)},...,w_{K^{(l)}}^{(l)}) \in \mathbbm R^{K^{(l)} \times K^{(l-1)}}$ for $l=0,...,L-1$ and $W^{(L)} \in \mathbbm R^{K^{(L)}}$.

\subsection{Recurrent Neural Network (RNN)}\label{app:lstm}

The LSTM is composed of a cell (the memory part of the LSTM unit) and three ``regulators'', called gates, of the flow of information inside the LSTM unit: an input gate, a forget gate, and an output gate. Intuitively, the cell is responsible for keeping track of the dependencies between the elements in the input sequence. The input gate controls the extent to which a new value flows into the cell, the forget gate controls the extent to which a value remains in the cell and the output gate controls the extent to which the value in the cell is used to compute the output activation of the LSTM unit.

We take $x_t=I_t$ as the input sequence of macroeconomic information, and the output is the state processes $h_t$. 
At each step, a new memory cell $\tilde{c}_t$ is created with current input $x_t$ and previous hidden state $h_{t-1}$
\begin{align*}
\tilde{c}_t&=\tanh(W_h^{(c)}h_{t-1}+W_x^{(c)}x_t + w^{(c)}_0).
\end{align*}
The input and forget gates control the memory cell, while the output gate controls the amount of information stored in the hidden state:
\begin{align*}
\text{input}_t&=\sigma(W_h^{(i)}h_{t-1}+W_x^{(i)}x_t + w^{(i)}_0 )\\
\text{forget}_t&=\sigma(W_h^{(f)}h_{t-1}+W_x^{(f)}x_t + w^{(f)}_0  )\\
\text{out}_t&=\sigma(W_h^{(o)}h_{t-1}+W_x^{(o)}x_t + w^{(o)}_0  ).
\end{align*}
The sigmoid function $\sigma$ is an element-wise non-linear transformation. Denoting the element-wise product by $\circ$, the final memory cell and hidden state are given by
\begin{align*}
c_t=\text{forget}_t\circ c_{t-1}+ \text{input}_t\circ\tilde{c}_t, \qquad h_t=\text{out}_t\circ\tanh(c_t).
\end{align*}
We use the state processes $h_t$ instead of the macroeconomic variables $I_t$ as input to our SDF network. 


\subsection{Implementation}\label{app:implementation}

For training deep neural networks the vanilla stochastic gradient descend method has proven to be not an efficient method. A better approach is to use optimization methods that introduce an adaptive learning rate.\footnote{See e.g. \cite{ruder2016} and \cite{kingma2014}. We use the leading algorithm Adam. Other adaptive gradient descent methods include Adagrad or Adadelta.} We use Adam which is an algorithm for gradient-based optimization of stochastic objective functions, based on adaptive estimates of lower-order moments to continuously adjust the learning rate. It is morel likely to escape saddle points and hence is more accurate, while also providing faster convergence. 

Regularization is crucial and prevents the model from over-fitting on the training sample. Although $l_1/l_2$ regularization might also be used in training other neural networks, Dropout is preferable and generally results in better performances. The term ``Dropout'' refers to dropping out units in a neural network. By dropping out a unit, we mean temporarily removing it from the network, along with all its incoming and outgoing connections with a certain probability. Dropout can be shown to be a form of ridge regularization and is only applied during the training.\footnote{See e.g. \cite{srivastava2014} for the better performance results for Dropout and \cite{wager2013} for the connection with ridge.} When doing out-of-sample testing, we keep all the units and their connections.

In summary, the hyperparameter selection works as follows: (1) First, for each possible combination of hyperparameters (384 models) we fit the GAN model. (2) Second, we select the four best combinations of hyperparameters on the validation data set. (3) Third, for each of the four combinations we fit 9 models with the same hyperparameters but different initialization. (4) Finally, we select the ensemble model with the best performance on the validation data set. Table \ref{tab:hyper-parameter} reports the tuning parameters of the best performing model. The feedforward network estimating the SDF weights has 2 hidden layers (HL) each of which has 64 nodes (HU). There are four hidden states (SMV) that summarize the macroeconomic dynamics in the LSTM network. The conditional adversarial network generates 8 moments (CHU) in a 0-layer (CHL) network. The macroeconomic dynamics for the conditional moments are summarized in 32 hidden states (CSMV). This conditional network essentially applies a non-linear transformation to the characteristics and the hidden macroeconomic states and then combines them linearly. The resulting moments can, for example, capture the pricing errors of long-short portfolios based on characteristic information or portfolios that only pay off under certain macroeconomic conditions. The FFN for the forecasting approach uses the optimal hyperparameters selected by \cite{gu2018} which is a 3-layer neural network with $[32,16,8]$ hidden units, dropout retaining probability of 0.95 and a learning rate of 0.001.\footnote{We have estimated our models on two GPU clusters where each cluster has two Intel Xeon E5-2698 v3 CPUs, 1TB memory and 8 Nvidia Titan V GPUs. We have used TensorFlow with Python 3.6 for the model fitting. A complete estimation of the GAN model with hyperparameter tuning takes around 3 days. We have confirmed that our estimation results are robust to using a larger hyperparameter space. As a full hyperparameter search on a larger hyperparameter space can easily take weeks or months even on our fast GPU cluster, we have selectively tested further hyperparameters.}

Our GAN network is inspired by machine learning GANs as proposed in Goodfellow et al. (\citeyear{goodfellow2014generative}) but implemented differently and specifically designed for our problem. The machine learning GAN is usually implemented with deeper neural networks and a simultaneous optimization, that is, each network is only doing some optimization steps without completing the optimization before iterating with the other network. This is often necessary, as the machine learning GANs are applied to huge data sets where it would be computationally too expensive to solve each optimization completely. In contrast, we run each optimization step until convergence. For our benchmark model, this means in the first step we find an SDF to price all 10,000 stocks without instrumenting them. Then, we completely solve the adversarial problem to generate an 8-dimensional vector of instruments. In the third step we completely estimate the SDF that can price all 80,000 instrumented stock returns. It turns out that this procedure is very stable and already converges after the first three steps in our empirical analysis as shown in Figure IA.1 in the Internet Appendix. In contrast a conventional machine learning GAN typically creates a smaller sets of adversarial data for computational reasons which then also leads to more iterations to exhaust the information set.

\section{Simulation Example}\label{sec:simulation}

We illustrate with simulations that (1) the no-arbitrage condition in GAN is necessary to find the SDF in a low signal-to-noise setup, (2) the flexible form of GAN is necessary to correctly capture the interactions between characteristics, and (3) the RNN with LSTM is necessary to correctly incorporate macroeconomic dynamics in the pricing kernel. On purpose, we have designed the simplest possible simulation setup to convey these points and to show that the forecasting approach or the simple linear model formulations cannot achieve these goals.\footnote{We have run substantially more simulations for a variety of different model formulations, where we reach the same conclusions. The other simulation results are available upon request.}  

\begin{figure}[h!]
\centering
\tcapfig{SDF Weights $\omega$ for the First Model with 2 Characteristics}\label{fig:sim_2char}
\begin{center}
\begin{minipage}{0.45\textwidth}
\centering
{\footnotesize  \bf Population Model}
\includegraphics[width=1\textwidth]{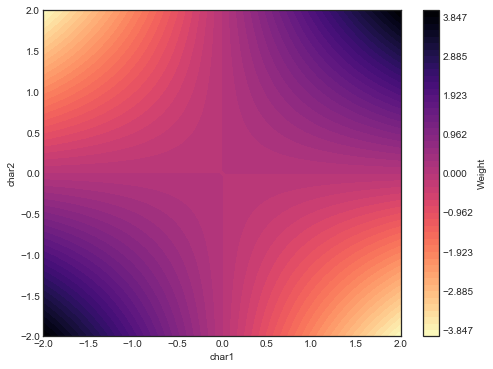}
\end{minipage}
\begin{minipage}{0.45\textwidth}
\centering
{\footnotesize  \bf GAN}
\includegraphics[width=1\textwidth]{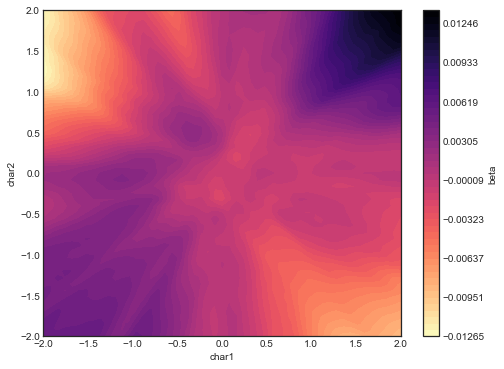}
\end{minipage}\\
\begin{minipage}{0.45\textwidth}
\centering
{\footnotesize \bf FFN}
\includegraphics[width=1\textwidth]{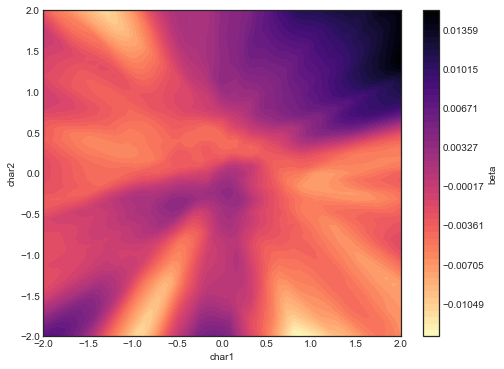}
\end{minipage}
\begin{minipage}{0.45\textwidth}
\centering
{\footnotesize  \bf LS}
\includegraphics[width=1\textwidth]{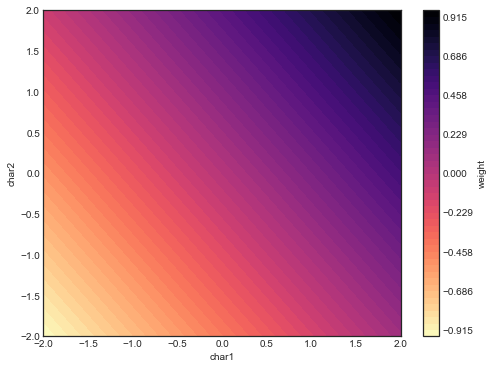}
\end{minipage}\\
\end{center}
\bnotefig{This figures shows the SDF weights $\omega$ as the function of the two characteristics estimated by different methods. Note that in our simple simulation the SDF weights $\omega$ coincide with the SDF loadings $\beta$.}
\end{figure}

Excess returns follow a no-arbitrage model $R_{t+1,i}^e = \beta_{t,i}F_{t+1} + \epsilon_{t+1,i}.$ In our simple model the SDF follows $F_t \overset{i.i.d.}{\sim}\mathcal{N}(\mu_F,\sigma_F^2)$ and the idiosyncratic component $\epsilon_{t,i}\overset{i.i.d.}{\sim}N(0,\sigma_e^2)$. We consider two different formulations for the risk-loadings:
\begin{enumerate}
    \item {\it Two characteristics:} The loadings are the multiplicative interaction of two characteristics
    \begin{align*}
        \beta_{t,i} =C_{t,i}^{(1)} \cdot C_{t,i}^{(2)} \qquad \text{with $C_{t,i}^{(1)},C_{t,i}^{(2)}\overset{i.i.d.}{\sim}\mathcal{N}(0,1)$}.
    \end{align*}
    \item {\it One characteristic and one macroeconomic state process:} The loading depends on one characteristic and a cyclical state process $h_t$:
    \begin{align*}
        \beta_{t,i} = C_{t,i} \cdot b(h_t), \qquad h_t=sin(\pi*t/24) + \epsilon^h_t, \qquad b(h)= \left \{
                \begin{array}{ll}
                  1 \qquad &\text{if $h>0$} \\
                  -1 \qquad &\text{otherwise.}
                \end{array}
              \right.
    \end{align*}
    We observe only the macroeconomic time series with trend $Z_t=\mu_M t +h_t$.
    All innovations are independent and normally distributed: $C_{t,i}\overset{i.i.d.}{\sim}\mathcal{N}(0,1)$ and $\epsilon^h_t\overset{i.i.d.}{\sim}\mathcal{N}(0,0.25)$. 
\end{enumerate}
The choice of the parameters is guided by our empirical results and summarized in Table \ref{tab:sim}. The panel data set is $N=500, T=600$, where the first $T_{train}=250$ are used for training, the next $T_{valid}=100$ observations are the validation and the last $T_{test}=250$ observations form the test data set. 
The first model setup with two characteristics has two distinguishing empirical features: (1) the loadings have a non-linear interaction effect for the two characteristics; (2) for many assets the signal-to-noise ratio is low. Because of the multiplicative form the loadings will take small values when two characteristics with values close to zero are multiplied. Figure \ref{fig:sim_2char} shows the form of the population loadings. The assets with loadings in the center are largely driven by idiosyncratic noise which makes it harder to extract their systematic component. 

Table \ref{tab:sim} reports the results for the first model. The GAN model outperforms the forecasting approach and the linear model in all categories. Note, that it is not necessary to include the elastic net approach as the number of covariates is only two and hence the regularization does not help. The Sharpe Ratio of the estimated GAN SDF reaches the same value as the population SDF used to generate the data. Based on the estimated loadings respectively the population loadings we project out the idiosyncratic component to obtain the explained variation and cross-sectional pricing errors. As expected the linear model is mis-specified for this setup and captures neither the SDF nor the correct loading structure. Note, that the simple forecasting approach can generate a high Sharpe Ratio but fails in explaining the systematic component.

\begin{table}[h!]
\centering
\tcaptab{~Performance of Different SDF Models in Two Simulation Setups}\label{tab:sim}
{\small
\begin{tabular}{lccc|ccc|ccc}
\toprule
& \multicolumn{3}{c}{Sharpe Ratio} & \multicolumn{3}{c}{EV} & \multicolumn{3}{c}{Cross-sectional $R^2$}\\
\cmidrule(l){2-10}
Model & Train & Valid & Test  & Train & Valid & Test & Train & Valid & Test\\
\midrule
\multicolumn{10}{c}{Two characteristics and no macroeconomic state variable}  \\
\midrule
Population & 0.96 & 1.09 & 0.94 &0.16 & 0.15 & 0.17 & 0.17 & 0.15 & 0.17  \\
GAN & 0.98 & 1.11 & 0.94 & 0.12 & 0.11 & 0.13 & 0.10 & 0.09 & 0.07 \\
FFN & 0.94 & 1.04 & 0.89 & 0.05 & 0.04 & 0.05 & -0.30 & -0.09 & -0.33  \\
LS & 0.07 & -0.10 & 0.01 & 0.00 & 0.00 & 0.00 & 0.00 & 0.01 & 0.01  \\
\midrule
\multicolumn{10}{c}{One characteristic and one macroeconomic state variable}  \\
\midrule
Population & 0.89 & 0.92 & 0.86 &0.18 & 0.18 & 0.17 & 0.19 & 0.20 & 0.15  \\
GAN & 0.79 & 0.77 & 0.64 & 0.18 & 0.18 & 0.17 & 0.19 & 0.20 & 0.15 \\
FFN & 0.05 & -0.05 & 0.06 & 0.02 & 0.01 & 0.02 & 0.01 & 0.01 & 0.02  \\
LS & 0.12 & -0.05 & 0.10 & 0.16 & 0.16 & 0.15 & 0.15 & 0.18 & 0.14  \\
\bottomrule
\end{tabular}
}
\bnotetab{This table reports the Sharpe Ratio (SR) of the SDF, explained time series variation (EV) and cross-sectional mean $R^2$ for the GAN,
FFN and LS model. EN is left out in this setup as there are only very few covariates. The data are generated with an SDF with Sharpe Ratio $SR=1$ and $\sigma_F^2=0.1$ and the idiosyncratic noise has $\sigma_e^2=1$. The macroeconomic time-series has the trend $\mu_M = 0.05$. The number of observations is $N=500, T=600$, $T_{train}=250$, $T_{valid}=100$ and $T_{test}=250$.\par}
\end{table}

Figure \ref{fig:sim_2char} explains why we observe the above performance results. Note, that the SDF has large positive respectively negative weights on the extreme corner combinations of the characteristics. The middle combinations are close to zero. The GAN network captures this pattern and assigns positive weights on the combinations of high/high and low/low and negative weights for high/low and low/high. The FFN on the other hand generates a more diffuse picture. It assigns negative weights for low/low combinations. The FFN SDF still loads mainly on the extreme portfolios which results in the high Sharpe Ratio. However, the FFN fails to capture the loadings correctly which leads to high unexplained variation and pricing errors. The linear model can obviously not capture the non-linear interaction. 

The second model setup with a macroeconomic state variable is designed to model the effect of a boom and recession cycle on the pricing model. In our model the SDF affects the assets differently during a boom and recession cycle. Note that in our example the macroeconomic variable can by construction only have a scaling effect on the loadings of the SDF factor, but not change its cross-sectional distribution which only depends on firm-specific information.

\begin{figure}[h!]
\centering
\caption{Dynamics of the Macroeconomic State Variable}\label{fig:sim_macro}
\begin{minipage}{\textwidth}
\centering
{\footnotesize  \bf Observed Macroeconomic Variable}
\includegraphics[width=0.75\textwidth]{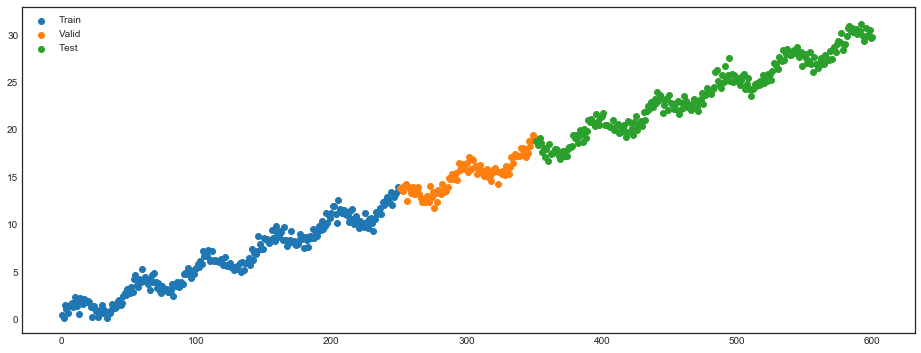}
\\
\centering
{\footnotesize  \bf First order difference of Macroeconomic Variable}
\includegraphics[width=0.75\textwidth]{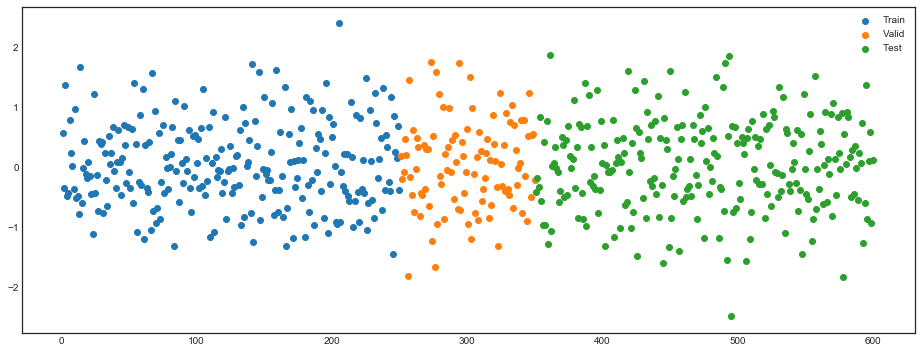}
\\
\centering
{\footnotesize  \bf True hidden Macroeconomic State}
\includegraphics[width=0.75\textwidth]{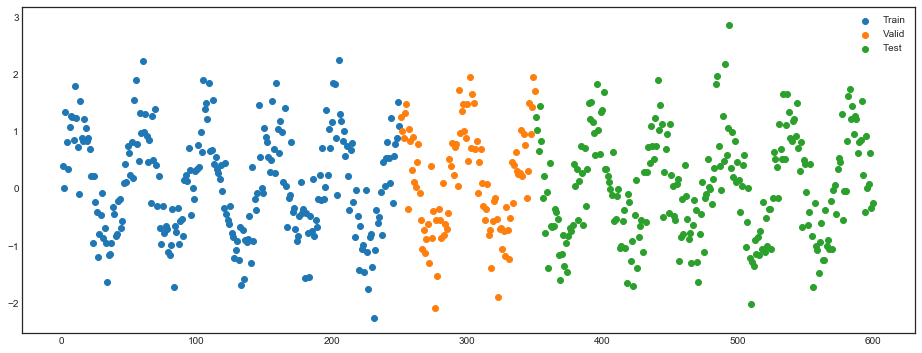}
\\
\centering
{\footnotesize  \bf Fitted Macroeconomic State by LSTM}
\includegraphics[width=0.75\textwidth]{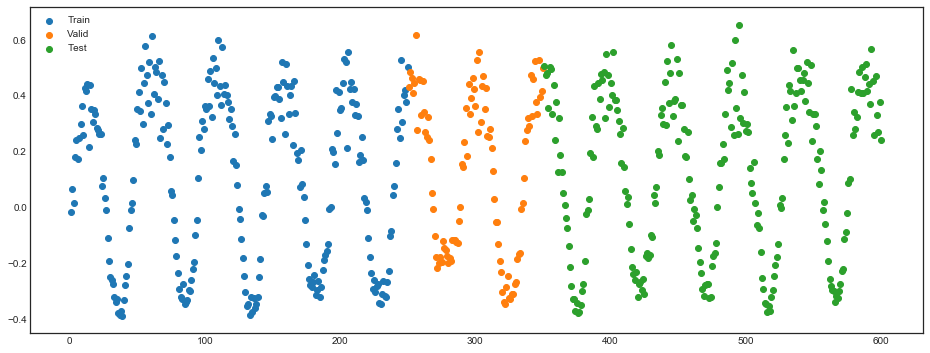}
\end{minipage}
\end{figure}

Figure \ref{fig:sim_macro} illustrates the path of the observed macroeconomic variable that has the distinguishing feature that we observe for most macroeconomic variables in our data set: (1) the macroeconomic process is non-stationary, i.e. it has a trend; (2) the process has a cyclical dynamic structure, i.e. it is influenced by business cycles. For example GDP level has a similar qualitative behaviour.  The conventional approach to deal with non-stationary data is to take first differences. Figure \ref{fig:sim_macro} shows that the differenced data does indeed look stationary but loses all information about the business cycle. The LSTM network in our GAN model can successfully extract the hidden state process. The models based on only the most recent first differences can by construction not infer any dynamics in the macroeconomic variables.

Table \ref{tab:sim} reports the results for the second model with macroeconomic state variable. As expected our GAN model strongly outperforms the forecasting and the linear model. Note, that the loading function here is linear and the macroeconomic state variable is only a time-varying proportionality constant for the loadings and SDF weights. As the projection on the systematic component is not affected by a proportionality constant, the linear model actually achieves the same explained variation and pricing errors as GAN. However, the Sharpe Ratio of the linear model collapses as for roughly half of the times it uses the wrong sign for the SDF weights.

The simulation section illustrates three findings: (1) All three evaluation metrics (SR, EV and XS-R2) are necessary to assess the quality of the SDF factor. (2) By conditioning only on the most recent macroeconomic observations, general macroeconomic dynamics are ruled out. (3) The no-arbitrage condition in the GAN model helps to deal with a low signal-to-noise ratio.

\section{Overview of Conditional SDF Models}\label{sec:appcondSDF}

We survey the most recent advances of relevant machine learning methods in asset pricing and explain their differences. All asset pricing models are captured by the general framework introduced in Section \ref{sec:noarbitrage}, which is based on the fundamental moment equation $\mathbbm E_t \left[ M_{t+1} R^e_{t+1,i}\right] = 0$, which implies the factor representation $\Ri= \bi^{\text{SDF}} F_{t+1} + \ei$. Different asset pricing model impose different structures on the SDF weights $\omega$ and SDF loadings $\beta^{\text{SDF}}$. The estimation challenge arises from modeling the conditional expectation $\Et[.]$ which can depend in a complex way on a large number of asset-specific and macroeconomic variables. This is where machine learning tools are essential to deal in a flexible way with the large dimensionality of the problem. Importantly, we need a model and estimator for both, the SDF weights $\omega$ and SDF loadings $\beta^{\text{SDF}}$, to explain individual stock returns.

\begin{subappendices}
\subsection{Characteristic Projection and Unconditional Models}\label{sec:unconditionalSDF}


The most common way is to translate the problem into an unconditional asset pricing model on sorted portfolios. 
Under additional assumptions one could obtain a valid SDF $M_{t+1}$ conditional on a set of asset-specific characteristics $I_{t,i}$ by its projection on the return space: 
\begin{align*}
M_{t+1} &= 1- \omega_{t}^{\top} \Rt  \qquad \text{with $\omega_{t,i} = f \left(I_{t,i}\right)$},
\end{align*}
where $I_{t,i}$ is a vector of $q$ characteristics observed for $N$ stocks and $f(\cdot)$ is a general, potentially nonlinear and non-separable function. Most of the reduced-form asset pricing models approximate this function by a (potentially very large) set of simple managed portfolios $f_j(\cdot)$, such that $f \left(I_{t,i}\right) \approx \sum_{j=1}^\Nb f_j \left(I_{t,i}\right) \tilde w_{j}$. The SDF then becomes a linear combination of these managed portfolios with constant weights $\tilde \omega_j$:
\begin{align}\label{eq:SDF}
M_{t+1} =1 - \sum_{j=1}^\Nb \tilde w_j \tilde R_{t+1,j} \qquad \text{with $\tilde R_{t+1,j} = \sum_{i=1}^N f_j \left(I_{t,i}\right) R^e_{t+1,i}$,}
\end{align}
where $\tilde R_{t+1}$ are the returns of $\Nb$ managed portfolios that correspond to different basis functions in the characteristic space. The number of basis portfolios increases by the complexity of the basis functions and the number of characteristics. The most common managed portfolios are sorted on characteristic quantiles, that is, they use indicator functions based on characteristic quantiles to approximate $f(I_{t,i})$. Popular sorts are the size and value double-sorted portfolios of \cite{10.2307/2329112}, that are also used to construct their long-short factors. Note, that these characteristic managed portfolios do not include macroeconomic variables which are not asset specific, in contrast to our GAN model. In order to simplify the exposition, we will focus now on the information set, which only conditions on firm-specific characteristics.

Importantly, the translation of the conditional model into an unconditional model implicitly imposes the following assumptions: First, the asset pricing modeler includes all basis functions $f_j \left(I_{t,i}\right)$ that are necessary to approximate the SDF weights $f \left(I_{t,i}\right) \approx \sum_{j=1}^\Nb f_j \left(I_{t,i}\right) \tilde w_{j}$. If the asset pricing modeler omits relevant basis functions, for example interactions between characteristics, the SDF will be misspecified. Second, the returns of the characteristics managed portfolios $\tilde R_{t+1,j}$ are mean and variance stationary, i.e. in particular they have constant mean and variance.

In this case the SDF can be obtained by solving an unconditional mean-variance optimization problem based on the characteristic managed portfolios:\footnote{In order to keep the notation closer to the literature we consider here the SDF $M_{t+1} =1 - \sum_{j=1}^J \tilde w_j (\tilde R_{t+1,j}-\mathbbm E[\tilde R_{t+1,j} ] )$ which yields the covariance matrix instead of the uncentered second moment matrix in the mean-variance problem. However, this normalization is inconsequential for the results.}
\begin{align}\label{SDFMV}
\tilde w =  \text{Var} \left( \tilde R_t \right )^{-1} \mathbbm E[ \tilde R_t] \qquad \tilde w \in \mathbbm R^{\tilde N^{\text{basis}}}.
\end{align}
The SDF weights for individual stocks are then given by the SDF weights for the characteristic managed portfolios and the weights of individual stocks in those portfolios, that is $w_{t,i} = \sum_{j=1}^\Nb f_{j}(I_{t,i}) \tilde w_j $.

Most asset pricing model fall into this category. Importantly, these models do not provide guidance on how to obtain the loadings for individual stock returns $\beta^{\text{SDF}}_t$, which is a separate problem. These unconditional models can only readily be used to explain the expected returns of the specific characteristic managed portfolios that have been used for their construction, but require an additional model to obtain the individual stock loadings $\beta^{\text{SDF}}_t$.

While Equation \ref{SDFMV} describes the population solution, we need additional assumptions to obtain a feasible estimator. If the number of characteristic managed portfolios $\Nb$ is large, either because the number of characteristics is large and/or because the functional form that should be approximated is complex and requires many basis functions, we cannot apply simple sample estimators and require regularization. The naive sample estimator for $\tilde w$ would be $\tilde w_{\text{MV}} = \hat \Sigma^{-1} \hat \mu$, where $\hat \Sigma^{-1} $ is the sample estimator of the covariance matrix and $\hat \mu$ is the estimator of the mean of the of characteristic managed portfolios $\tilde R_t$. \cite{kozak2017} (KNS) propose to apply a modified ridge penalty and lasso penalty to this regression problem to estimate $\tilde w$:
\begin{align*}
\tilde w_{\text{KNS}} = \text{arg min}_{\tilde w} \left[ \left( \hat \mu - \hat \Sigma \tilde w \right)^{\top} \hat \Sigma^{-1} \left(  \hat \mu - \hat \Sigma \tilde w \right) + 2 \nu_1 \sum_{j=1}^{\tilde N} |\tilde w_i | +  \nu_2 \tilde w^{\top} \tilde w  \right],
\end{align*}
where $\nu_1$ is a lasso penalty and $\nu_2$ a ridge-type penalty. \cite{kozak2017} advocate to use their estimator in the PCA space (which implies a diagonal matrix for $\hat \Sigma$), which as shown by \cite{lettaupelger2018} yields a closed-form solution:
\begin{align*}
\hat w_{\text{KNS},i} = \begin{cases}
    \frac{\hat \mu_i - \nu_1}{\hat \sigma_i^2 + \nu_2}& \text{if } \hat \mu_i \geq \nu_1 \\
    0 &\text{if } \hat \mu_i < \nu_1,
  \end{cases}
\end{align*}
where $\hat \sigma_i^2$ are the variances of the PCA factors and $\hat \mu_i$ their mean returns. The choice of the lasso penalty maps into a sparse representation of a small number of basis assets, which are selected based on their means. Importantly, the KNS model does not readily provide a framework to explain the mean return for individual stocks or assets, which are not used as basis assets.

The Asset-Pricing Trees (AP-Trees) of \cite{bryzgalova2019} generalize the SDF approach of KNS among several dimensions. First, given a choice of basis assets, they deal with the large uncertainty in mean estimation, by introducing mean-shrinkage in addition to variance shrinkage. Second and more importantly, they use recursive tree basis functions to generate the characteristic projected portfolios. These more flexible basis functions include the conventional univariate sorts as special cases, but allow also to deal with interaction effects between multiple characteristics. A particularly appealing element is to leverage the overlapping structure in trees and to use the robust SDF recovery to select the tree basis functions to span the SDF. This has the crucial advantage of obtaining a low dimensional representation similar to PCA factors, while retaining interpretability and capturing more general patterns. The primary focus of the basis functions obtained by \cite{bryzgalova2019} is to use them as informative test assets for unconditional asset pricing models.

\subsection{Inversion of Unconditional Models}

In order to use an SDF, which is estimated as an unconditional model from conditional portfolio sorts or projections, we need to invert the conditional projection. In other words, the unconditional SDF model on the projected returns $\tilde R_{t+1}$ needs to be mapped back into the individual stock returns $R_{t+1}^e$. The fundamental challenge is that this requires the modeling of conditional covariances of the individual stocks with the unconditional SDF model. Hence, specifying a set of portfolio sorts and estimating an unconditional SDF model on those portfolio sorts, is not sufficient. We need to estimate the conditional covariance of each stock with the SDF portfolios:
\begin{align*}
\bi^{\text{SDF}} =\frac{\cov(\Ri,F_{t+1})}{\var(F_{t+1})} = \frac{\cov(\Ri, \tilde w^{\top} \tilde R_t )}{\var (\tilde w^{\top} \tilde R_t )}. 
\end{align*}
Obviously, using an unconditional covariance is not internally consistent if stocks have time-varying characteristics. In other words, the assumption of a mean and covariance stationary model on portfolio sorts, implies a conditional model for individual stock returns. An internally consistent model would use the same basis functions $f_j \left(I_{t,i}\right)$, which are used to span the conditional SDF weights $w_{t,i}$ to span the conditional SDF loadings $\beta^{\text{SDF}}_{t,i}$. This requires an additional non-parametric regression with the corresponding challenges in a high-dimensional setup. 



\subsection{Unconditional Factor Models}
The linear factor model literature imposes the additional assumption that a small number of risk factors based on characteristic managed portfolios should span the SDF.   
The majority of the literature studies unconditional factor models on characteristic managed portfolios. This requires the same assumptions as in the unconditional SDF models in Section \ref{sec:unconditionalSDF}. In addition, it imposes that the excess returns of characteristic managed portfolios follow a factor structure
\begin{align}
\tilde R_{t,i}&= \tilde F_t \tilde \beta_i^{\top} + e_{t,i} \qquad i=1,...,\Nb, \;\; t=1,...,T 
\end{align}
The factors can be observed fundamental factors, for example the Fama-French factor model of \cite{fama20151}, or latent asset pricing factors estimated from the unconditional moments of $\tilde R_t$ by PCA or its improvement RP-PCA (\cite{lettaupelger2018}). The tangency portfolio, that is spanned by ${\tilde F_t}$, has the factor weights
\begin{align}\label{eqn:factorsdf}
    \tilde \omega_{{\tilde F}} = \Sigma_{\tilde F}^{-1}  \mu_{\tilde F},
\end{align}
where $\*{\mu}_{\tilde F}$ and $\Sigma_{\tilde F}$ are the mean and variance-covariance matrix of ${\tilde F_t}$. The implied SDF is given by $M_t  = 1- \tilde \omega_{{\tilde F}}^{\top} \left({\tilde F}_{t} - \E[{\tilde F}_{t}] \right)$. The SDF weights of the factors in Equation \ref{eqn:factorsdf} follow from pricing the factors with the factors themselves, that is the factors serve as basis and test assets to obtain the SDF weight. Importantly, the assumption of constant loadings $\tilde \beta$ is only reasonable on the characteristic managed portfolios, but individual stock returns must have time-varying loadings if characteristics are time-varying. Hence, estimating any factor model on portfolio sorts does not readily imply a factor model and an SDF on individual stock returns. The loadings for individual stock returns are given by the conditional loadings $\bi^{\text{SDF}}$: 
\begin{align*}
\bi^{\text{SDF}} =\frac{\cov(\Ri,F_{t+1})}{\var(F_{t+1})} =\frac{\cov(\Ri,   \tilde \omega_{\text{$\tilde F$}}^{\top}  \tilde F^{t+1} )}{\var ( \tilde \omega_{\text{$\tilde F$}}^{\top}  \tilde F^{t+1} )} .
\end{align*}
If for example one of the risk factors is a ``size'' factor, which is a long-short factor based on size sorted quantiles, such a risk factor should have a larger loading $\tilde \beta$ for a portfolio that includes only small stocks compared to a portfolio that includes only large stocks. As the market capitalization of stocks is time-varying, this mechanically implies that the loadings of a size factor have to be time-varying for individual stocks. More specifically, as the weights to construct the size factor are a function of the market capitalization, it mechanically implies that the loadings for individual stocks have to be a function of the market capitalization as well.

In practice, unconditional factors, which have been constructed based on a specific choice of portfolio sorts, e.g. the 3 Fama-French factors in \cite{10.2307/2329112} based on double-sorted size and book-to-market sorted quantiles or PCA factors estimated from a large set of univariate quantile sorts, have been applied to test assets which are not the basis assets or to individual stock returns. The usual application runs unconditional time-series regressions on the set of factors. This is only valid if the loadings to the factors and the moments of the factors are constant over time, which by construction cannot be the case for individual stocks with time-varying characteristics. A common practice is to run rolling window regressions to obtain local loadings on individual stocks. While this can potentially address the time-variation, it does not readily result in a conditional factor model which can be evaluated on different stocks out-of-sample. 

The extreme unbalancedness in individual stock returns severely limits the use of local window regressions with unconditional factors. The stocks in the first part of the data are too a large extent not available in the second part, which restricts the number of stocks that can be used for an out-of-sample evaluation. In a conditional model we estimate the conditional SDF weights $\omega$ and loadings $\beta^{\text{SDF}}$ as a function of characteristics on different stocks for the estimation and evaluation. Therefore, we can evaluate the model on stocks that are not available in the first part of the data. Importantly, the conditional model also allows us to directly study the economic sources of risk in terms of firm-characteristics.

\subsection{Conditional Factor Models}

A conditional factor model assumes that the SDF is spanned by a linear combination of conditional risk factors, and hence it restricts the basis assets that span the SDF. In contrast to the unconditional model, the SDF weights and loadings of the conditional risk factors are a function of the characteristics and hence time-varying. We use the Instrumented Principal Component Analysis (IPCA) of \cite{kelly2018} to illustrate this setup. This conditional factor model directly models individual stock returns as a function of characteristics given by $\Ri= b_{t,i}^{\top} f_{t+1}^{\text{IPCA}} + \ei$ with $b_{t,i}= I_{i,t}^{\top} \Gamma_b.$ Instead of allowing the SDF to be $\M=1-  \sum_{i=1}^N \wi \Ri $, it is restricted to $\M=1-  \sum_{k=1}^K w^{\text{IPCA}}_{t,k} f_{t+1,k}^{\text{IPCA}}$, where the IPCA factors are estimate with IPCA. Even after estimating the conditional factors, we still need to solve a conditional GMM problem.  The fundamental moment equation becomes
\begin{align}\label{eqn:IPCA}
\mathbbm E_t \left[ \left(1-  \sum_{k=1}^K w^{\text{IPCA}}_{t,k} f_{t+1,k}^{\text{IPCA}}\right) R^e_{t+1,i}\right] = 0 .
\end{align}
If we assume no misspecification and that the $\ei$ are independent of the IPCA factors with a conditional mean of zero, then population solution for the SDF weights to the equation \ref{eqn:IPCA} becomes 
\begin{align*}
w^{\text{IPCA}}_t = \text{Cov}_t \left(f_{t+1}^{\text{IPCA}} \right)^{-1} \mathbbm E_t [f_{t+1}^{\text{IPCA}} ],
\end{align*}
which requires the estimation of conditional moments. These conditional moments can be estimated by a prediction of $f_{t+1}^{\text{IPCA}}$ and ${f_{t+1}^{\text{IPCA}}}^2$ using the information set at time $t$. Given the SDF weights $w^{\text{IPCA}}_t$, the SDF loadings are a linear combination of the conditional factor loadings $\beta^{\text{SDF}}_t= b_{t,i}^{\top} \nu_t^{\text{IPCA}}$ such that
\begin{align*}
\Ri= b_{t,i}^{\top}\nu_t^{\text{IPCA}} {w^{\text{IPCA}}_t}^{\top}    f_{t+1}^{\text{IPCA}} + \ei. 
\end{align*}
Under additional assumptions it holds that $\nu_t^{\text{IPCA}}=w^{\text{IPCA}}_t \cdot c_t$ up to a proportionality constant $c_t$, but this does not need to hold in the general case. However, the choice of $w^{\text{IPCA}}_t$ uniquely pins down $\nu_t^{\text{IPCA}}$, that is, selecting a combination of the conditional factors results in a specific combination of the conditional loadings.

Several papers with conditional factors assume unconditional SDF weights, that is, they set 
\begin{align*}
w^{\text{IPCA}} = \text{Cov} \left(f_{t+1}^{\text{IPCA}} \right)^{-1} \mathbbm E [f_{t+1}^{\text{IPCA}} ].
\end{align*}
However, this choice of SDF weights also restrict the SDF loadings to be a specific linear combination of the conditional factors loadings, that is, $w^{\text{IPCA}}$ determines $\nu_t^{\text{IPCA}}$. The residuals relative to a conditional multivariate cross-sectional regression on $b_{t}$ are not the same residuals as those relative to the SDF loading $\beta^{\text{SDF}}_t$, when the SDF weights are set to the unconditional tangency portfolio weights $w^{\text{IPCA}} = \text{Cov} \left(f_{t+1}^{\text{IPCA}} \right)^{-1} \mathbbm E [f_{t+1}^{\text{IPCA}} ]$. A coherent asset pricing model needs to impose the same constraints on how it combines the conditional factors and conditional loadings into a one-factor representation. In Section \ref{sec:IPCA}, we show that for the IPCA model the unconditional mean-variance combination of the conditional factors does not explain the mean returns of stocks, while the conditional combination of the conditional factors estimated with the GAN framework results in a better asset pricing model.

\subsection{Adversarial Estimation and Mean-Variance Optimization}

We compare the adversarial estimation with the mean-variance estimation of the SDF. In order to provide intuition, we start with the special case of pre-specified basis functions to estimate the conditional moments. Intuitively, we assume that a pre-specified set of portfolios sorts, for example univariate sorting based on characteristics, is sufficiently rich to span the SDF. Our starting point is the conditional moment equation $\mathbbm E_t \left[ M_{t+1} R^e_{t+1,i}\right] = 0$, which implies the unconditional moments
\begin{align}
\mathbbm E[\M \Ri g(I_{t,i})] = 0
\end{align}
for any function $g(.)$. In the case of pre-specified basis functions, we assume that the SDF weight and the conditioning functions, that are sufficient to identify the SDF weights, are spanned by those basis functions:
\begin{align*}
w(I_{t,i}) = \sum_{j=1}^{N^{\text{basis}}} f_j(I_{t,i}) \tilde w_j, \qquad  \qquad  g(I_{i,t}) = \sum_{j=1}^{N^{\text{test}}} g_j(I_{t,i}) \tilde w^{\text{test}}_j.
\end{align*}
The $N^{\text{basis}}$ characteristic managed portfolios $\tilde R_{t+1,j}^{\text{basis}}=\sum_{i=1}^{N} f_j(I_{t,i} )R^e_{t+1,i}$ can be interpreted as the basis assets to span the SDF. The $N^{\text{test}}$ characteristic managed portfolios $\tilde R_{t+1,j}^{\text{test}}=\sum_{i=1}^{N} g_j(I_{t,i}) R^e_{t+1,i}$ correspond to the test assets, which are used to identify and estimate the constant weights $\tilde w$. For the pre-specified basis functions, we impose the additional assumption that the basis and test assets have stationary first and second moments. 

First, we consider exact identification in the GMM problem. If we set the basis assets identical to the test assets, that is $\tilde R_{t}^{\text{basis}}=\tilde R_{t}^{\text{test}}$, then the unconditional moment equation
\begin{align}
\mathbbm E \left[ \left( 1 - \tilde w^{\top} \tilde R_{t}^{\text{basis}} \right) \tilde R_{t}^{\text{test}} \right ] = 0 \label{eqn:GMM}
\end{align}
has the solution
\begin{align}
\tilde w = \mathbbm E \left[  \tilde R_{t}^{\text{basis}}  { \tilde R_{t}^{\text{basis}\top}}  \right]^{-1} \mathbbm E \left[  \tilde R_{t}^{\text{basis}}  \right] .\label{eqn:meanvar}
\end{align}

If we approach this problem from an adversarial perspective, the solution would be the same, that is, the problem
\begin{align}
\min_{\tilde w} \max_{\tilde R_{t}^{\text{test}}} \left \|  \mathbbm E \left[ \left( 1 - \tilde w^{\top} \tilde R_{t}^{\text{basis}} \right) \tilde R_{t}^{\text{test}} \right]  \right \|_2^2 \label{eqn:advGMM}
\end{align}
has the solution in Equation \ref{eqn:meanvar}. This is because we have a GMM problem with exact identification, where the number of parameters is exactly the same as the number of moments. While the discussion so far is based on population moments, it becomes more complex when we estimate it empirically. When the number of basis and test assets $N^{\text{basis}}=N^{\text{test}}$ is large relative to the number of time-series observations $T$, then we need to regularize the second (and potentially also first) sample moment of $\tilde R_{t}^{\text{basis}}$ by using some form of regularization. The possible regularizations include using the PCA factors based on the covariance matrix of $\tilde R_{t}^{\text{basis}}$, or a ridge, lasso or elastic penalty in the mean-variance optimization problem. These regularizations have in common that they either hard-threshold (in the case of PCA) or soft-threshold (for ridge penalties) the components that explain less variation in the relatively arbitrarily chosen set of basis and test assets. One problem of such an approach is that it misses weak factors, which are relevant for explaining mean returns, but would be down-weighted by the regularization. 

In general, we have an overidentified GMM problem, where the basis assets for the SDF and the test assets do not need to be the same. This means that the space of test assets can be larger than the set of basis assets and $N^{\text{test}}>N^{\text{basis}}$. Under appropriate assumptions, the solution to the unconditional moments in \ref{eqn:GMM} becomes
\begin{align}
\tilde w = \left( \mathbbm  E \left[ \tilde R_{t}^{\text{basis}} {\tilde R_{t}^{\text{test}\top}} \right] \Omega_{\text{GMM}} \mathbbm E \left[ \tilde R_{t}^{\text{test}} {\tilde R_{t}^{\text{basis}\top}} \right]  \right)^{-1} \mathbbm E \left[  \tilde R_{t}^{\text{basis}} {\tilde R_{t}^{\text{test}\top}} \right] \Omega_{\text{GMM}}\mathbbm E \left[ \tilde R_{t}^{\text{test}} \right], \label{eqn:GMMgeneral}
 \end{align}
where $\Omega_{\text{GMM}}$ is the GMM weighting matrix for the moments. In the special case of an identity matrix, that is $\Omega_{\text{GMM}}=I_{N^{\text{test}}}$, all test assets receive the same weight. The sample solution obviously requires some form of regularization similar to the case with exact identification. 

If the number of test assets is larger than the number of basis assets, then solution of the adversarial GMM in \ref{eqn:advGMM} is more complex and can in general not be mapped into the form of Equation \ref{eqn:GMMgeneral}. The adversarial approach has a number of advantages relative to the conventional regularized GMM framework:
\begin{enumerate}
\item It is more robust to misspecification as shown in \cite{hansen1997}. More specifically, if the set of basis functions is chosen too restrictive or the form or regularization is not appropriate, the adversarial approach provides a more robust pricing kernel.
\item The adversarial GMM can address the issue of weak asset pricing factors. The test assets in most asset pricing applications are to some degree arbitrarily chosen (usually by using the same set of double-sorted portfolios). A weak asset pricing factor might only explain a small amount of variation for the specific choice of test assets, but might be important to explain a certain component of the SDF, that is relevant for the risk premium. The adversarial GMM will upweight this component, while a conventional regularization based on ridge or PCA will neglect these weak factors.
\item The adversarial GMM will ensure identification of all parameters of the SDF. If certain components of the SDF are based on weak factors, the conventional GMM with regularization might not identify those components.
\end{enumerate}


The general problem considered in this paper is substantially harder as we do not use pre-specified basis functions for the basis and test assets, but learn them from the data. The set of potential non-parametric basis functions that can be generated by the neural networks is extremely large and goes into many millions, when considering all possible non-linear interactions. Under this level of generality and when we allow for this degree of flexibility, there is no meaningful way to solve this problem in a conventional regularized mean-variance framework. Here, the adversarial GMM framework is essential to obtain a feasible solution. Hence, in addition to the benefits outlined above, the adversarial approach provides a feasible solution to the general problem that cannot be solved otherwise. Furthermore, the non-parametric adversarial GMM approach is based on weaker assumptions as it does not require the stationarity of the first and second moments of a large number of pre-specified sorted portfolios, which is assumed in \ref{eqn:GMMgeneral}. The non-parametric adversarial GMM only requires the orthogonality of $\left( 1 - \sum_{i=1}^N w(I_{t,i})  R^e_{t+1,i} \right)$ from $R^e_{t+1,i} g(I_{t,i})$, which allows for time-varying moments.

\end{subappendices}

\clearpage
\section{List of the Firm-Specific Characteristics}

\begin{table}[htb]
\tcaptab{~\textbf{Firm Characteristics by Category}}
\bigskip
\label{tab:category}
\footnotesize
\begin{tabular}{lllllll}

\toprule
     & \multicolumn{2}{l}{{\ul \textbf{Past Returns}}}                                                        &  &      & \multicolumn{2}{l}{{\ul \textbf{Value}}}                                                                       \\
(1)  & r2\_1                        & Short-term momentum                                   &  & (26) & A2ME                             & Assets to market cap                                     \\
(2)  & r12\_2                       & Momentum                                              &  & (27) & BEME                             & Book to Market Ratio                                     \\
(3)  & r12\_7                       & Intermediate momentum                                 &  & (28) & C                                & Ratio of cash and short-term \\
  &                        &                                &  & &                                 & investments to total assets \\
(4)  & r36\_13                      & Long-term momentum                                    &  & (29) & CF                               & Free Cash Flow to Book Value                             \\
(5)  & ST\_Rev                      & Short-term reversal                                   &  & (30) & CF2P                             & Cashflow to price                                        \\
(6)  & LT\_Rev                      & Long-term reversal                                    &  & (31) & D2P                              & Dividend Yield                                           \\
     &                              &                                                       &  & (32) & E2P                              & Earnings to price                                        \\
     & \multicolumn{2}{l}{{\ul \textbf{Investment}}}                                                      &  & (33) & Q                                & Tobin's Q                                                \\
(7)  & Investment                   & Investment                                            &  & (34) & S2P                              & Sales to price                                           \\
(8)  & NOA                          & Net operating assets                                  &  & (35) & Lev                              & Leverage                                                 \\
(9)  & DPI2A                        & Change in property, plants, and              &  &      &                                  &                                                          \\
  &                         &  equipment             &  &      &                                  &                                                          \\
(10) & NI                           & Net Share Issues                                      &  &      & \multicolumn{2}{l}{{\ul \textbf{Trading Frictions}}}                                                          \\
     &                              &                                                       &  & (36) & AT                               & Total Assets                                             \\
     & \multicolumn{2}{l}{{\ul \textbf{Profitability}}}                                                       &  & (37) & Beta                             & CAPM Beta                                                \\
(11) & PROF                         & Profitability                                         &  & (38) & IdioVol                          & Idiosyncratic volatility                                 \\
(12) & ATO                          & Net sales over lagged net operating assets            &  & (39) & LME                              & Size                                                     \\
(13) & CTO                          & Capital turnover                                      &  & (40) & LTurnover                        & Turnover                                                 \\
(14) & FC2Y                         & Fixed costs to sales                                  &  & (41) & MktBeta                          & Market Beta                                              \\
(15) & OP                           & Operating profitability                               &  & (42) & Rel2High                         & Closeness to past year high                              \\
(16) & PM                           & Profit margin                                         &  & (43) & Resid\_Var                       & Residual Variance                                        \\
(17) & RNA                          & Return on net operating assets                        &  & (44) & Spread                           & Bid-ask spread                                           \\
(18) & ROA                          & Return on assets                                      &  & (45) & SUV                              & Standard unexplained volume                              \\
(19) & ROE                          & Return on equity                                      &  & (46) & Variance                         & Variance                                                 \\
(20) & SGA2S                        & Selling, general and administrative  &  &      &                                  &                                                          \\
 &                         &  expenses to sales &  &      &                                  &                                                          \\
(21) & D2A                          & Capital intensity                                     &  &      &                                  &                                                          \\
     &                              &                                                       &  &      &                                  &                                                          \\
     & \multicolumn{2}{l}{{\ul \textbf{Intangibles}}}                                                          &  &      &                                  &                                                          \\
(22) & AC                           & Accrual                                               &  &      &                                  &                                                          \\
(23) & OA                           & Operating accruals                                    &  &      &                                  &                                                          \\
(24) & OL                           & Operating leverage                                    &  &      &                                  &                                                          \\
(25) & PCM                          & Price to cost margin                                  &  &      &                                  &                                                     \\
\bottomrule    
\end{tabular}
\bnotetab{This table shows the 46 firm-specific characteristics sorted into six categories. More details on the construction are in the Internet Appendix.}
\end{table}

\clearpage
\newgeometry{top=0.8in, bottom=0.9in,hmargin=0.85in}

\section{Asset Pricing Results for Sorted Portfolios}

\begin{table}[H]
{\small
\tcaptab{~Explained Variation and Pricing Errors for Decile Sorted Portfolios}\label{tab:decile4}
\centering
\begin{tabular}{cccc|ccc||ccc|ccc}
\toprule
 & EN & FFN & GAN & EN& FFN & GAN  & EN & FFN & GAN & EN& FFN & GAN \\
 \cmidrule{2-13}
 &  \multicolumn{6}{c||}{Short-Term Reversal} & \multicolumn{6}{c}{Momentum} \\
\midrule
 Decile & \multicolumn{3}{c|}{Explained Variation} & \multicolumn{3}{c||}{Alpha} &  \multicolumn{3}{c|}{Explained Variation} & \multicolumn{3}{c}{Alpha}\\
1 & 0.84 & 0.74 & 0.77 & -0.18 & -0.21 & -0.13 & 0.04 & -0.06 & 0.33 & 0.37 & 0.39 & 0.11 \\
2 & 0.86 & 0.81 & 0.82 & 0.00 & -0.05 & 0.00 & 0.12 & 0.10 & 0.52 & 0.25 & 0.18 & -0.01 \\
3 & 0.80 & 0.82 & 0.84 & 0.13 & 0.04 & 0.06 & 0.19 & 0.25 & 0.66 & 0.14 & 0.05 & -0.06 \\
4 & 0.69 & 0.80 & 0.82 & 0.16 & 0.03 & 0.03 & 0.28 & 0.34 & 0.73 & 0.15 & 0.08 & -0.02 \\
5 & 0.58 & 0.68 & 0.71 & 0.13 & -0.03 & -0.04 & 0.37 & 0.46 & 0.80 & 0.19 & 0.09 & 0.02 \\
6 & 0.43 & 0.66 & 0.75 & 0.22 & 0.05 & 0.01 & 0.45 & 0.58 & 0.78 & 0.02 & -0.03 & -0.09 \\
7 & 0.23 & 0.64 & 0.77 & 0.20 & 0.03 & -0.02 & 0.62 & 0.69 & 0.68 & 0.01 & 0.01 & -0.05 \\
8 & -0.07 & 0.49 & 0.67 & 0.23 & 0.03 & -0.05 & 0.58 & 0.71 & 0.64 & -0.03 & -0.04 & -0.09 \\
9 & -0.25 & 0.29 & 0.58 & 0.30 & 0.09 & -0.01 & 0.55 & 0.70 & 0.58 & 0.08 & 0.04 & -0.03 \\
10 & -0.24 & -0.04 & 0.35 & 0.47 & 0.38 & 0.18 & 0.51 & 0.53 & 0.53 & 0.24 & 0.29 & 0.19 \\
\midrule
 & \multicolumn{3}{c|}{Explained Variation} & \multicolumn{3}{c||}{Cross-Sectional $R^2$} &  \multicolumn{3}{c|}{Explained Variation} & \multicolumn{3}{c}{Cross-Sectional $R^2$}\\
All & 0.43 & 0.58 & 0.70 & 0.45 & 0.79 & 0.94 & 0.26 & 0.27 & 0.54 & 0.66 & 0.71 & 0.93 \\
\midrule
&  \multicolumn{6}{c||}{Book-To-Market} & \multicolumn{6}{c}{Size} \\
\midrule
 Decile & \multicolumn{3}{c|}{Explained Variation} & \multicolumn{3}{c||}{Alpha} &  \multicolumn{3}{c|}{Explained Variation} & \multicolumn{3}{c}{Alpha}\\
1 & 0.38 & 0.66 & 0.70 & 0.03 & -0.12 & -0.08 & 0.80 & 0.75 & 0.79 & 0.09 & -0.00 & 0.10 \\
2 & 0.48 & 0.73 & 0.78 & 0.10 & -0.05 & -0.04 & 0.89 & 0.89 & 0.90 & -0.11 & -0.09 & -0.06 \\
3 & 0.71 & 0.84 & 0.86 & 0.07 & -0.03 & -0.01 & 0.91 & 0.80 & 0.91 & -0.07 & 0.02 & -0.02 \\
4 & 0.76 & 0.88 & 0.89 & 0.00 & -0.07 & -0.07 & 0.90 & 0.77 & 0.91 & -0.05 & 0.04 & -0.01 \\
5 & 0.82 & 0.87 & 0.88 & 0.05 & 0.02 & 0.01 & 0.90 & 0.78 & 0.91 & 0.01 & 0.10 & 0.04 \\
6 & 0.77 & 0.82 & 0.88 & 0.06 & 0.04 & 0.02 & 0.88 & 0.80 & 0.91 & 0.03 & 0.09 & 0.02 \\
7 & 0.81 & 0.81 & 0.87 & 0.03 & 0.08 & 0.03 & 0.84 & 0.81 & 0.89 & 0.04 & 0.05 & -0.01 \\
8 & 0.71 & 0.59 & 0.78 & 0.03 & 0.12 & 0.06 & 0.84 & 0.85 & 0.88 & 0.06 & 0.03 & -0.02 \\
9 & 0.80 & 0.72 & 0.80 & -0.02 & 0.11 & 0.07 & 0.77 & 0.81 & 0.82 & 0.06 & -0.01 & -0.04 \\
10 & 0.68 & 0.73 & 0.79 & -0.05 & -0.00 & 0.00 & 0.32 & 0.28 & 0.49 & -0.04 & -0.15 & -0.10 \\
\midrule
 & \multicolumn{3}{c|}{Explained Variation} & \multicolumn{3}{c||}{Cross-Sectional $R^2$} &  \multicolumn{3}{c|}{Explained Variation} & \multicolumn{3}{c}{Cross-Sectional $R^2$}\\
All & 0.70 & 0.75 & 0.82 & 0.97 & 0.94 & 0.98 & 0.83 & 0.78 & 0.86 & 0.96 & 0.95 & 0.97 \\
\bottomrule
\end{tabular}
\bnotetab{ This table shows the out-of-sample explained variation and pricing errors for decile-sorted portfolios based on Short-Term Reversal (\texttt{ST\_REV}), Momentum (\texttt{r12\_2}), Book to Market Ratio (\texttt{BEME}) and Size (\texttt{LME}).}
}
\end{table}

\begin{table}[H]
{\footnotesize
\tcaptab{~Explained Variation and Pricing Errors for Double-Sorted Portfolios based on Short-Term Reversal /Momentum and Size/Book-to-Market Ratio}\label{tab:double_sort_mom}
\centering
\begin{tabular}{ccccc|ccc||ccc|ccc}
\toprule
 \multicolumn{8}{c||}{Short-Term Reversal and Momentum}& \multicolumn{6}{c}{Size and Book-To-Market Ratio}\\
\midrule
 &  & EN & FFN & GAN & EN & FFN & GAN &   EN & FFN & GAN & EN & FFN & GAN \\
 \midrule
 \texttt{ST\_REV}/ & \texttt{r12\_2}/ & \multicolumn{3}{c|}{Explained Variation} & \multicolumn{3}{c||}{Alpha} &  \multicolumn{3}{c|}{Explained Variation} & \multicolumn{3}{c}{Alpha}\\
  \texttt{LME}& \texttt{BEME} & \multicolumn{3}{c|}{ } & \multicolumn{3}{c||}{}  & \multicolumn{3}{c|}{ } & \multicolumn{3}{c}{}\\
\midrule
1 & 1 & 0.35 & 0.32 & 0.62 & 0.16 & 0.13 & 0.08  & 0.55 & 0.47 & 0.63 & -0.01 & -0.00 & -0.06 \\
1 & 2 & 0.55 & 0.48 & 0.72 & -0.02 & -0.04 & -0.05  & 0.66 & 0.62 & 0.74 & 0.01 & 0.00 & -0.04 \\
1 & 3 & 0.66 & 0.61 & 0.74 & -0.06 & -0.07 & -0.05  & 0.74 & 0.70 & 0.76 & 0.04 & 0.01 & 0.01 \\
1 & 4 & 0.74 & 0.62 & 0.67 & -0.06 & -0.05 & -0.02 & 0.77 & 0.69 & 0.75 & 0.01 & -0.02 & 0.01 \\
1 & 5 & 0.69 & 0.58 & 0.58 & -0.10 & -0.06 & -0.03 & 0.70 & 0.66 & 0.76 & -0.01 & -0.03 & 0.02 \\
2 & 1 & 0.17 & 0.16 & 0.53 & 0.22 & 0.19 & 0.11 & 0.58 & 0.20 & 0.68 & 0.01 & 0.11 & -0.02 \\
2 & 2 & 0.32 & 0.39 & 0.67 & 0.18 & 0.11 & 0.08 & 0.68 & 0.48 & 0.81 & 0.02 & 0.07 & -0.01 \\
2 & 3 & 0.59 & 0.61 & 0.71 & 0.08 & 0.03 & 0.01 &  0.82 & 0.74 & 0.86 & 0.04 & 0.06 & 0.03 \\
2 & 4 & 0.72 & 0.74 & 0.59 & 0.00 & -0.03 & -0.02 &  0.81 & 0.75 & 0.85 & -0.03 & -0.00 & -0.01 \\
2 & 5 & 0.56 & 0.61 & 0.54 & 0.08 & 0.05 & 0.06 &  0.77 & 0.79 & 0.85 & -0.04 & 0.00 & 0.02 \\
3 & 1 & -0.02 & -0.01 & 0.48 & 0.18 & 0.16 & 0.01 &  0.53 & 0.25 & 0.73 & 0.08 & 0.12 & 0.02 \\
3 & 2 & 0.13 & 0.33 & 0.65 & 0.12 & 0.02 & -0.03 &  0.70 & 0.59 & 0.85 & 0.10 & 0.11 & 0.05 \\
3 & 3 & 0.41 & 0.62 & 0.66 & 0.13 & 0.02 & -0.00 &  0.86 & 0.82 & 0.90 & 0.06 & 0.08 & 0.05 \\
3 & 4 & 0.46 & 0.60 & 0.48 & 0.03 & -0.06 & -0.07 &  0.86 & 0.82 & 0.88 & 0.01 & 0.05 & 0.02 \\
3 & 5 & 0.39 & 0.53 & 0.42 & 0.08 & -0.01 & -0.02 &  0.79 & 0.76 & 0.81 & -0.04 & 0.02 & 0.01 \\
4 & 1 & -0.24 & -0.27 & 0.31 & 0.26 & 0.24 & 0.06 &  0.53 & 0.50 & 0.79 & 0.12 & 0.09 & 0.01 \\
4 & 2 & -0.24 & 0.15 & 0.58 & 0.14 & 0.05 & -0.04 &  0.74 & 0.78 & 0.85 & 0.07 & 0.04 & -0.00 \\
4 & 3 & 0.02 & 0.51 & 0.68 & 0.11 & -0.02 & -0.06 &  0.80 & 0.84 & 0.83 & 0.05 & 0.02 & 0.00 \\
4 & 4 & 0.19 & 0.53 & 0.51 & 0.11 & -0.01 & -0.04 &  0.83 & 0.81 & 0.85 & 0.02 & 0.03 & 0.01 \\
4 & 5 & 0.17 & 0.47 & 0.51 & 0.14 & 0.02 & -0.01 &  0.73 & 0.77 & 0.79 & -0.05 & -0.02 & -0.01 \\
5 & 1 & -0.58 & -0.88 & 0.08 & 0.13 & 0.17 & -0.08 &  0.28 & 0.29 & 0.44 & 0.01 & -0.09 & -0.06 \\
5 & 2 & -0.41 & -0.12 & 0.42 & 0.14 & 0.06 & -0.06 &  0.54 & 0.53 & 0.58 & 0.00 & -0.08 & -0.05 \\
5 & 3 & -0.28 & 0.23 & 0.53 & 0.16 & 0.03 & -0.03 &  0.51 & 0.56 & 0.57 & -0.01 & -0.04 & -0.04 \\
5 & 4 & -0.06 & 0.31 & 0.44 & 0.12 & -0.00 & -0.05 &  0.54 & 0.60 & 0.67 & -0.01 & -0.00 & -0.02 \\
5 & 5 & -0.01 & 0.27 & 0.36 & 0.29 & 0.16 & 0.11 &  0.37 & 0.52 & 0.56 & -0.04 & -0.03 & -0.03 \\
\midrule
 & & \multicolumn{3}{c|}{Explained Variation} & \multicolumn{3}{c||}{Cross-Sectional $R^2$}  & \multicolumn{3}{c|}{Explained Variation} & \multicolumn{3}{c}{Cross-Sectional $R^2$}\\
\multicolumn{2}{c}{All} & 0.20 & 0.26 & 0.53 & 0.50 & 0.77 & 0.92  & 0.67 & 0.60 & 0.76 & 0.94 & 0.91 & 0.98 \\
\bottomrule
\end{tabular}
\bnotetab{ This table shows the out-of-sample explained variation and pricing errors for double sorted portfolios based on short-term reversal (\texttt{ST\_REV}) and momentum (\texttt{r12\_2}) respectively size (\texttt{LME}) and book-to-market ratio (\texttt{BEME}).}
}
\end{table}

\clearpage

\section{Variable Importance}

\begin{figure}[H]
\tcapfig{Macroeconomic Variable Importance for GAN SDF}\label{fig:VI-GAN-Macro}
\centering
\includegraphics[width=0.9\linewidth]{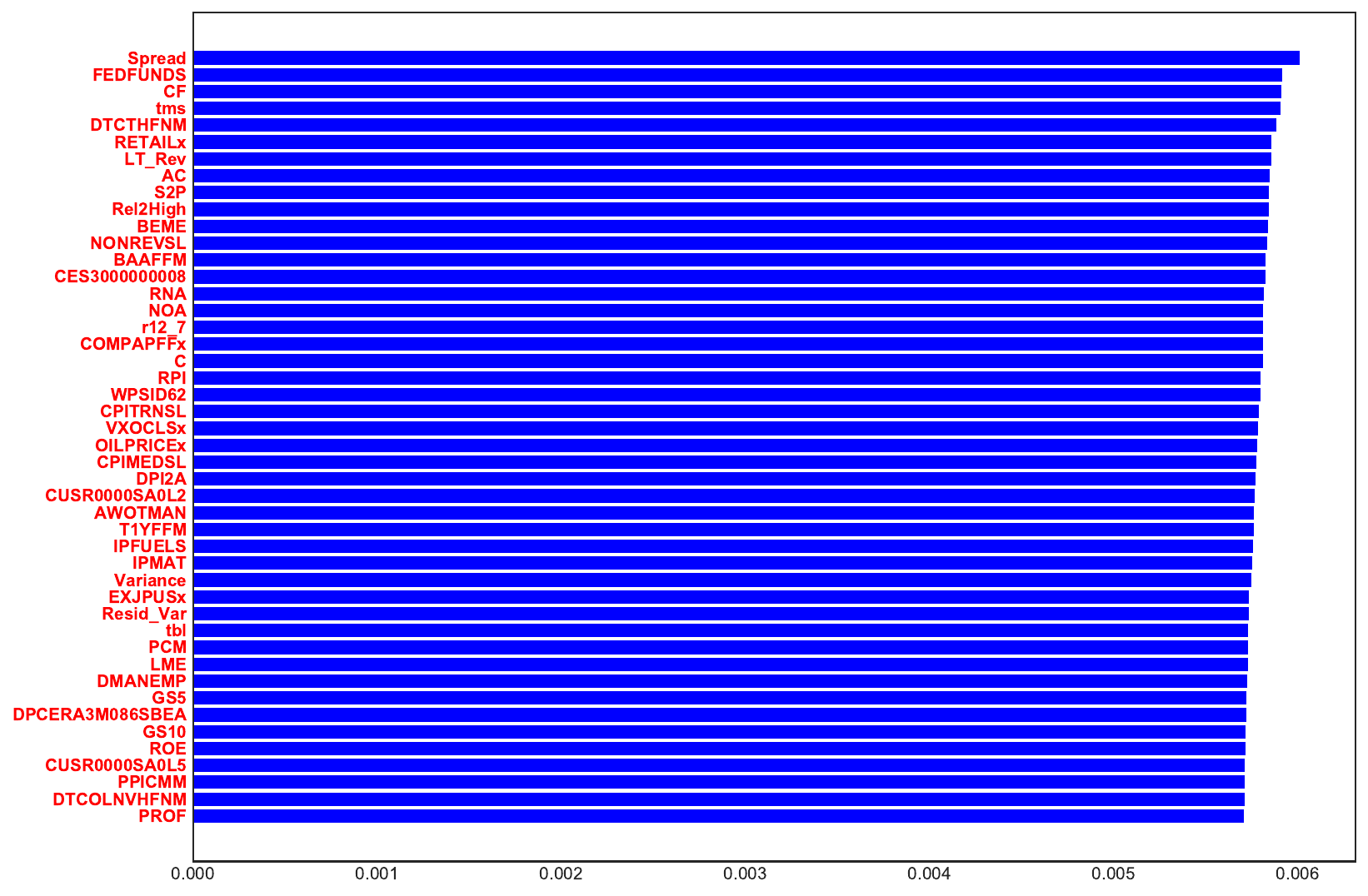}
\bnotefig{This figures shows the GAN variable importance ranking of the 178 macroeconomic variables in terms of average absolute gradient on the test data. The values are normalized to sum up to one.}
\end{figure}

\begin{figure}[H]
\tcapfig{Characteristic Importance for EN}\label{fig:VI-EN}
\centering
\includegraphics[width=0.75\linewidth]{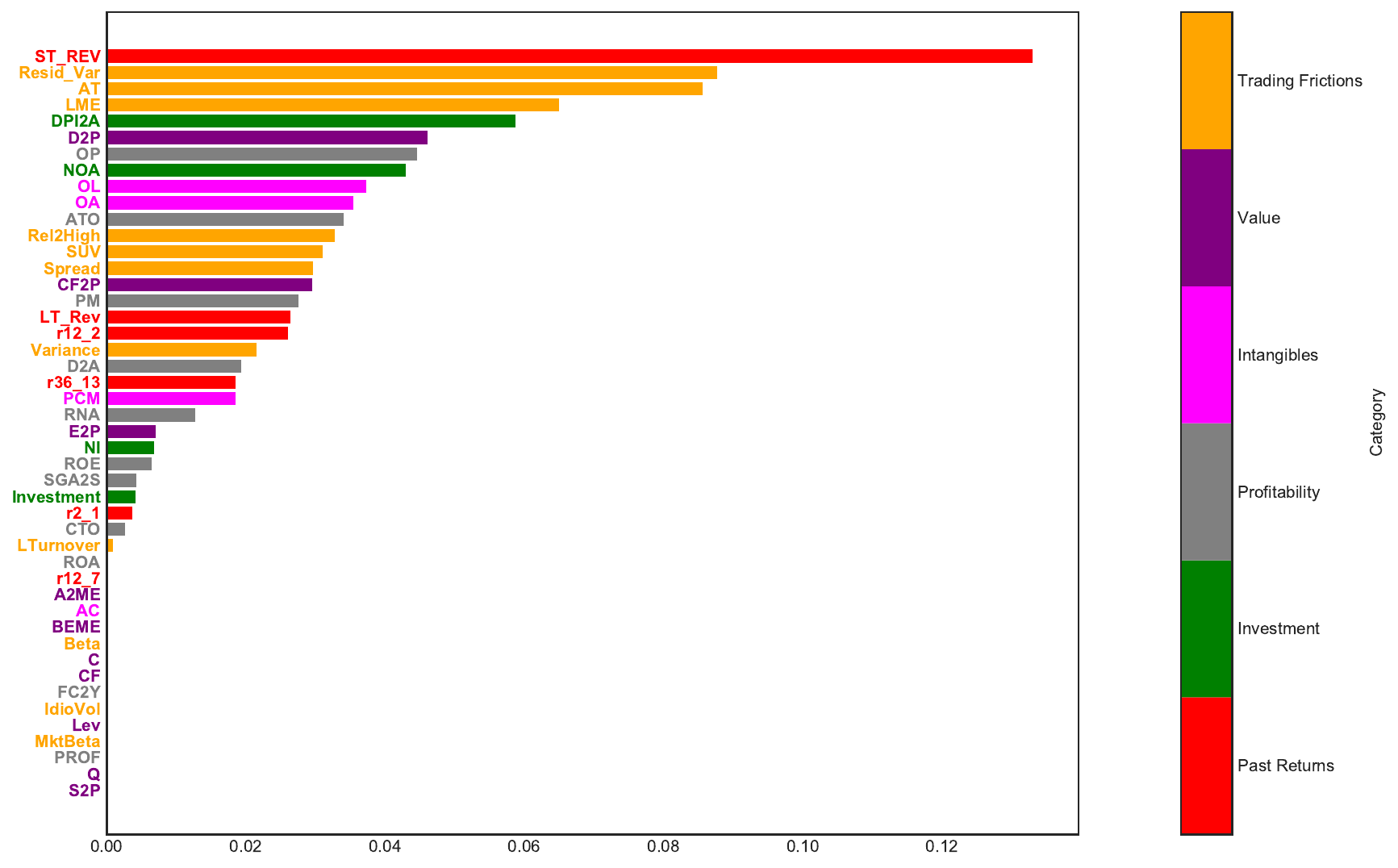}
\bnotefig{The figure shows the Elastic Net variable importance ranking of the 46 firm-specific characteristics in terms of average absolute gradient on the test data. The values are normalized to sum up to one. \par}
\end{figure}

\begin{figure}[H]
\tcapfig{Characteristic Importance for LS}\label{fig:VI-LS}
\centering
\includegraphics[width=0.75\linewidth]{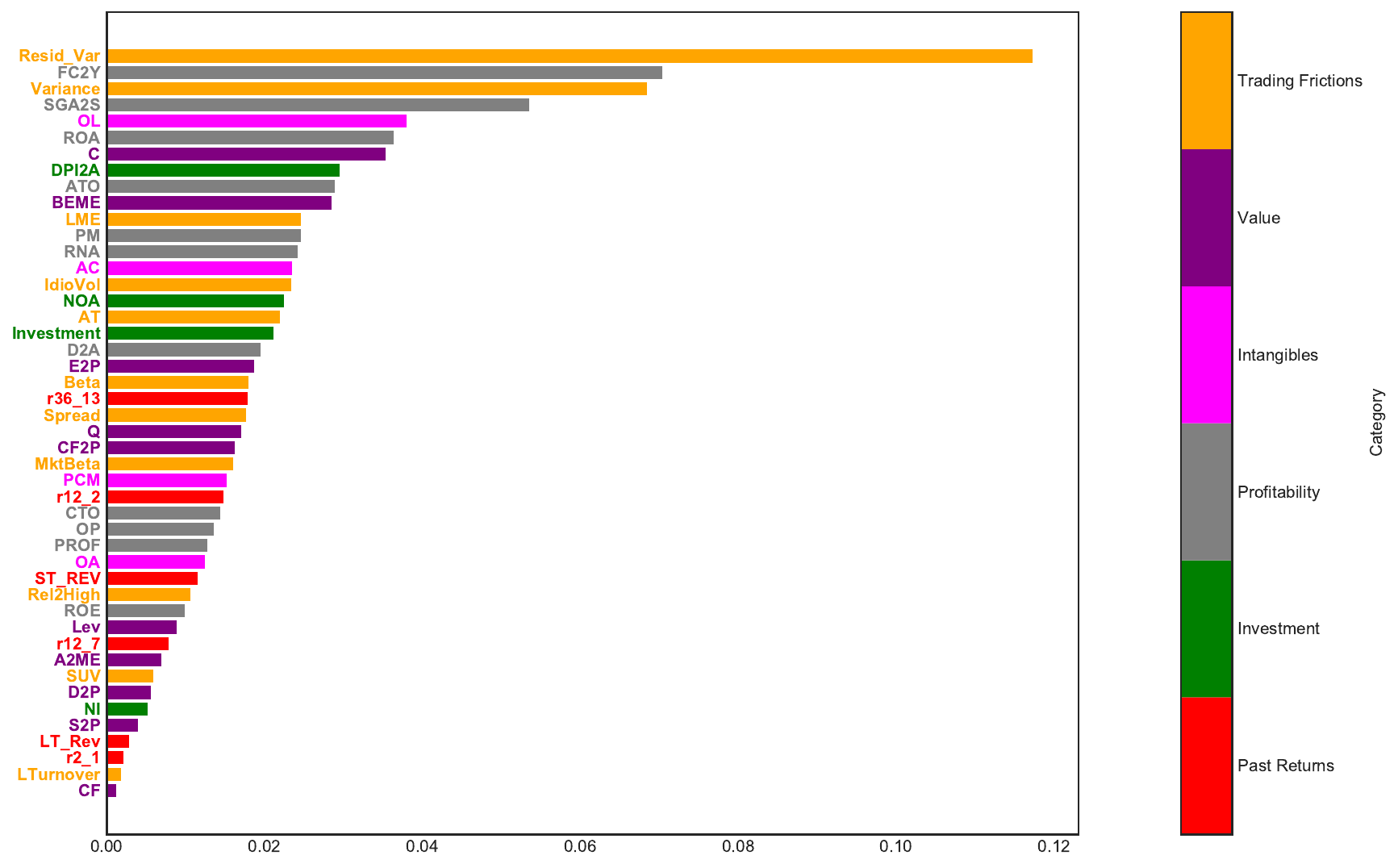}
\bnotefig{The figure shows the LS variable importance ranking of the 46 firm-specific characteristics in terms of average absolute gradient on the test data. The values are normalized to sum up to one. \par}
\end{figure}

\begin{figure}[H]
\tcapfig{Characteristic Importance for Conditioning Function $g$ for GAN}\label{fig:VI-GAN-g}
\centering
\includegraphics[width=0.75\linewidth]{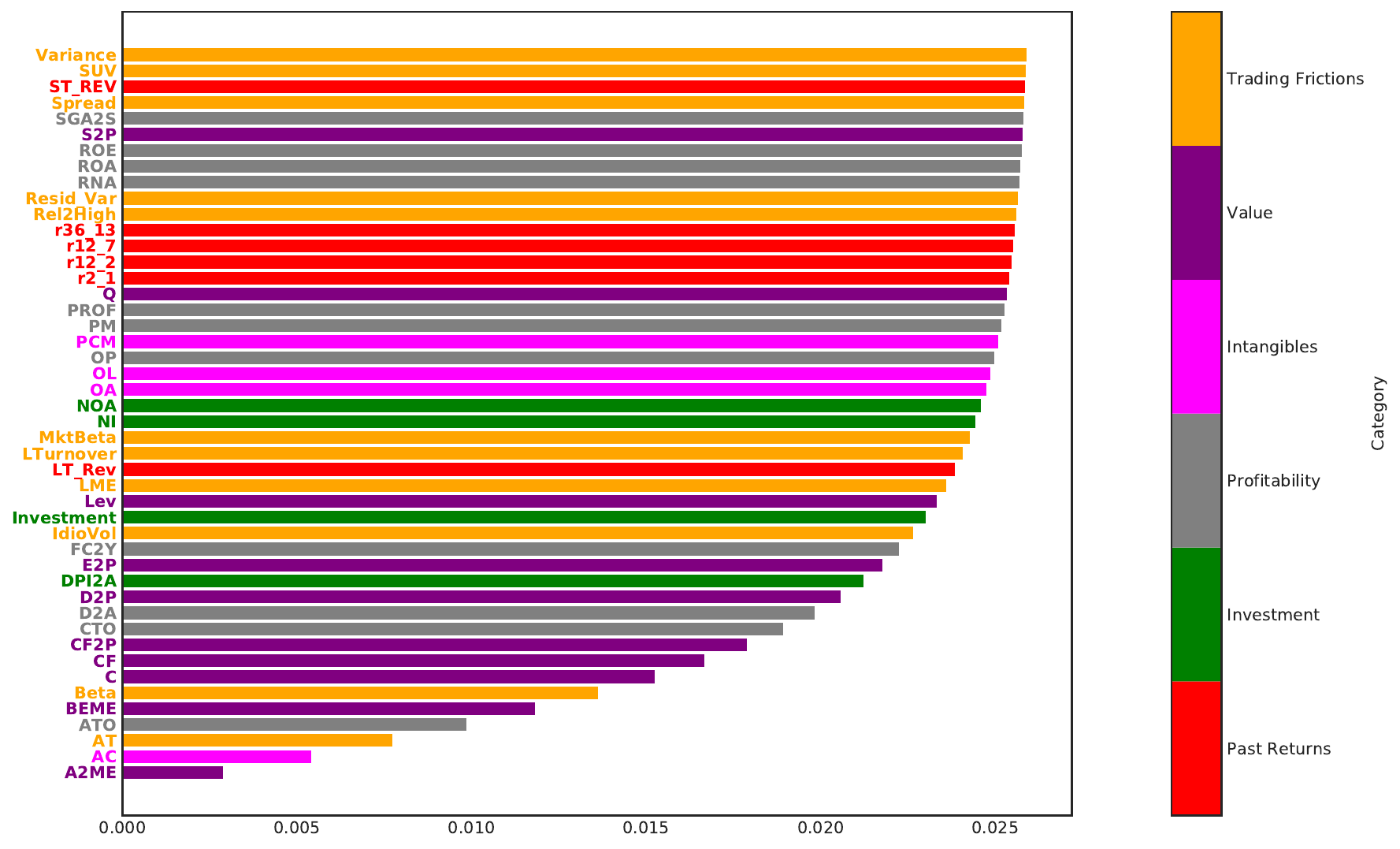}
\bnotefig{The figure shows the GAN variable importance ranking for the conditioning function $g$ of the 46 firm-specific characteristics in terms of average absolute gradient on the test data. The ranking is the average of the absolute gradients for the nine ensemble fits. The values are normalized to sum up to one. }
\end{figure}

\begin{figure}[H]
\tcapfig{Characteristic Importance for Conditioning Function $g$ for GAN No Frict}\label{fig:VI-GAN-NF}
\centering
\includegraphics[width=0.75\linewidth]{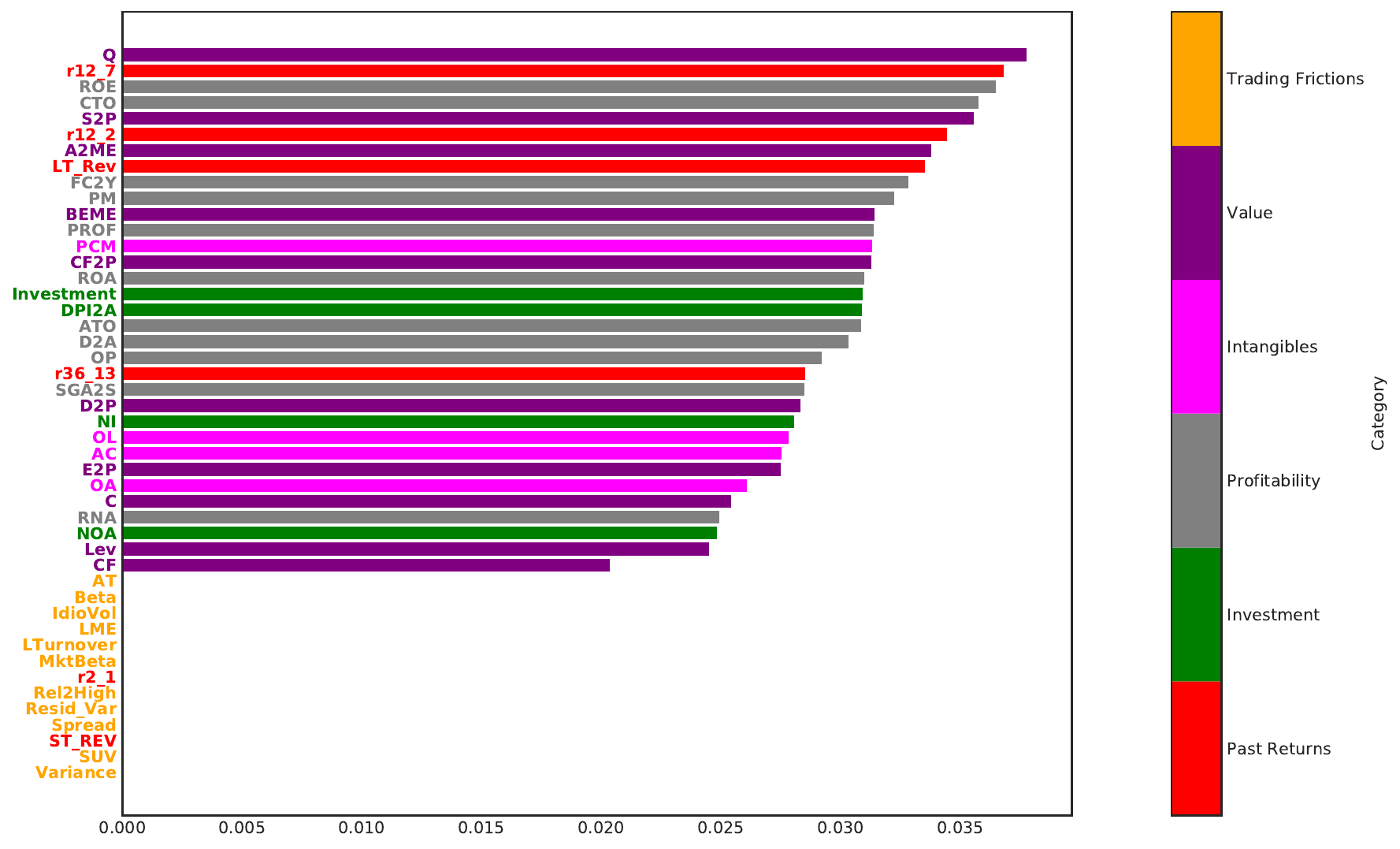}
\bnotefig{The figure shows the GAN variable importance ranking for the conditioning function $g$ of the 46 firm-specific characteristics in terms of average absolute gradient on the test data. The ranking is the average of the absolute gradients for the nine ensemble fits. This GAN model excludes trading frictions and past returns in the construction of the conditioning function $g$. The values are normalized to sum up to one. }
\end{figure}

\section{SDF Structure}

\begin{table}[H]
\tcaptab{Correlation of GAN SDF with Fama-French 5 Factors and Time-Series Regression}\label{tab:corr}
{\small
\centering
  \begin{tabular}{l|c|c|c|c|c|c}
\toprule
& Mkt-RF & SMB & HML & RMW & CMA & intercept \\
\midrule
Regression Coefficients & 0.00 & 0.00 & -0.04 & 0.08*** & 0.04 & 0.76*** \\
& (0.02) & (0.02) & (0.03) & (0.03) & (0.04) & (0.06) \\
Correlations & -0.10 & -0.09 & 0.01 & 0.17 & 0.05 & - \\
\bottomrule
\end{tabular}
}
\bnotetab{This table shows out-of-sample correlations and the time-series regression of the GAN SDF on the Fama-French 5 factors. Standard errors are in parenthesis. 
The regression intercept is the monthly time series pricing error of the SDF portfolio. 
}
\end{table}

\begin{figure}[H]
\centering
\tcapfig{SDF weight $\omega$ as a Function of Characteristics for Different Models}\label{fig:SDF_weight_1D}
\begin{minipage}{\textwidth}
\begin{center}
GAN\\
\includegraphics[width=0.3\textwidth]{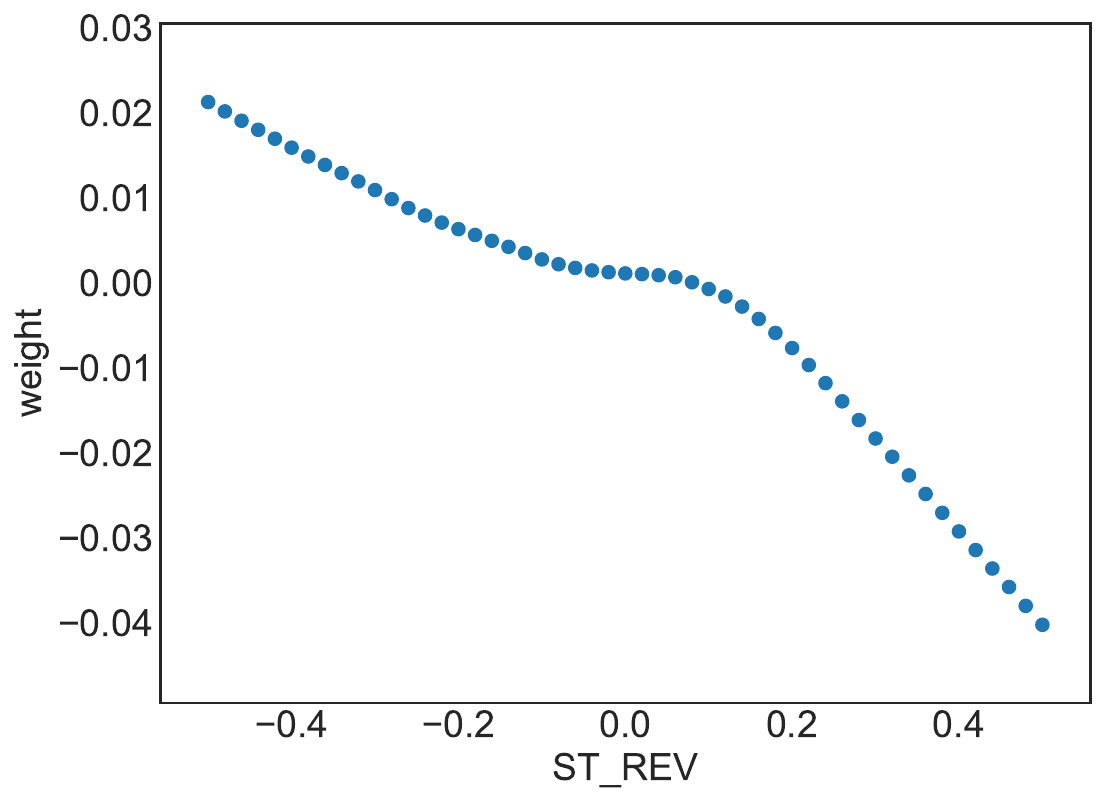}
\includegraphics[width=0.3\textwidth]{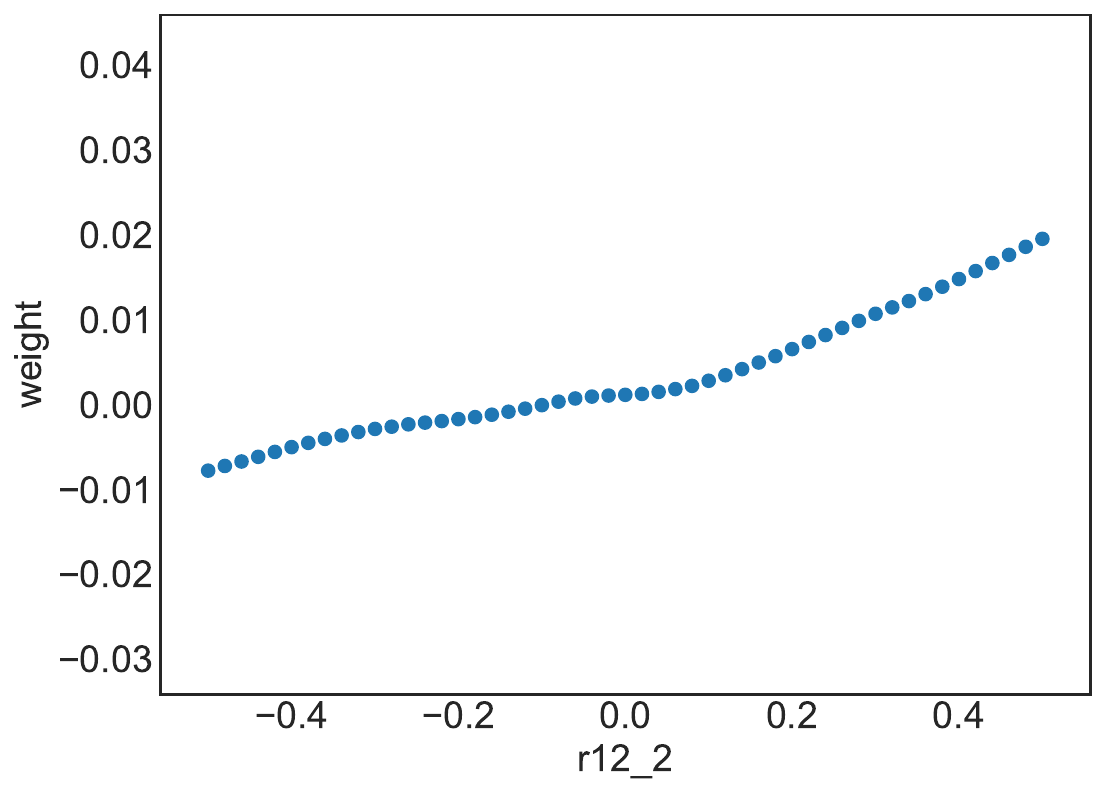}
\includegraphics[width=0.3\textwidth]{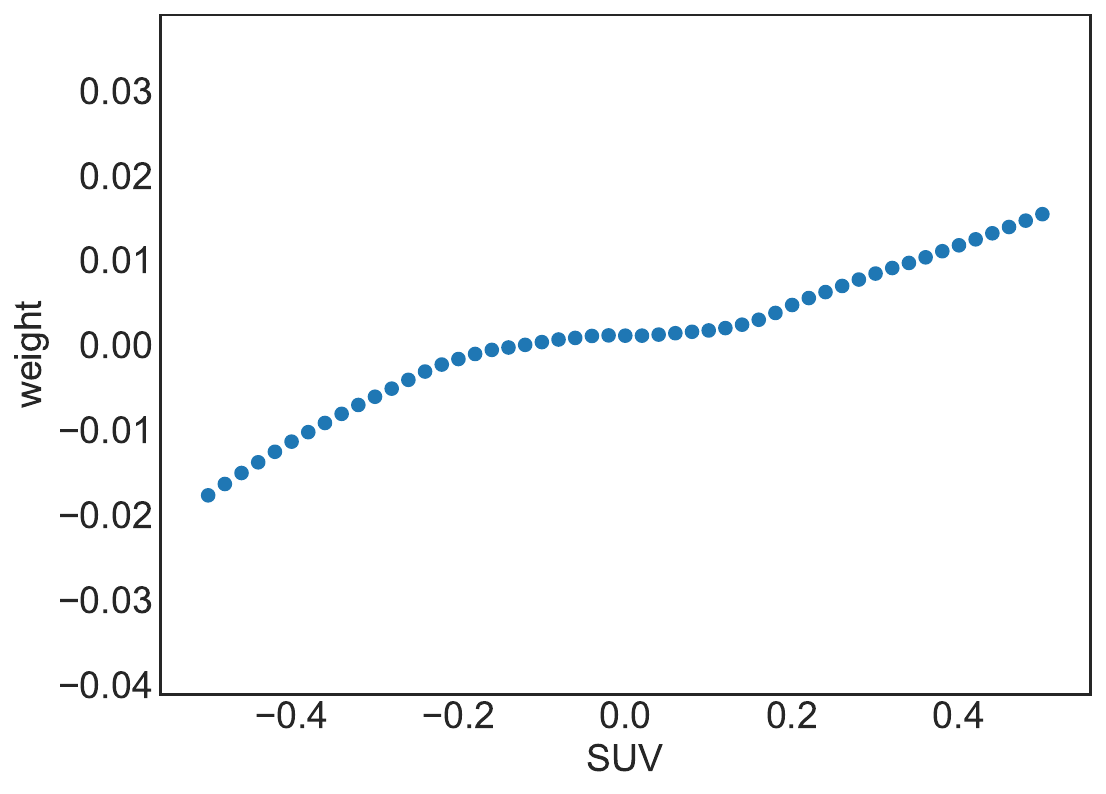}\\
FFN\\
\includegraphics[width=0.3\textwidth]{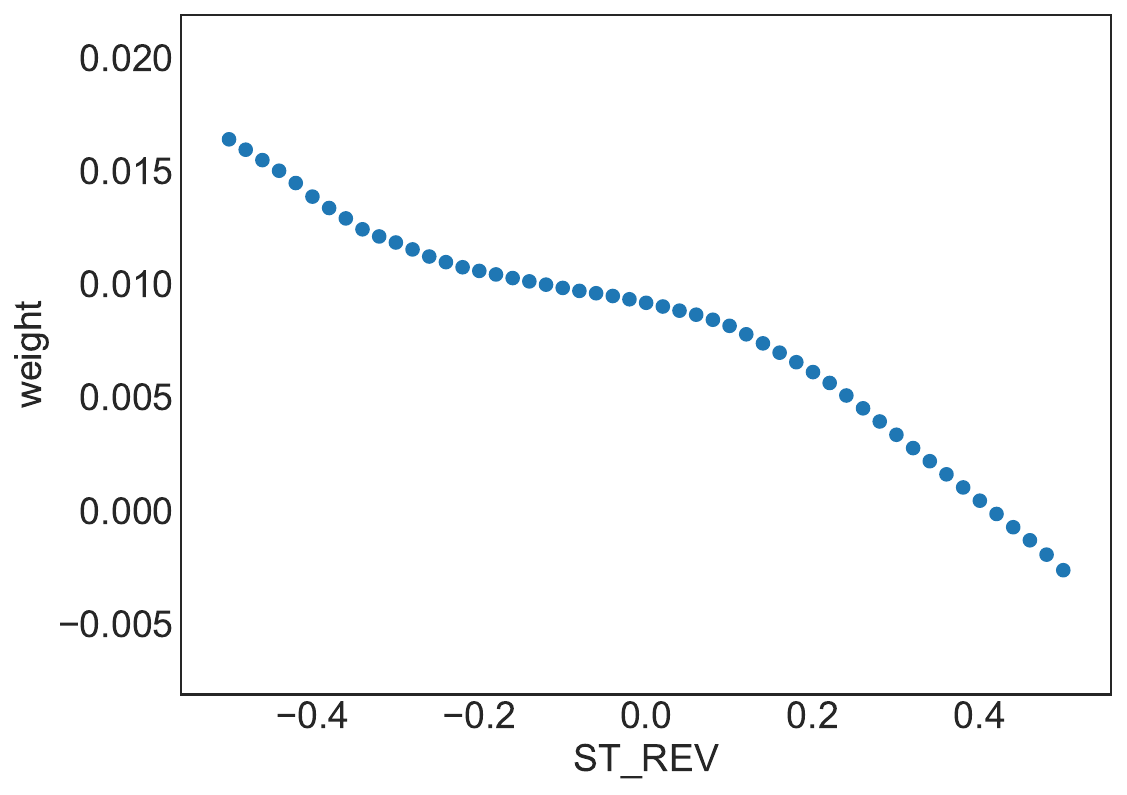}
\includegraphics[width=0.3\textwidth]{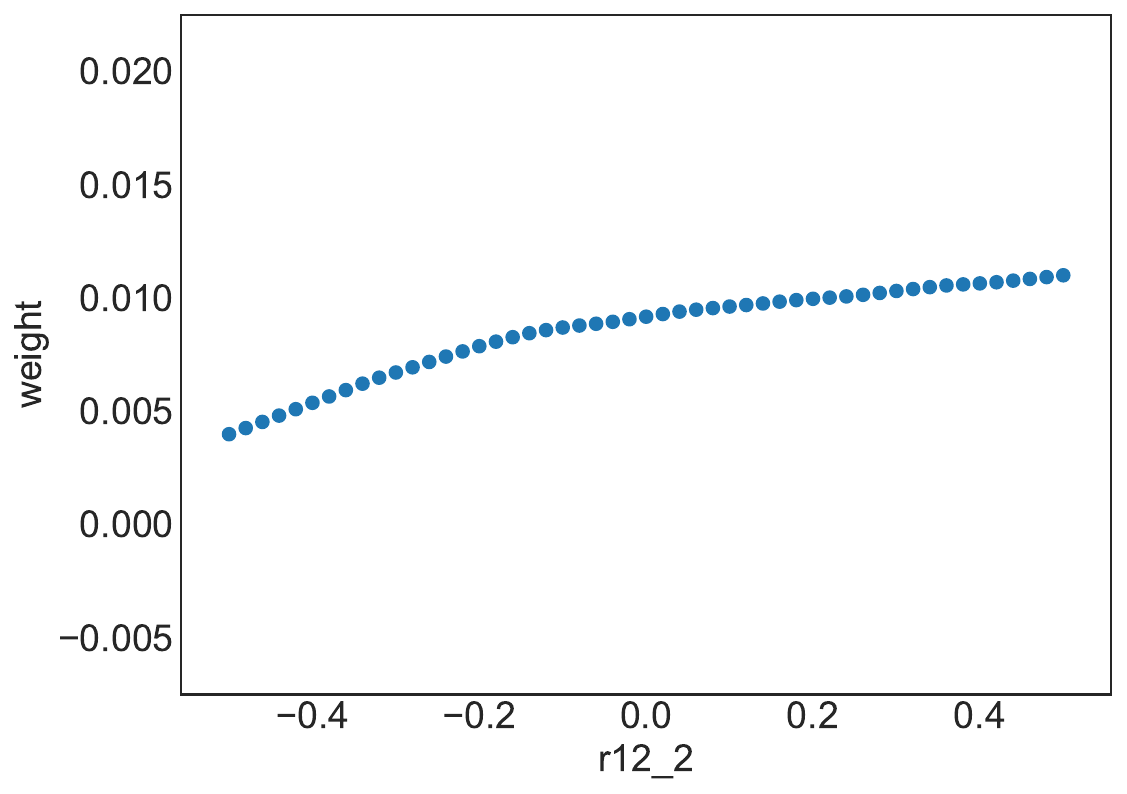}
\includegraphics[width=0.3\textwidth]{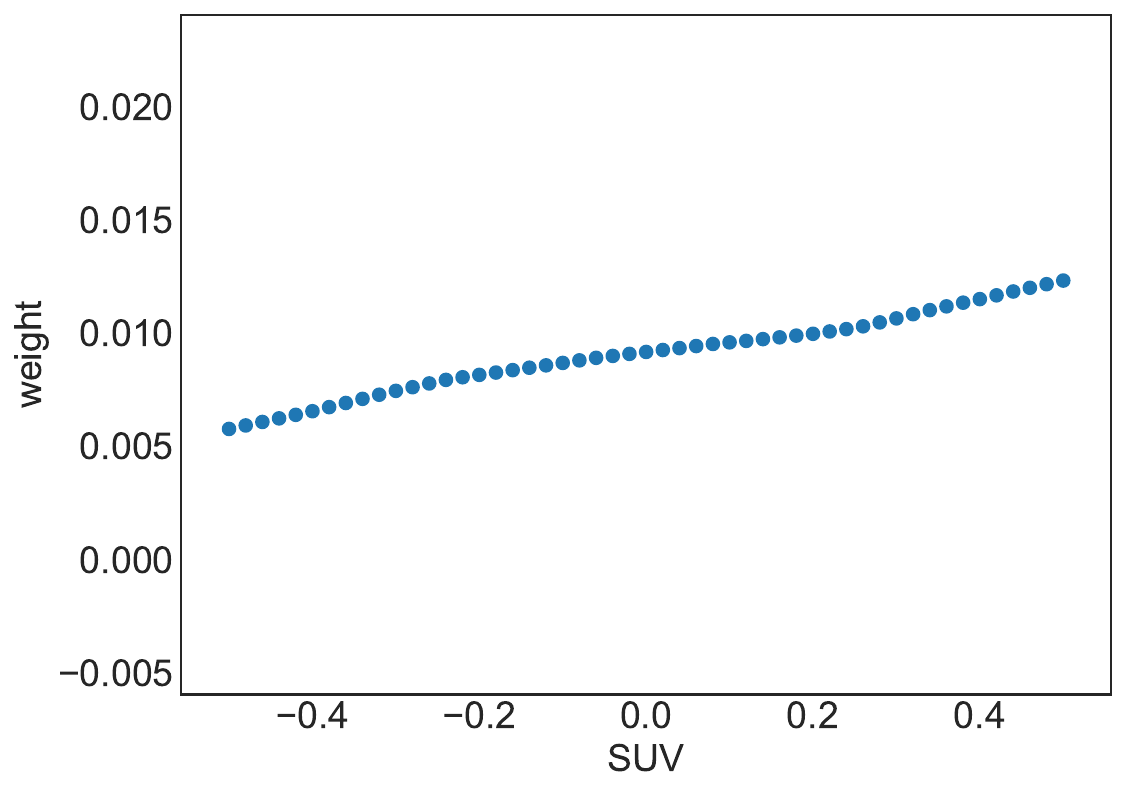}\\
EN\\
\includegraphics[width=0.3\textwidth]{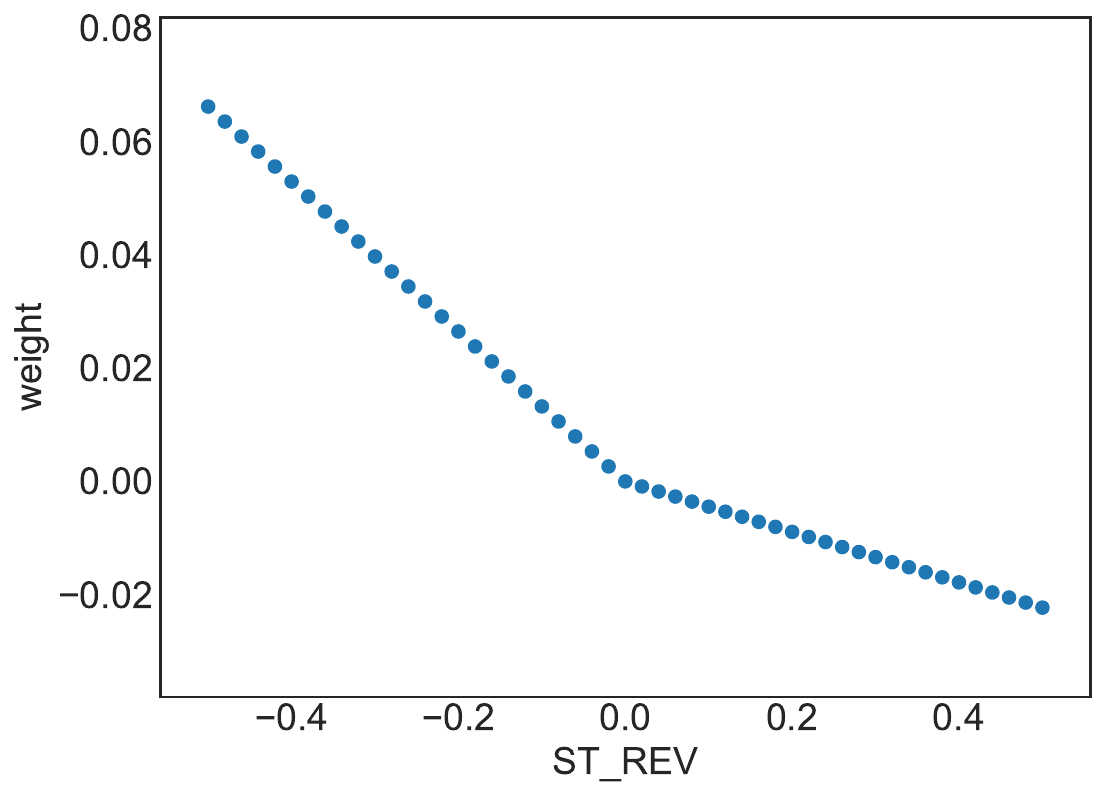}
\includegraphics[width=0.3\textwidth]{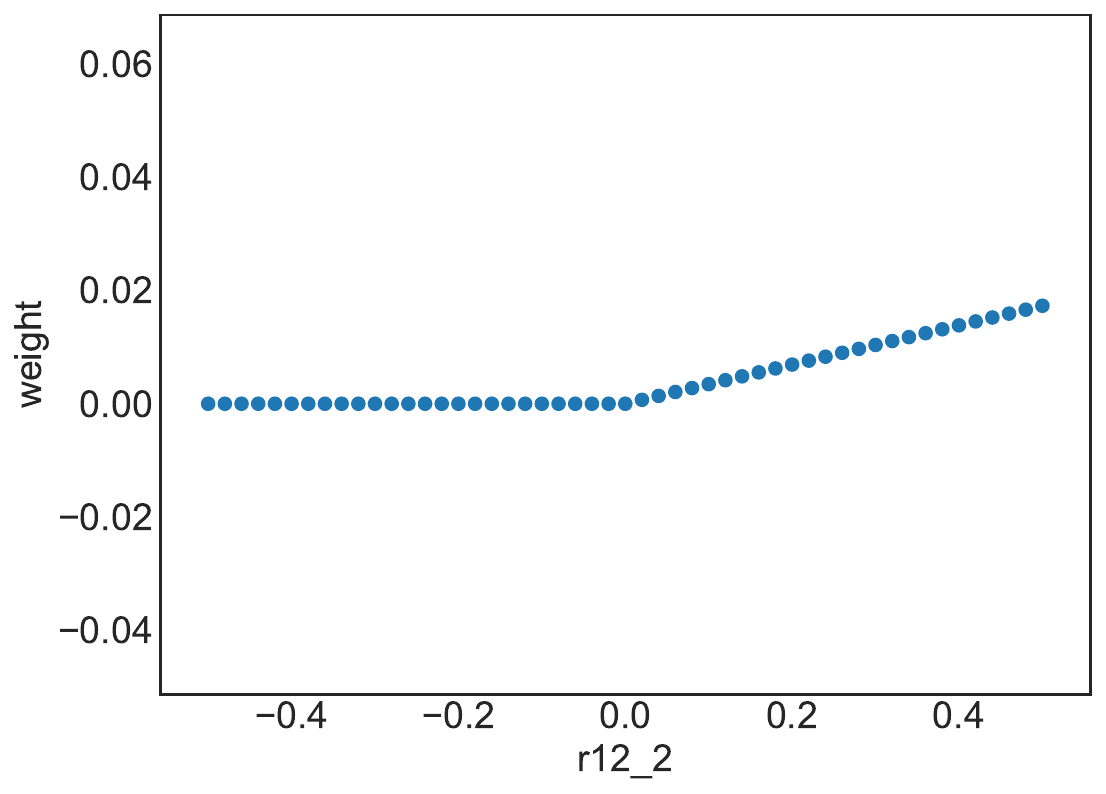}
\includegraphics[width=0.3\textwidth]{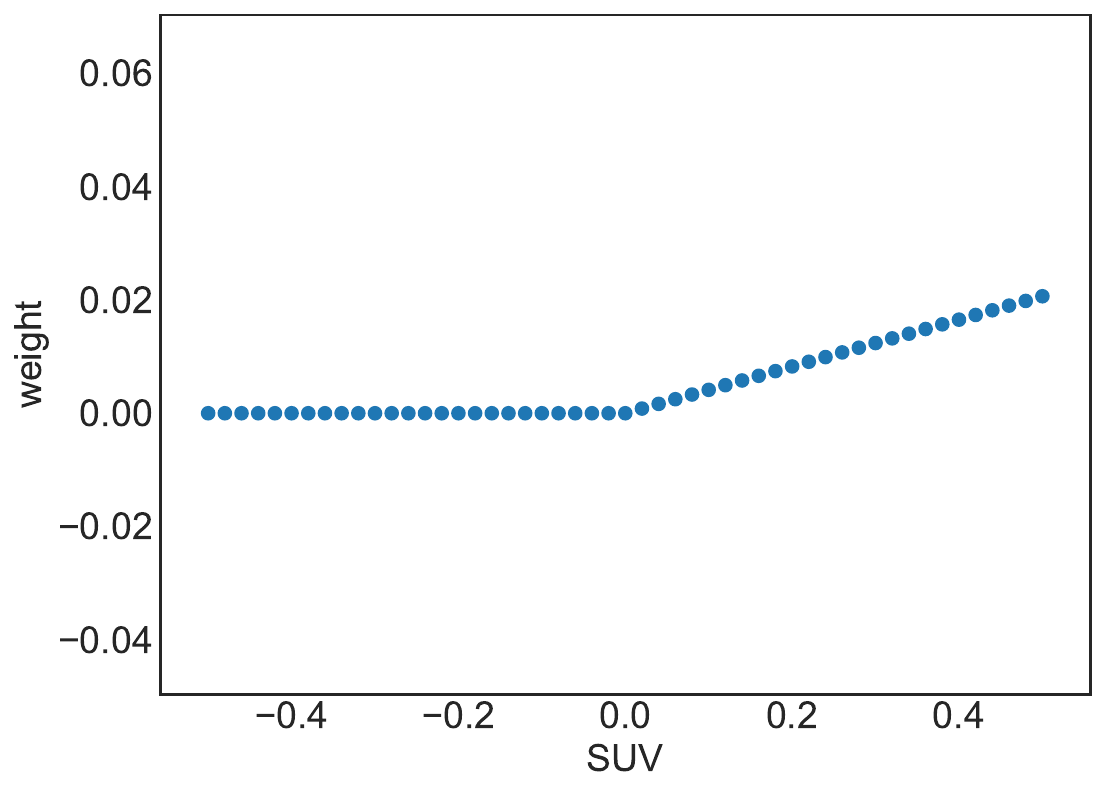}\\
LS\\
\includegraphics[width=0.3\textwidth]{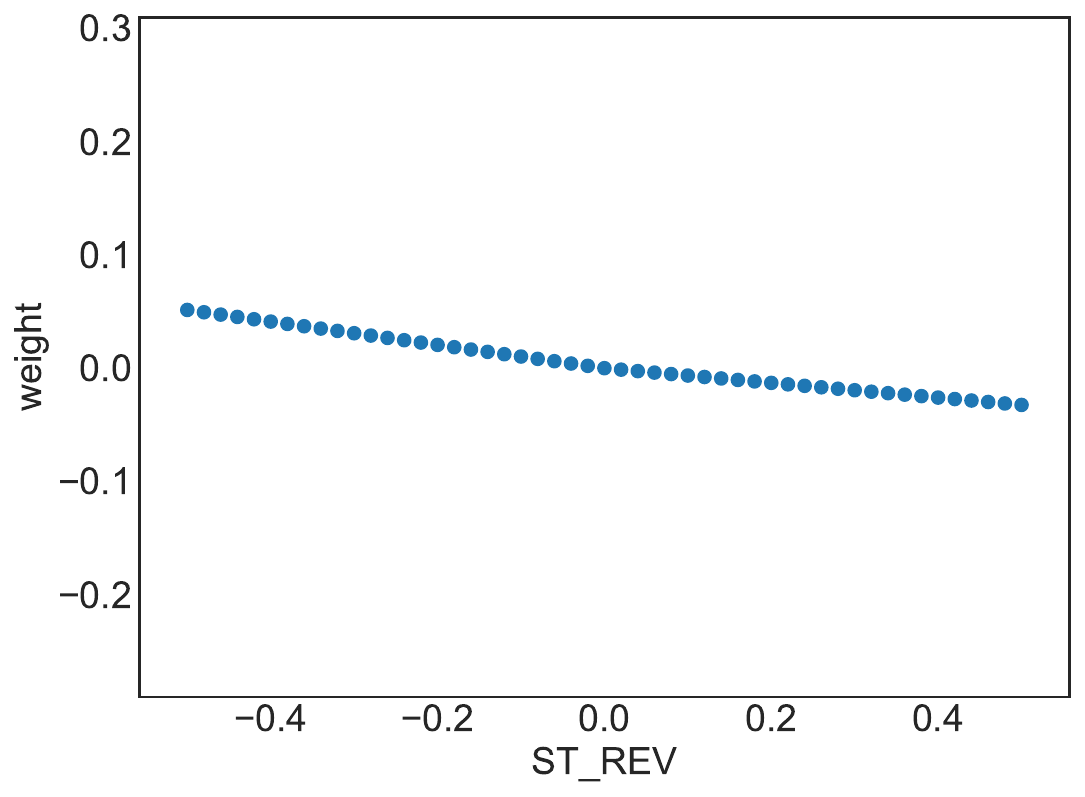}
\includegraphics[width=0.3\textwidth]{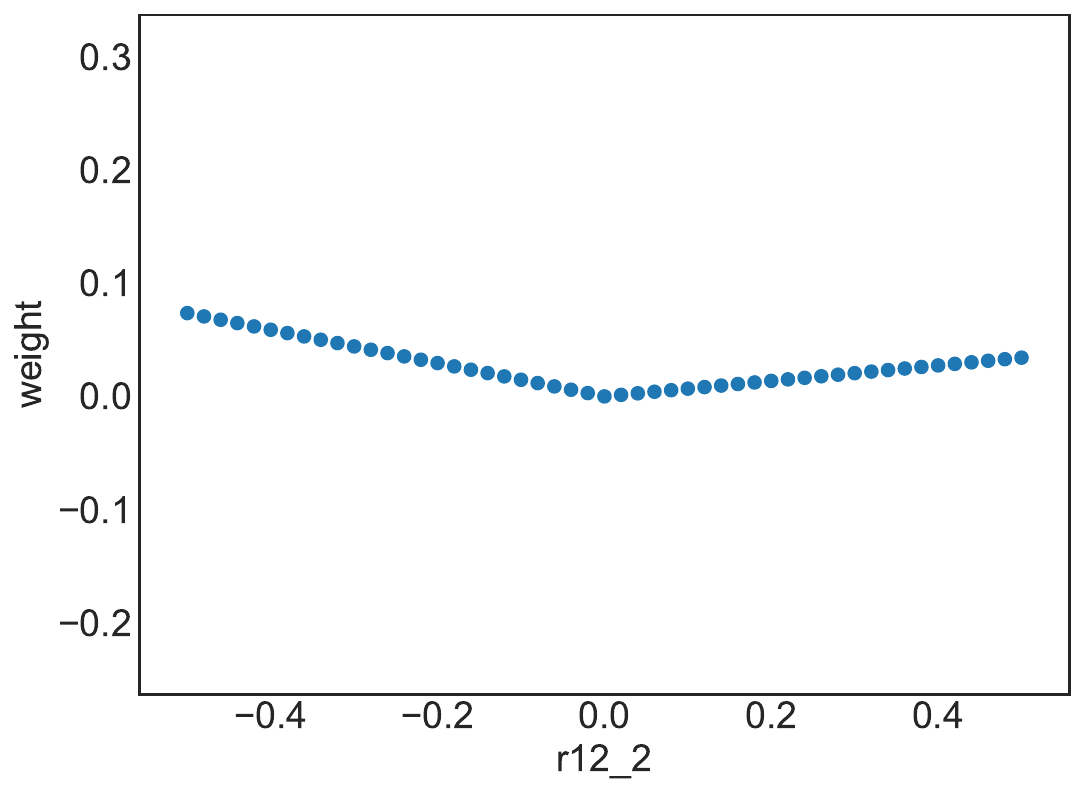}
\includegraphics[width=0.3\textwidth]{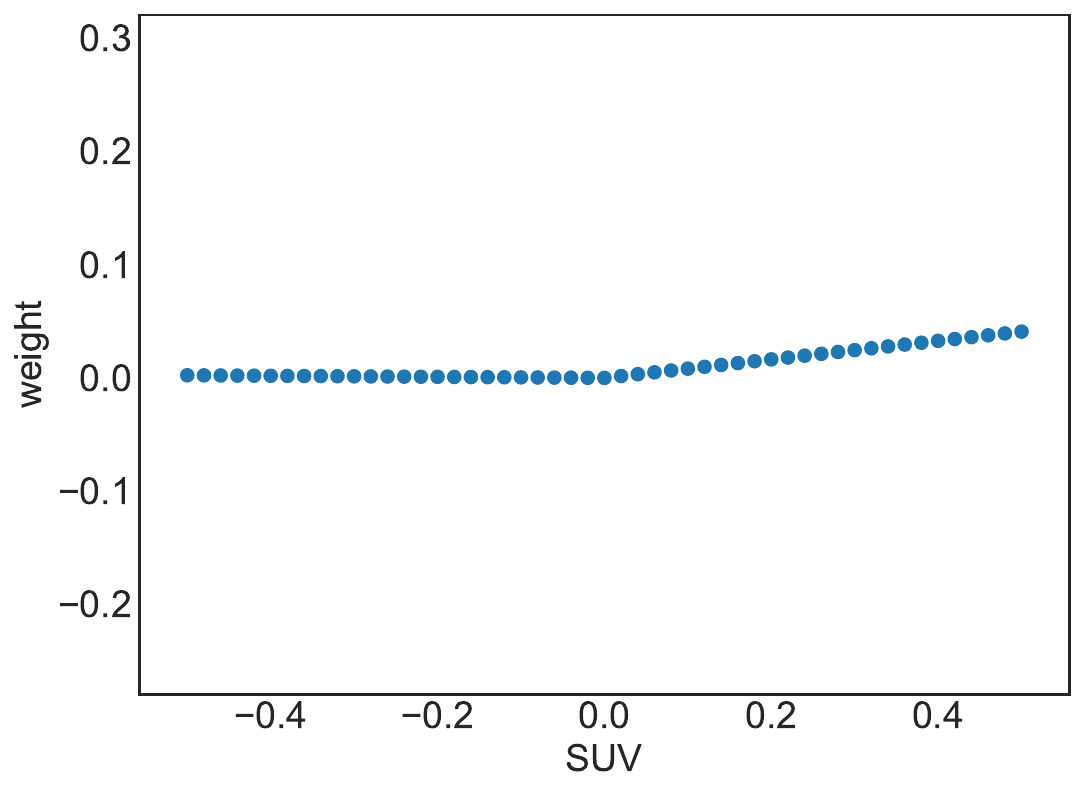}\\
\end{center}
\bnotefig{This figure shows the SDF weight $\omega$ as a one-dimensional function of covariates keeping the other covariates at their mean level. The covariates are Short-Term Reversal (\texttt{ST\_REV}), Momentum (\texttt{r12\_2}) and Standard Unexplained Volume (\texttt{SUV}).}
\end{minipage}
\end{figure}

\section{Machine Learning Investment}

\begin{table}[H]
\centering
\tcaptab{~SDF Portfolio Risk Measures}\label{tab:SDF-Factor}
{\small
\begin{tabular}{cccc|ccc|ccc}
\toprule
& \multicolumn{3}{c}{SR} & \multicolumn{3}{c}{Max Loss} & \multicolumn{3}{c}{Max Drawdown}\\
\cmidrule(l){2-10}
Model & Train & Valid & Test & Train & Valid & Test & Train & Valid & Test \\
\midrule
FF-3 & 0.27 & -0.09 & 0.19 & -2.45 & -2.85 & -4.31 & 7 & 10 & 10 \\
FF-5 & 0.48 & 0.40 & 0.22 & -2.62 & -2.33 & -4.90 & 4 & 3 & 7 \\
LS & 1.80 & 0.58 & 0.42 & -1.96 & -1.87 & -4.99 & 1 & 3 & 4 \\
EN & 1.37 & 1.15 & 0.50 & -2.22 & -1.81 & -6.18 & 1 & 3 & 5 \\
FFN & 0.45 & 0.42 & 0.44 & -3.30 & -4.61 & -3.37 & 6 & 3 & 5 \\
\midrule
GAN & 2.68 & 1.43 & 0.75 & 0.38 & -0.28 & -5.76 & 0 & 1 & 5 \\
\bottomrule
\end{tabular}
}
\bnotetab{This table reports the Sharpe ratio, maximum 1-month loss and maximum drawdown of the SDF portfolios. We include the mean-variance efficient portfolio based on the 3 and 5 Fama-French factors.}
\end{table}

\begin{table}[H]
\centering
\tcaptab{~Turnover by Models}\label{tab:turnover}
{\small
\begin{tabular}{cccc|ccc}
\toprule
& \multicolumn{3}{c}{Long Position} & \multicolumn{3}{c}{Short Position}\\
\cmidrule(l){2-7}
Model & Train & Valid & Test & Train & Valid & Test \\
\midrule
LS & 0.25 & 0.22 & 0.24 & 0.64 & 0.55 & 0.61 \\
EN & 0.36 & 0.35 & 0.35 & 0.83 & 0.83 & 0.84 \\
FFN & 0.69 & 0.63 & 0.65 & 1.38 & 1.29 & 1.27\\
\midrule
GAN & 0.47 & 0.40 & 0.40 & 1.05 & 1.04 & 1.02 \\
\bottomrule
\end{tabular}
}
\bnotetab{This table reports the turnover for positions with positive and negative weights for the SDF portfolios. It is defined as $\frac{1}{T}\sum_{t = 1}^T (\sum_i |(1+R_{P, t+1})w_{i, t+1} - (1+R_{i,t+1})w_{i,t}|)$, where $w_{i,t}$ is the portfolio weight of stock $i$ at time $t$, and $R_{P,t+1}=\sum_iR_{i,t+1}w_{i,t}$ is the corresponding portfolio return. Long and short positions are calculated separately, and the portfolio weights are normalized to $\lVert w_t\rVert_1=1$.\par}
\end{table}

\section{Implementation and Robustness Results}

\subsection{Tuning Parameters and Robustness}

\begin{table}[H]
\centering
\tcaptab{~Selection of Hyperparameters for GAN in the Empirical Analysis}
\label{tab:hyper-parameter}
{\small
\begin{tabular}{r|l|c|c}
\toprule
Notation & Hyperparameters & Candidates & Optimal \\
\midrule
HL & Number of layers in SDF Network & 2, 3 or 4 & 2 \\
HU & Number of hidden units in SDF Network & 64 & 64 \\
SMV & Number of hidden states in SDF Network & 4 or 8 & 4 \\
CSMV & Number of hidden states in Conditional Network & 16 or 32 & 32 \\
CHL & Number of layers in Conditional Network & 0 or 1 & 0 \\
CHU & Number of hidden units in Conditional Network & 4, 8, 16 or 32 & 8 \\
LR & Initial learning rate & 0.001, 0.0005, 0.0002 or 0.0001 & 0.001 \\
DR & Dropout & 0.95 &0.95 \\
\bottomrule
\end{tabular}
\bnotetab{ This table shows the optimal tuning parameters of the benchmark GAN. The optimal SDF network has 2 layers each with 64 nodes and uses 4 hidden macroeconomic state processes. The optimal adversarial network is a generalized linear model creating 8 instruments and uses 32 hidden macroeconomic states. The Dropout parameter denotes the probability of keeping a node.}
}
\end{table}

\begin{table}[H]
\centering
\tcaptab{Best Performing GAN Models on the Validation Data}\label{tab:tuning}
{\small
\begin{tabular}{ccccccccccc}
\toprule
Model & SMV & CSMV & HL & CHL & CHU & LR & SR (Train) & SR (Valid) & SR (Test)\\
\midrule
GAN 1 & 4 & 32 & 4 & 0 & 32 & 0.001  & 2.78 & 1.47 & 0.72 \\
GAN 2 & 4 & 32 & 2 & 0 & 8  & 0.001  & 3.02 & 1.39   & 0.77 \\
GAN 3 & 4 & 32 & 4 & 0 & 16 & 0.0005 & 2.55 & 1.38 & 0.74 \\
GAN 4  & 4 & 16 & 3 & 1 & 16 & 0.0005 & 2.44 & 1.38 & 0.77 \\
\bottomrule
\end{tabular}}
\bnotetab{This tables shows a re-estimation of the GAN model that is independent of the benchmark GAN model and reports the network structure. GAN 1 has 4 layers, 32 instruments and 4 hidden states. GAN 2 has 2 layers, 8 instruments and 4 hidden states and the hence the same architecture as our benchmark model. GAN 3 has 4 layers, 16 instruments and 4 hidden states, while GAN 4 has 3 layers, 16 instruments and 4 hidden states.}
\end{table}

\begin{table}[H]
\tcaptab{Correlation of Benchmark GAN SDF with SDF of Alternative GAN Estimations}\label{tab:corralt}
{\small
\sisetup{table-format=-1.2}   
\centering
\begin{tabular}{c|ccccccc}
\toprule
             & GAN    & GAN 1  & GAN 2  & GAN 3  & GAN 4  & GAN Rolling & GAN No Frict \\
\midrule
GAN          & 1      & 0.84   & 0.87   & 0.84   & 0.80   & 0.70        & 0.78     \\
GAN 1        & 0.84   & 1      & 0.88   & 0.92   & 0.89   & 0.79        & 0.89     \\
GAN 2        & 0.87   & 0.88   & 1      & 0.87   & 0.88   & 0.73        & 0.83     \\
GAN 3        & 0.84   & 0.92   & 0.87   & 1      & 0.89   & 0.74        & 0.86     \\
GAN 4        & 0.80   & 0.89   & 0.88   & 0.89   & 1      & 0.78        & 0.84     \\
GAN Rolling  & 0.70   & 0.79   & 0.73   & 0.74   & 0.78   & 1           & 0.78     \\
GAN No Frict & 0.78   & 0.89   & 0.83   & 0.86   & 0.84   & 0.78        & 1        \\
\bottomrule
\end{tabular}
}
\bnotetab{This  table reports the factor correlations for alternative estimations of GAN. GAN is the benchmark model, GAN 1, 2, 3 and 4 are the top four performing models from a re-estimation. GAN Rolling is estimated on a rolling window of 240 months. GAN No Frict is an estimation of the GAN model without trading friction variables and past returns for the conditioning instruments $g$.}
\end{table}

\subsection{GAN Estimation on Rolling Window}

\begin{figure}[H]
\centering
\caption{Characteristic Importance for Rolling Window GAN}\label{fig:VI_window}
\begin{minipage}{\textwidth}
\centering
\includegraphics[width=0.75\linewidth]{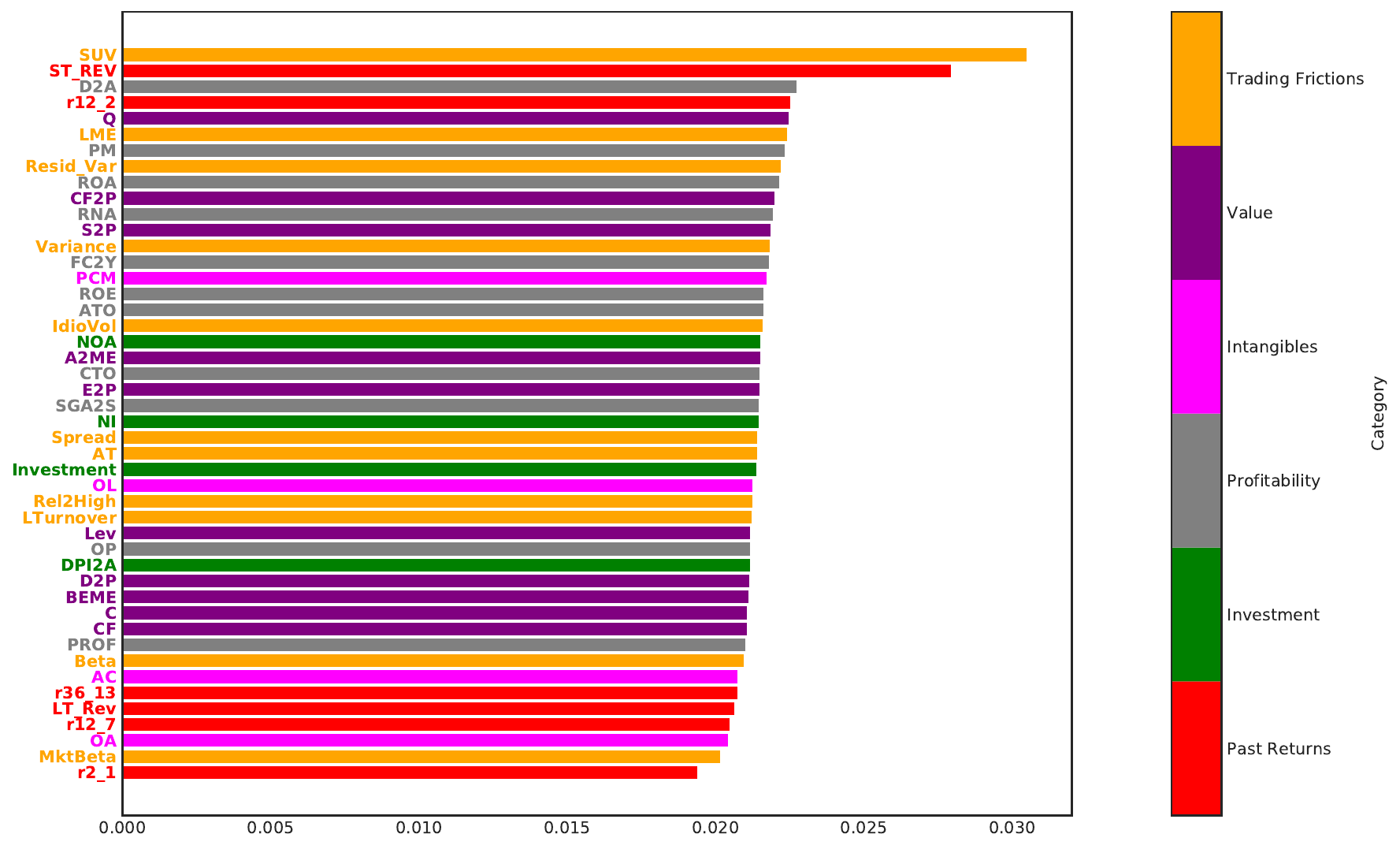}
{\footnotesize  \par}
\end{minipage}
\end{figure}

\begin{figure}[H]
\centering
  \tcapfig{SDF weight $\omega$ as a Function of Covariates for Rolling Window Fit}\label{fig:SDF_window}
\begin{subfigure}[t]{1\textwidth}
\begin{minipage}{\textwidth}
\begin{center}
\includegraphics[width=0.3\textwidth]{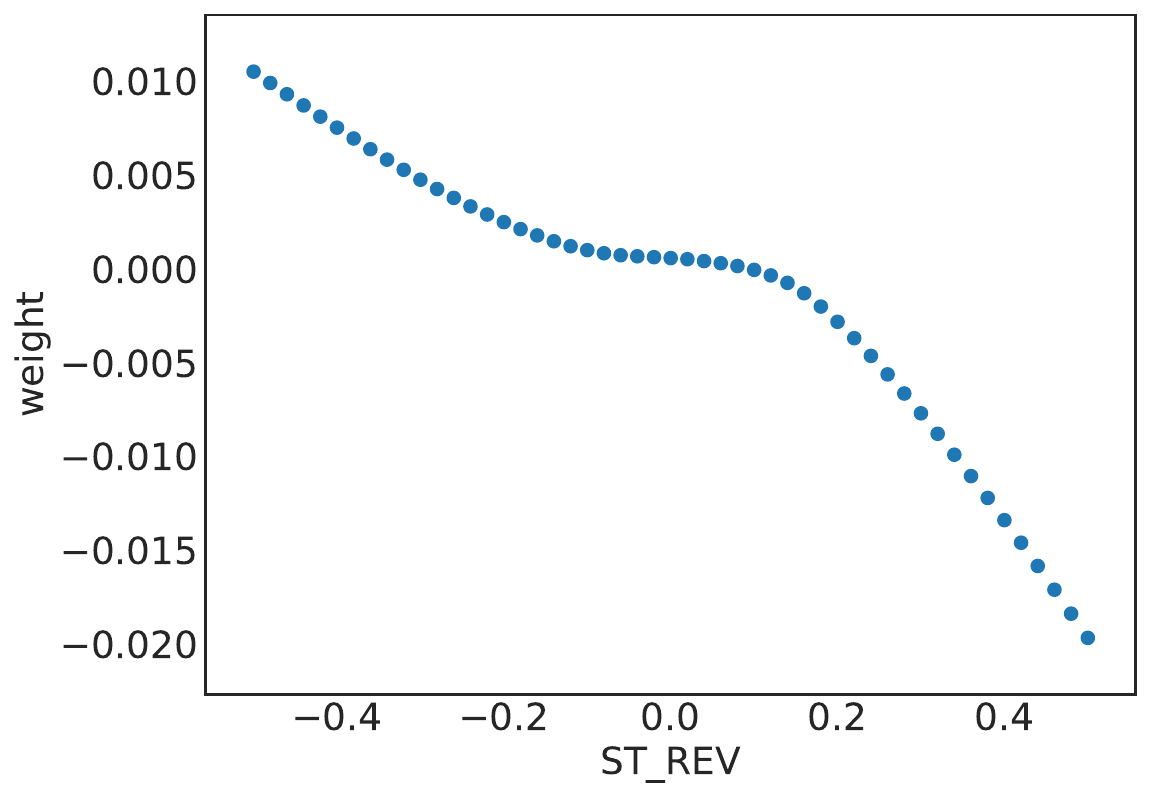}
\includegraphics[width=0.3\textwidth]{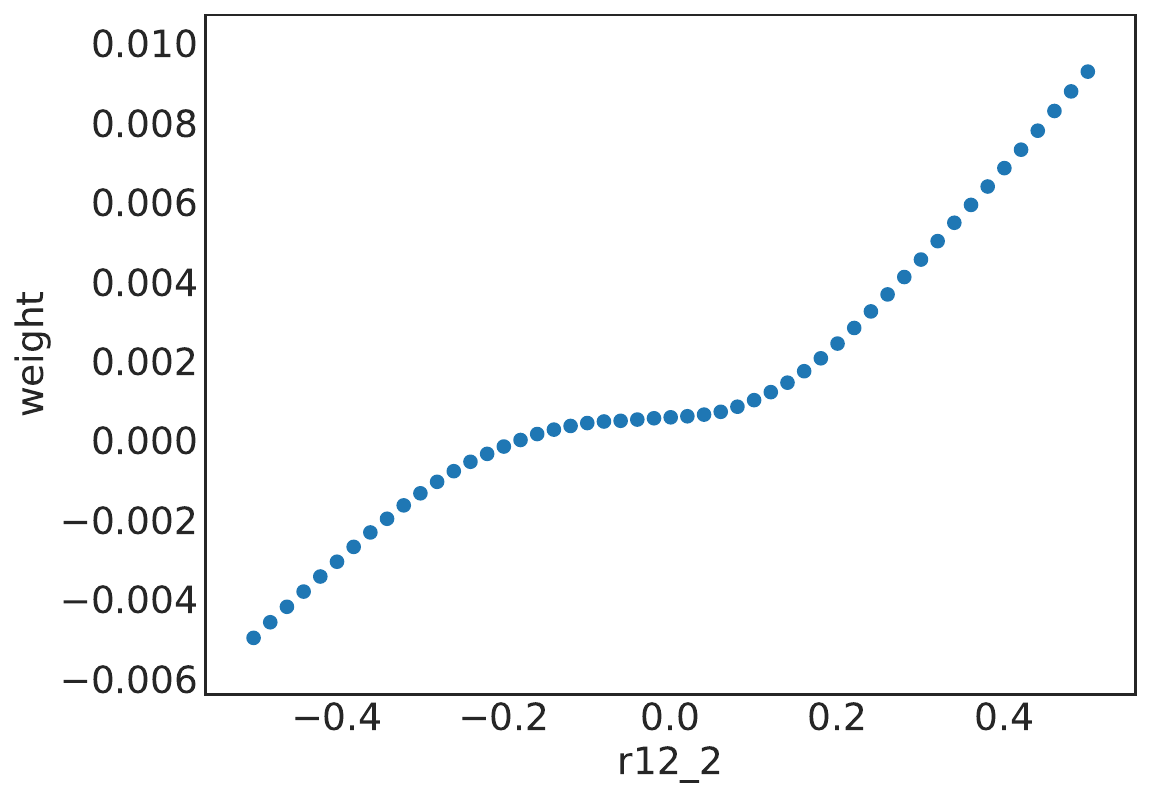}
\includegraphics[width=0.3\textwidth]{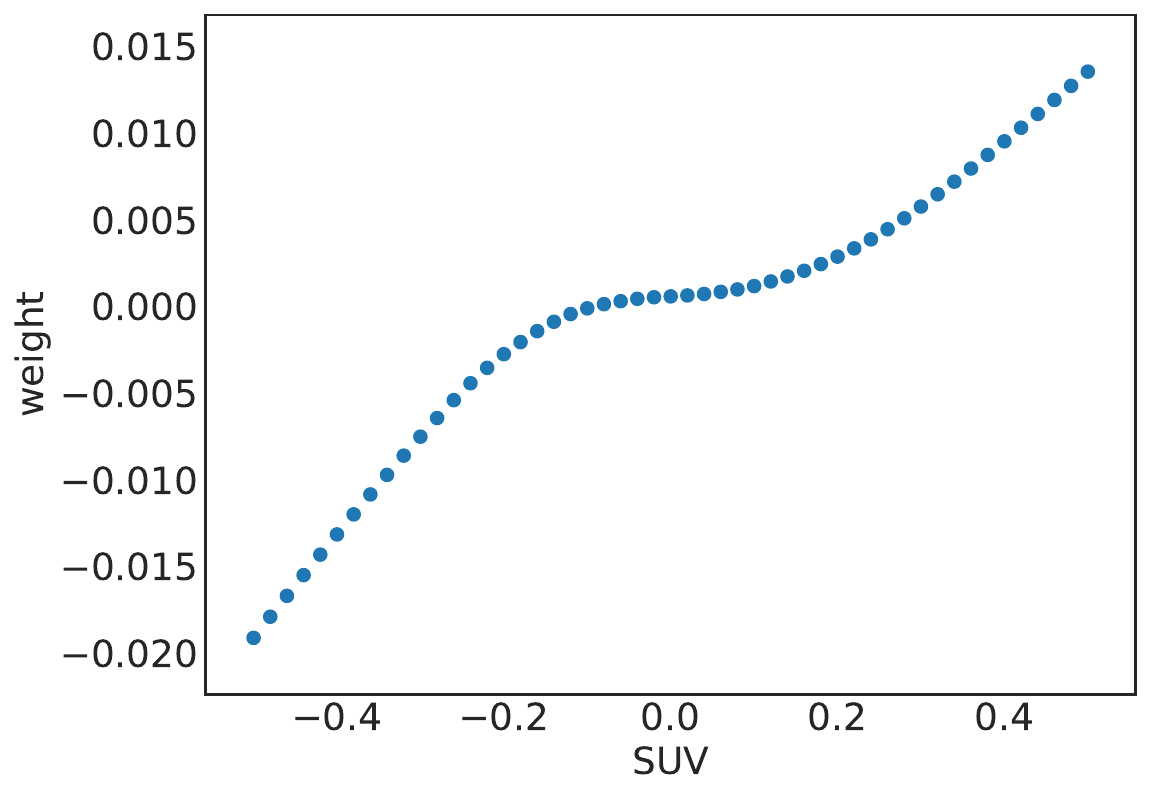}\\
\end{center}
\end{minipage}
\subcaptab{SDF weight $\omega$ as a function of one characteristic}
  \end{subfigure}\hfill

\begin{subfigure}[t]{.45\textwidth}
  \centering
    \includegraphics[width=1\textwidth]{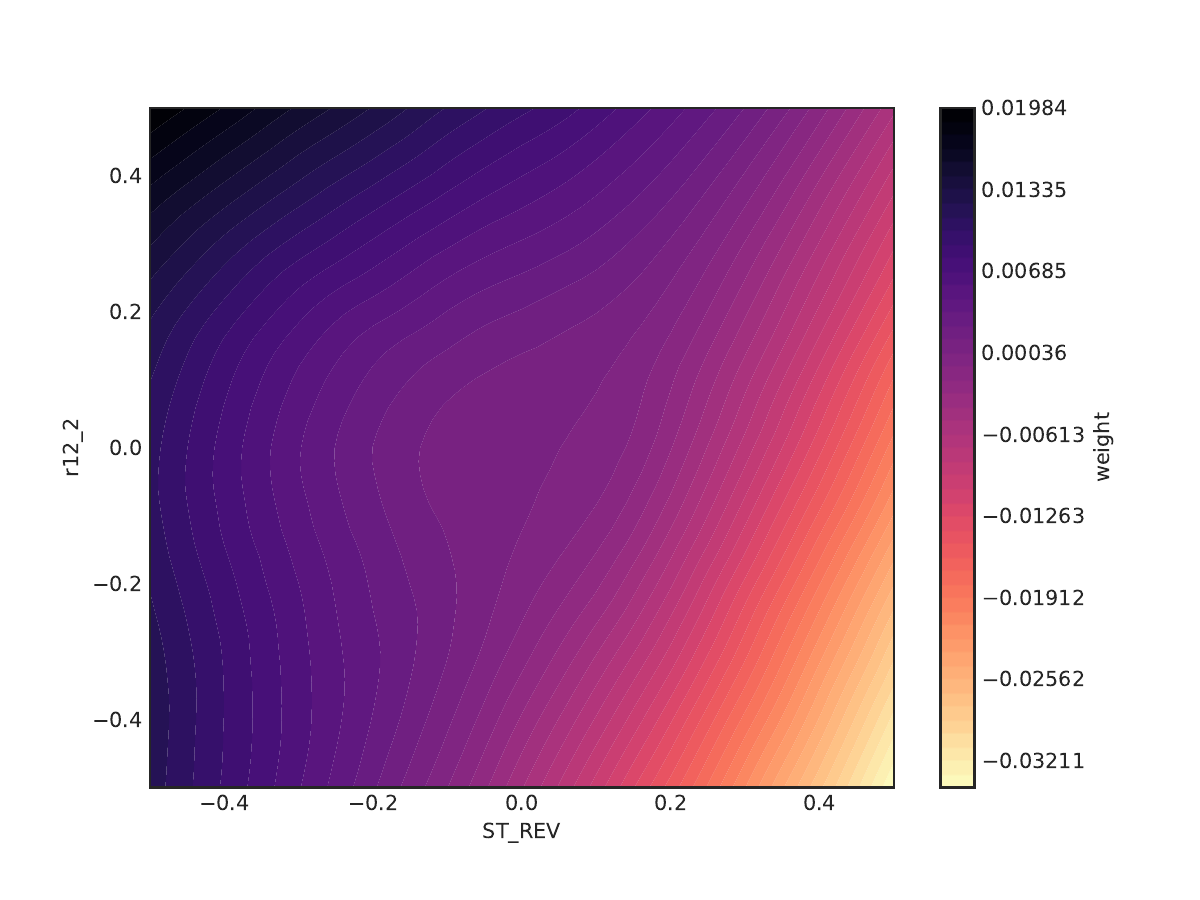}
    \subcap{Interaction between Short-Term Reversal (ST\_REV) and Momentum (r12\_2)}
  \end{subfigure}\hfill
  \begin{subfigure}[t]{.45\textwidth}
  \centering
    \includegraphics[width=1\textwidth]{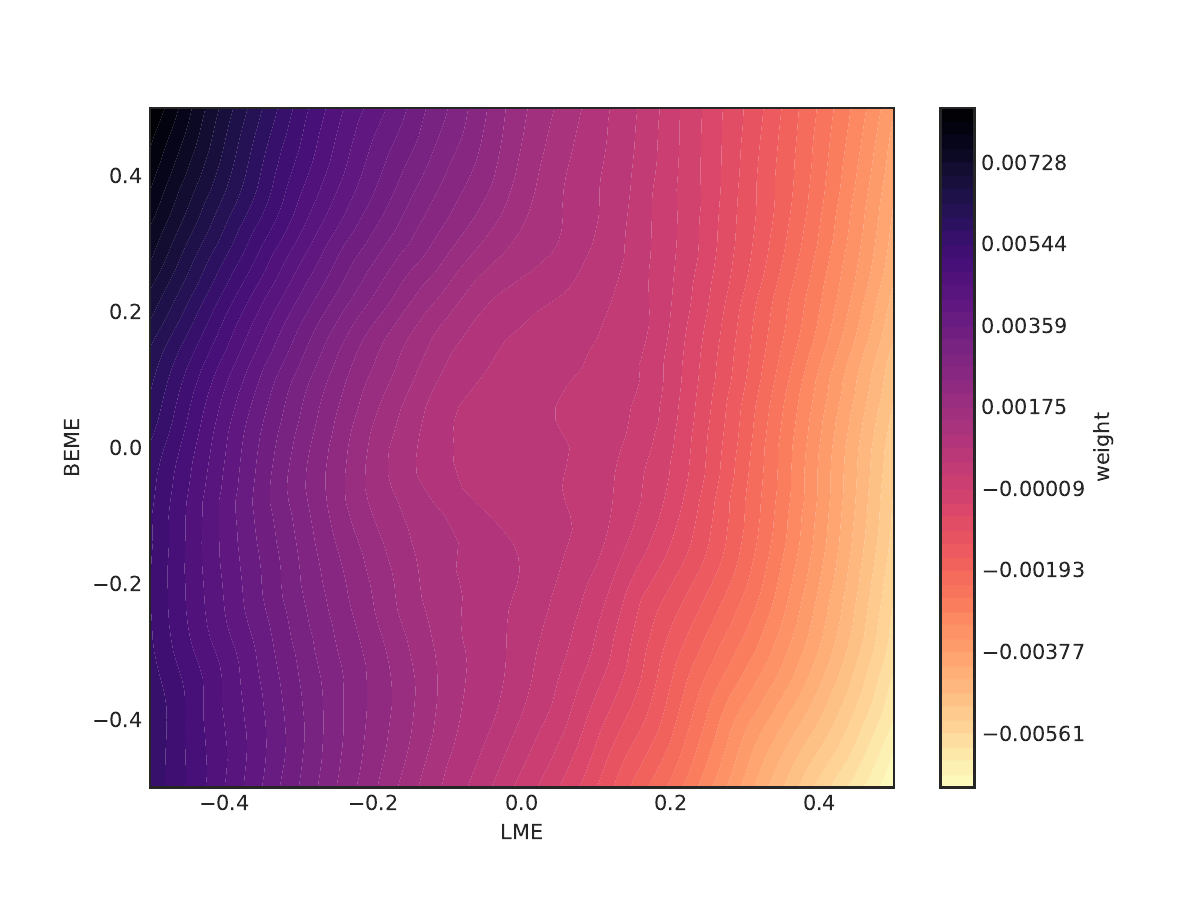}
    \subcap{Interaction between Size (LME) and Book to Market Ratio (BEME)}
  \end{subfigure}
\bnotefig{These figures show the variable importance and functional form of the SDF estimated on a rolling window of 240 months. The sensitivities and SDF weights $\omega$ are the average over those rolling window estimates.}
\end{figure}

%
%
%

\end{document}